\documentclass[11pt,journal, onecolumn]{IEEEtran}
\makeatletter
\long\def\@makecaption#1#2{\ifx\@captype\@IEEEtablestring%
\footnotesize\begin{center}{\normalfont\footnotesize #1}\\
{\normalfont\footnotesize\scshape #2}\end{center}%
\@IEEEtablecaptionsepspace
\else
\@IEEEfigurecaptionsepspace
\setbox\@tempboxa\hbox{\normalfont\footnotesize {#1.}~~ #2}%
\ifdim \wd\@tempboxa >\hsize%
\setbox\@tempboxa\hbox{\normalfont\footnotesize {#1.}~~ }%
\parbox[t]{\hsize}{\normalfont\footnotesize \noindent\unhbox\@tempboxa#2}%
\else
\hbox to\hsize{\normalfont\footnotesize\hfil\box\@tempboxa\hfil}\fi\fi}
\makeatother

\usepackage{amssymb,latexsym,amsfonts}
\usepackage{graphicx}
\usepackage{verbatim}
\usepackage{bm}	
\usepackage[cmex10]{amsmath}
\usepackage{subeqnarray}
\usepackage{comment}
\usepackage{amsthm}

\newtheorem{defn}{Definition}
\newtheorem{thm}{Theorem}
\newtheorem{cor}{Corollary}
\newtheorem{lem}{Lemma}
\newtheorem{rem}{Remark}

\begin{document}
\def \x {\mathbf{x}}
\def \X {\mathbf{X}}
\def \y {\mathbf{y}}
\def \Y {\mathbf{Y}}
\def \W {\mathbf{W}}
\def \k {\mathbf{k}}
\def \fx {f_{\scriptscriptstyle X}}
\def \fy {f_{\scriptscriptstyle Y}}
\def \fw {f_{\scriptscriptstyle W}}
\def \hx {h_{\scriptscriptstyle X}}
\def \hy {h_{\scriptscriptstyle Y}}
\def \hw {h_{\scriptscriptstyle W}}
\def \subx {\scriptscriptstyle X}
\def \suby {\scriptscriptstyle Y}
\def \subw {\scriptscriptstyle W}
\def \subg {\scriptscriptstyle G}
\def \subt {\scriptscriptstyle T}
\def \fxstar {f_{\scriptscriptstyle X^*}}
\def \fystar {f_{\scriptscriptstyle Y^*}}
\def \fone {f_{\scriptscriptstyle 1}}
\def \ftwo {f_{\scriptscriptstyle 2}}
\def \fxhat {f_{\scriptscriptstyle\hat{X}}}
\def \fyhat {f_{\scriptscriptstyle\hat{Y}}}
\title{Corrected Version of `A Unifying Variational Perspective on Some Fundamental Information Theoretic Inequalities'}
\author{Sangwoo~Park,~Erchin~Serpedin, and Khalid Qaraqe
\thanks{S. Park and E. Serpedin are with the Department
of Electrical and Computer Engineering, Texas A\&M University, College Station,
 USA. email: serpedin@ece.tamu.edu}
\thanks{K. Qaraqe is with Texas A\&M University at Qatar, Doha, Qatar.}
\thanks{This represents a corrected version of the paper ``A Unifying Variational Perspective on Some Fundamental Information Theoretic Inequalities", IEEE Transactions on Information Theory, Vol. 59, no. 11, Nov. 2013.  This work was supported by NSF Grant: CCF-1318338. A small part of this paper was presented at ISIT 2013.}
\thanks{Copyright (c) 2011 IEEE. Personal use of this material is permitted.  However, permission to use this material for any other purposes must be obtained from the IEEE by sending a request to pubs-permissions@ieee.org.} }

\maketitle

\begin{abstract}
This paper proposes  a unifying variational approach for proving  some fundamental information theoretic inequalities.  Fundamental information theory results  such as maximization of differential entropy, minimization of  Fisher information (Cram\'{e}r-Rao inequality), worst additive noise lemma, %entropy power inequality (EPI),
and extremal entropy inequality (EEI) are interpreted as  functional problems and proved within the framework of calculus of variations. Several applications and possible extensions  of the proposed results are briefly mentioned.
\end{abstract}

\IEEEpeerreviewmaketitle

\section{Introduction}\label{INT}
\IEEEPARstart{I}{n} the information theory realm, it is well-known that given the second-order moment (or variance), a Gaussian density function maximizes the differential entropy. Similarly, given the second-order moment, the Gaussian density function minimizes  the Fisher information, a result which is referred to as the Cram\'{e}r-Rao inequality in the signal processing literature. Surprisingly, the proofs proposed in literature for  these fundamental results are  quite diverse, and no unifying feature exists. Since differential entropy or Fisher information is a functional with respect to a probability density function, the most natural way to establish these results is by approaching them from  the perspective of functional analysis. This paper presents a unifying variational framework to address these results as well as numerous other fundamental information theoretic results. A challenging information theoretic inequality, referred to as the extremal entropy inequality (EEI) \cite{Extremal:Liu}, can be dealt with successfully in the proposed functional framework. Furthermore,  the proposed variational calculus perspective is useful in establishing  other  novel results, applications and  extensions of  the existing information theoretic inequalities.

The main theme of this paper is to illustrate how some  tools from calculus of variations can be used successfully to prove some of the fundamental information theoretic inequalities, which have been widely used in information theory and other fields, and to establish some applications. The proposed variational approach provides alternative proofs for some of the fundamental information theoretic inequalities and enables finding novel extensions of the existing results. This
statement is strengthened  by the fact that the proposed variational framework is quite general and powerful, and it allows easy integration of various
 linear and inequality constraints into the functional that is to be optimized. Therefore, we believe that  a large number of applications could benefit of these tools.  The proposed variational approach offers also a potential  guideline for finding the optimal solution for many  open problems. %In addition, as a general feature,  the proposed variational approach enables to find simpler proofs for some quite challenging results such as EPI and EEI.

Variational calculus techniques have been used with great success in solving important problems in image processing and computer vision \cite{image} such as image reconstruction (denoising, deblurring), inverse problems, and image segmentation. Recently, variational techniques were also advocated for optimization of multiuser communication systems \cite{palomar0},  for deriving analytical wireless channel models using the maximum entropy principle when only limited information about the environment is available \cite{debbah1}, and for designing optimal training sequences for radar and sonar applications \cite{radar1}-\cite{radar2}.  Maximum entropy principle found also applications in spectral estimation (e.g.,  Burg's maximum entropy spectral density estimator \cite{inf:cover}) and Bayesian statistics  \cite{jaynes}.

The major results of this paper are enumerated as follows. First, using calculus of variations, the maximizing differential entropy and minimizing Fisher information theorems are proved under the classical (standard) assumptions found in the literature as well as under a different set of assumptions. It is shown that a Gaussian density function maximizes the differential entropy but it minimizes the Fisher information, given the second-order moment.  It is also shown that  a half normal density function maximizes the differential entropy over the set of non-negative random variables, given the second-order moment.  Furthermore, it is shown that  a half normal density function minimizes the Fisher information over the set of  non-negative random variables, provided that the regularity condition\footnote{The regularity condition is defined in Theorems \ref{EF_thm5} and \ref{EF_thm6}.}  is ignored and the second-order moment is given.  It is also shown that  a chi  density function minimizes the Fisher information over the set of non-negative random variables, under the assumption that  the regularity condition  holds and the second-order moment is given.

Second, a novel proof of the worst additive noise lemma \cite{WANoise:Cover} is provided in the proposed functional framework.  Previous proofs of the worst additive noise lemma were based on Jensen's inequality or data processing inequality \cite{WANoise:Cover}, \cite{EPI:Rioul}.  Unlike the previous proofs, our approach is purely based on calculus of variations techniques, and the vector version of the lemma is treated.

%Third, EPI is proved based on calculus of variations. We first re-cast EPI as a functional problem. Then, the necessary optimal solutions for the functional problem are found using Euler's equation and by exploiting  the necessary conditions for the existence of the optimal solution for the considered  functional. In the scalar version of EPI, the necessarily optimal solution, which is the Gaussian density function, is actually sufficient since only the Gaussian density function satisfies the Euler's equation. This is one of the main benefits using calculus of variations since it allows finding global optimal solutions simply by checking the set of solutions imposed by the set of necessary conditions.  In the vector version of EPI, Euler's equation only shows that the Gaussian density functions are necessarily optimal, since the covariance matrices of the optimal solutions are not determined. However, this information alone--i.e., the fact that the  optimal solutions are Gaussian--is enough to prove EPI.

%Fourth,
 Third, EEI is studied from the perspective of a functional problem. The main advantage of the  proposed new proof is that neither the channel enhancement technique and the entropy power inequality (EPI),  adopted in \cite{Extremal:Liu}, nor the equality condition of data processing inequality and the technique based on  the moment generating functions,  used in \cite{Extrem:SPark}, are required. Using a technique based on calculus of variations, an alternative proof of  EEI is provided. Finally, several applications and extensions of the proposed results are discussed.

The rest of this paper is organized as follows. Some variational calculus preliminary results and their corollaries  are first reviewed in Section \ref{CAL}. Maximizing differential entropy theorem and minimizing Fisher information  theorem (Cram\'{e}r-Rao inequality) are proved in Section \ref{EF}.  In Section \ref{WA}, the worst additive noise lemma is introduced and proved based on variational arguments. %EPI and EEI are proved in Sections \ref{EPI} and \ref{EI}, respectively.
EEI is proved in Section \ref{EI}. In Section \ref{APP}, some additional applications of the proposed variational techniques are briefly mentioned. Finally, Section \ref{CON} concludes this paper.

\section{Some Preliminary Calculus of Variations Results}\label{CAL}

In this section, we will review some of the fundamental results from variational calculus, and establish the concepts, notations and results that
will be used constantly throughout the rest of the paper. These results are standard and therefore will be described briefly without
further details. Additional details can be found in calculus of variations books such as \cite{CalVariations:Gelfand}-\cite{Cal:Hans}.

% definition of a functional (def1)
\begin{defn}\label{CAL_def1}
A functional $U[f]$ might be defined as
\begin{eqnarray}
\label{CAL_eq1_1}
U[f] = \int_{a}^{b} K(x,f(x), f'(x)) dx,
\end{eqnarray}
which is defined on the set of continuous functions $(f(x))$ with continuous first-order derivatives $(f'(x)=df(x)/dx)$ on the interval $[a, \; b]$.  The function $f(x)$ is assumed to satisfy the boundary conditions $f(a) = A$ and $f(b) = B$. The functional $K(\cdot, \cdot, \cdot)$ is also assumed to have continuous first-order and second-order (partial) derivatives with respect to (wrt)  all of its arguments.
% Also, notation $f'(x)$ denotes the first-order derivative wrt $x$.
\end{defn}

\begin{defn}\label{CAL_def2}
The increment of a functional $U[f]$ is defined as
\begin{eqnarray}
\label{CAL_eq2_1}
\Delta U[t] = U[f + t] - U[f],
\end{eqnarray}
where the function $t(x)$, that satisfies the boundary conditions $t(a)=t(b)=0$, represents the admissible increment of $f(x)$, and it is assumed independent of the function $f(x)$ and twice differentiable.
\end{defn}

\begin{defn}\label{CAL_def3}
Suppose that given $f(x)$, the increment in (\ref{CAL_eq2_1}) is expressed as
\begin{eqnarray}
\label{CAL_eq3_1}
\Delta U\left[t\right] = \varphi\left[t\right] + \epsilon \|t\|,
\end{eqnarray}
where $\varphi\left[t\right]$ is a linear functional, $\epsilon$ goes to zero as $\|t\|$ approaches zero, and $\|\cdot\|$ denotes a norm  defined in the case of a function $f(x)$ as:
\begin{eqnarray}
\label{CAL_eq4_1}
\|f \| = \sum_{i=0}^n \max_{a\leq x \leq b} \left|f^{(i)}(x)\right|,
\end{eqnarray}
where $f^{(i)}(x) = d^i  f(x)/dx^{i}$ are assumed to exist and be continuous for $i=0,\ldots,n$ on the interval $[a, \; b]$,  and the summation upper index $n$ might vary depending on the normed linear space considered (e.g., if the normed linear space consists of all continuous functions $f (x)$, which have continuous first-order derivative on the interval $[a,b]$, $\|f \| = \max_{a\leq x \leq b} \left|f (x)\right| + \max_{a\leq x \leq b} \left|f' (x)\right|$, and in this case $n=1$; see e.g., \cite{CalVariations:Gelfand} for further details). Under the above assumptions, the functional $U\left[f\right]$ is said to be differentiable, and the major part of the increment $\varphi\left[t\right]$ is called the (first-order) variation of the functional $U\left[f\right]$ and it is expressed as $\delta U\left[f\right]$.
\end{defn}

Based on Definitions \ref{CAL_def1}, \ref{CAL_def2}, \ref{CAL_def3} and Taylor's theorem (see e.g., \cite{CalVariations:Gelfand}-\cite{Cal:Hans} for additional justifications), the first-order and the second-order variations of a functional $U\left[f\right]$ can be  expressed as
\begin{eqnarray}
\label{CAL_eq5_1}
\delta U\left[f\right] \hspace*{-3mm}& = & \hspace*{-4mm}\int \hspace*{-1mm} \left[K'_{f }\left(x,f ,f' \right) t (x)+ K'_{f' }\left(x, f , f' \right) t' (x)\right] dx\\
\label{CAL_eq5_2}
\delta^2 U\left[f \right] \hspace*{-3mm}& = & \hspace*{-3mm}\frac{1}{2} \int \Big[K''_{f  f }\left(x,f ,f' \right) {t (x)}^2 + 2
K''_{f  f' }\left(x,f ,f' \right) t (x) t' (x) + K''_{f'  f' } \left(x,f ,f' \right){t' (x)}^{2} \Big] dx\nonumber\\
\hspace*{-3mm}& = & \hspace*{-3mm}\frac{1}{2} \int \left[K''_{f'  f' } {t' }^{2}  + \left( K''_{f  f }  - \frac{d}{dx} K''_{f  f' } \right) {t }^2 \right] dx,
\end{eqnarray}
where $K'_{f }$ and $K'_{f' }$ stand  for the first-order partial derivatives wrt $f $ and $f' $, respectively, $K''_{f  f' }$ denotes  the second-order partial derivative   wrt $f $ and $f' $, $K''_{f  f }$ represents the second-order partial derivative wrt $f $, and $K''_{f'  f' }$ is the second-order partial derivative    wrt $f' $. Throughout the paper to simplify the exposition,  the arguments of functionals or functions are omitted unless the arguments are ambiguous or confusing. Also, the range of integration in various integrals will not be explicitly marked   unless the range is ambiguous.

% first variation is zero (Extremum) (thm1)
\begin{thm}[\cite{CalVariations:Gelfand}]\label{CAL_thm1}
A necessary condition for the functional $U[f ]$ in (\ref{CAL_eq1_1}) to have an extremum (or local optimum) for a given function $f=f^*$ is that its first variation vanishes at $f=f^*$:
\begin{eqnarray}
\label{CAL_eq6_1}
\delta U[f^*] = 0,
\end{eqnarray}
for all admissible  increments.%\footnote{The increments that satisfy the constraints of a given functional will be referred to as admissible.}, which are assumed to be twice differentiable and to satisfy the boundary conditions.
This implies
\begin{eqnarray}
\label{CAL_eq7_1}
K'_{f^*} - \frac{d}{dx}K'_{f'^{*}} = 0,
\end{eqnarray}
a result which is known as Euler's equation. When the functional in (\ref{CAL_eq1_1}) includes multiple functions (e.g., $f_{1}, \ldots, f_{m}$) and multiple integrals wrt $x_1, \ldots, x_n$, i.e.,
\begin{eqnarray}
\int \cdots \int K\left(x_1, \ldots, x_n,f_1, \ldots, f_m ,f'_1, \ldots, f'_m \right) dx_1 \cdots dx_n,\nonumber
\end{eqnarray}
then Euler's equation in (\ref{CAL_eq7_1}) takes the form of the system of equations:
\begin{eqnarray}
\label{CAL_eq7_4}
K'_{f^*_{i}} - \sum_{j=1}^n \frac{d}{dx_j}K'_{f'^*_{i}} = 0,\hspace{10mm} i=1,\ldots,m.
\end{eqnarray}
In particular, when the functional  does not depend on the first-order derivative of the functions $f_{1}, \ldots, f_{m}$, the equations in (\ref{CAL_eq7_4}) reduce  to
\begin{eqnarray}
\label{CAL_eq7_6}
 K'_{f^*_{i}} = 0,\hspace{10mm} i=1,\ldots,m.
\end{eqnarray}
\begin{proof}
Details of the proof of this theorem can be found, e.g., in \cite{CalVariations:Gelfand}-\cite{Cal:Hans}.
\end{proof}
\end{thm}

% second variation is non-negative (Maximum) (thm2)
\begin{thm}[\cite{CalVariations:Gelfand}]\label{CAL_thm2}
A necessary condition for the functional $U[f]$ in (\ref{CAL_eq1_1}) to have a minimum for a given $f=f^*$ is that the second variation of functional $U[f]$ be nonnegative:
\begin{eqnarray}
\label{CAL_eq8_1}
\delta^2 U[f^*] \geq 0,
\end{eqnarray}
for all admissible  increments. This implies
\begin{eqnarray}
\label{CAL_eq8_3}
K''_{f'^* f'^*} \geq 0.
\end{eqnarray}
In particular, when the functional in (\ref{CAL_eq1_1}) does not depend on the first-order derivative of the function $f$,  (\ref{CAL_eq8_3}) simplifies to
\begin{eqnarray}
\label{CAL_eq8_5}
K''_{f^* f^*} \geq 0.
\end{eqnarray}
When the functional in (\ref{CAL_eq1_1}) includes multiple functions (e.g., $f_{1}, \ldots, f_{m}$) and multiple integrals wrt $x_1, \ldots, x_n$, i.e.,
$$\int \cdots \int K\left(x_1, \ldots, x_n,f_1, \ldots, f_m \right) dx_1 \cdots dx_n,$$
 then the condition in (\ref{CAL_eq8_5}) is expressed in terms of  the positive semi-definiteness of the matrix:
\begin{eqnarray}
\label{CAL_eq8_7}
\left[\begin{array}{ccc}
K''_{f_{1} f_{1}} & \cdots & K''_{f_{1} f_{m}}\\
\vdots & \ddots & \vdots\\
K''_{f_{m} f_{1}} & \cdots & K''_{f_{m} f_{m}}\\
\end{array}\right]  \succeq  0 .
\end{eqnarray}
\begin{proof}
The inequality in (\ref{CAL_eq8_5}) is easily derived from the inequality in (\ref{CAL_eq8_3}) since $K''_{f'_{\scriptscriptstyle X} f'_{\scriptscriptstyle X}}$ and $K''_{f_{\scriptscriptstyle X} f'_{\scriptscriptstyle X}}$ are vanishing in (\ref{CAL_eq5_2}) when the functional in (\ref{CAL_eq1_1}) does not depend on the first-order derivative of the function $f_{\scriptscriptstyle X}$. The remaining details of the proof can be tracked  in  \cite{CalVariations:Gelfand}.
\end{proof}
\end{thm}

% With constraints in the form of integral (thm3)
\begin{thm}[\cite{CalVariations:Gelfand}]\label{CAL_thm3}
Given the functional
\begin{eqnarray}
\label{CAL_eq9_1}
U[f_{1}, f_{2}] = \int_{a}^{b} K(x,f_{1}, f_{2},f'_{1}, f'_{2}) dx,
\end{eqnarray}
assume that the admissible functions satisfy the following boundary conditions:
\begin{eqnarray}
\label{CAL_eq10_1}
\hspace*{-3mm} &&f_{1} (a) = A_{1}, \; f_{1} (b) = B_{1}, \; f_{2} (a) = A_{2}, \; f_{2} (b) = B_{2}, \nonumber\\
\label{CAL_eq10_2}
\hspace*{-3mm} && {k}(x,f_{1} ,f_{2} )  = 0,\\
\hspace*{-3mm} && L[f_{1} , f_{2} ] = \int_{a}^{b} \tilde{L}(x,f_{1} , f_{2} , f'_{1} , f'_{2} ) dx = l,
\end{eqnarray}
where $a$, $b$, $A_{1} $, $B_{1} $, $A_{2} $, $B_{2} $, and $l$ are constants, ${k}(x,f_{1} ,f_{2} )$ is a functional wrt $f_1$ and $f_2$, and $U[f_{1} ,f_{2} ]$ is assumed to have an extremum for $f_{1} =f^*_{1}$ and $f_{2} =f^*_{2}$.

If $f^*_{1}$ and $f^*_{2}$ are not extremals of $L[f_{1} , f_{2} ]$, or $k'_{\scriptscriptstyle f^*_{1}}$ and $k'_{\scriptscriptstyle f^*_{2}}$ do not vanish simultaneously at any point in (\ref{CAL_eq10_2}), there exist a constant $\lambda$ and  a function $\lambda(x)$ such that $f^*_{1}$ and $f^*_{2}$ are extremals of the functional
\begin{eqnarray}
\int_{a}^{b} ( K(x,f_{1} , f_{2} , f'_{1} , f'_{2} ) + \lambda \tilde{L}(x,f_{1} , f_{2} , f'_{1} , f'_{2} ) + \lambda(x) k(x,f_{1} , f_{2} ) ) dx.  \label{CAL_eq11_1}
\end{eqnarray}
\end{thm}

Based on Theorem \ref{CAL_thm3}, the following corollary is derived.
% With constraints & multiple integral (thm5)
\begin{cor}\label{CAL_cor1}
Given the functional
\begin{eqnarray}
\label{CAL_eq12_1}
U[\fx , \fy ] = \int_{a}^{b} \int_{a}^{b} K(x, y, \fx , \fy ) dxdy,
\end{eqnarray}
assume  that  the admissible functions satisfy the following boundary conditions:
\begin{eqnarray}
\hspace*{-3mm}&& \hspace*{-3mm}\fx(a) = A_{\subx}, \; \fx(b) = B_{\subx}, \; \fy(a) = A_{\suby}, \; \fy(b) = B_{\suby}, \nonumber\\
\hspace*{-3mm}&&\hspace*{-3mm} k(y,\fx ,\fy )=g(y, \fy ) -\int_{a}^{b}\tilde{k}(x,y,\fx ) dx = 0,\nonumber\\
\hspace*{-3mm}&& \hspace*{-3mm} L_i[\fx , \fy ] = \int_{a}^{b} \int_{a}^{b} \tilde{L}_i(x,y,\fx , \fy ) dx dy= l_i,~~ i=1,2,\cdots,n,   \label{CAL_eq13_1}
\end{eqnarray}
where $a$, $b$, $A_{\subx} $, $B_{\subx} $, $A_{\suby} $, and $B_{\suby}$ stand for some constants,  $\fx $ is a function of $x$, $\fy $ is a function of $y$,  $g(y, \fy )$ is a function of $\fy$,  and $\tilde{k}(x,y,\fx )$ is a function of $\fx$. The functional $U[\fx , \fy ]$ is assumed to have an extremum at $\fx =\fxstar$ and $\fy  = \fystar$.

Unless  $\fxstar$ and $\fystar$ are extremals of $L_i[\fx , \fy ]$, or $k'_{\fxstar }$ and $k'_{\fystar }$ simultaneously vanish at any point of $k(y,\fx ,\fy )$, there exist constants   $\lambda_i, i=1,2,\cdots,n,$ and a function $\lambda(y)$ such that $\fx  = \fxstar$ and $\fy  = \fystar$ is an extremal of the functional
\begin{eqnarray}
\int_{a}^{b} \Big\{ \Big[\int_{a}^{b}  (K(x,y,\fx , \fy )+ \sum\limits_{i=1}^n \lambda_i \tilde{L}_i(x,y, \fx , \fy )
 - \lambda(y) \tilde{k}(x,y, \fx ))dx \Big] +\lambda(y)g(y,\fy) \Big\}dy. \label{CAL_eq15_1}
\end{eqnarray}
\begin{proof}
See Appendix \ref{sec:appendix_a}.
\end{proof}
\end{cor}

Based on Theorems \ref{CAL_thm1}, \ref{CAL_thm2} and Corollary \ref{CAL_cor1}, we can derive the following corollary, which will be repeatedly used throughout this paper.
\begin{cor}\label{CAL_cor2}
Based on the functional defined in (\ref{CAL_eq15_1}), the following necessary conditions are derived for the optimal solutions $\fxstar$ and $\fystar$:
\begin{eqnarray}\label{CAL_eq16_1}
\int  K'_{\fxstar}(x,y,\fxstar , \fystar ) + \sum\limits_{i=1}^n \lambda_i \tilde{L_i}'_{\fxstar }(x,y, \fxstar , \fystar )  -\lambda(y) \tilde{k}'_{\fxstar}(x,y, \fxstar ) dy \hspace*{-2mm}&=&\hspace*{-2mm} 0,  \\
\label{CAL_eq16_2}
\int K'_{\fystar }(x,y,\fxstar,\fystar) + \sum\limits_{i=1}^n \lambda_i \tilde{L_i}'_{\fystar }(x,y, \fxstar , \fystar )dx
 + \lambda(y)g'_{\fystar }(y,\fystar ) \hspace*{-2mm}&=&\hspace*{-2mm} 0,
\end{eqnarray}
and the matrix
\begin{eqnarray}
\label{CAL_eq17_1}
\left[\begin{array}{cc}
G''_{\fxstar  \fxstar } & G''_{\fxstar  \fystar }\\
G''_{\fystar  \fxstar } & G''_{\fystar  \fystar }
\end{array}\right],
\end{eqnarray}
is positive semi-definite. The functional $G$ is defined as
\begin{eqnarray}
G(x,y,\fxstar , \fystar ) = K(x,y,\fxstar , \fystar ) + \sum\limits_{i=1}^N \lambda_i \tilde{L}_i(x,y, \fxstar , \fystar )
-\lambda(y)\tilde{k}(x,y, \fxstar )+ \lambda(y)g(y,\fystar ) q(x),\nonumber
\end{eqnarray}
and $q(x)$ is a (arbitrary but fixed) function which satisfies $\int_{a}^b q(x) dx =1$, and it is introduced to homogenize the functional in (\ref{CAL_eq15_1}). In particular, if function $g(y,\fy)$ only involves first order component of $\fy$, i.e., $g(y,\fy)=\fy$, the necessary condition reduces to check the positive semi-definiteness of the matrix
\begin{eqnarray}
\notag
\left[\begin{array}{cc}
H''_{\fxstar  \fxstar } & H''_{\fxstar  \fystar }\\
H''_{\fystar  \fxstar } & H''_{\fystar  \fystar }
\end{array}\right],
\end{eqnarray}
where 
\begin{eqnarray}
H(x,y,\fxstar , \fystar ) = K(x,y,\fxstar , \fystar ) + \sum\limits_{i=1}^N \lambda_i \tilde{L}_i(x,y, \fxstar , \fystar )
-\lambda(y)\tilde{k}(x,y, \fxstar ).\nonumber
\end{eqnarray}
\begin{proof}
See Appendix \ref{sec:appendix_a}.
%The equations in (\ref{CAL_eq16_1}) and (\ref{CAL_eq16_2}) are derived from the first-order variation condition in Theorem \ref{CAL_thm1}.  The equations  in (\ref{CAL_eq16_1}) and (\ref{CAL_eq16_2}) are Euler's equations for multiple integrals. The positive semi-definiteness of the matrix in (\ref{CAL_eq17_1}) is derived from the second-order variation condition in Theorem \ref{CAL_thm2}.  This condition is the same as the one  in (\ref{CAL_eq8_7}). %Since the proof is straightforward,
%The details of the proof are omitted here.
\end{proof}
\end{cor}
%%%%%

%%%%%

% Maximize entropy, minimize Fisher Information (thm6)
\section{MAX Entropy and MIN Fisher Information}\label{EF}
This simple but significant result--given the second-order moment (or  variance) of a random vector, a Gaussian random vector maximizes the differential entropy--is well-known. %However, its complete rigorous proof can hardly be found.
In this section,  a completely rigorous and general derivation of the distribution achieving the maximum entropy will be first provided.  This proof sets up the variational framework for establishing a second important result in this section, namely the Cram\'{e}r-Rao bound, which states that for a given mean and correlation matrix,  a normally distributed random vector  minimizes the Fisher information matrix.

%Similar to Theorem \ref{EF_thm1}, given a correlation matrix (or a covariance matrix), a multi-variate Gaussian density function maximizes the differential entropy as shown by the following theorem.
% Maximizing entropy (Vector)
\begin{thm}[\cite{inf:cover}, \cite{EPI:Rioul}]\label{EF_thm2}
Given (a vector mean $\boldsymbol{\mu}_{\scriptscriptstyle X}$ and)  a correlation matrix $\boldsymbol{\Omega}_{\scriptscriptstyle X}$, a Gaussian random vector $\mathbf{X}_{\scriptscriptstyle G}$ with the correlation matrix $\boldsymbol{\Omega}_{\scriptscriptstyle X}$ (and the vector mean $\boldsymbol{\mu}_{\scriptscriptstyle X}$) maximizes the differential entropy, i.e.,
\begin{eqnarray}
\label{EF_eq8_1}
h(\mathbf{X}) & \leq & h(\mathbf{X}_{\scriptscriptstyle G}),
\end{eqnarray}
where $h(\cdot)$ denotes differential entropy, $\mathbf{X}$ is an arbitrary (but fixed)  random vector with the correlation matrix $\boldsymbol{\Omega}_{\scriptscriptstyle X}$. % and $\mathbf{X}_{\scriptscriptstyle G}$ is a Gaussian random vector whose correlation matrix is identical to % the one of $\mathbf{X}$.
\begin{proof}
We first construct a functional, which represents the inequality in (\ref{EF_eq8_1}) and required constraints, as follows:
\begin{eqnarray}
\label{EF_eq30_1}
&&\min_{f_{\scriptscriptstyle X}} \quad \int f_{\scriptscriptstyle X}(\mathbf{x}) \log f_{\scriptscriptstyle X}(\mathbf{x}) d\mathbf{x},\\
\label{EF_eq30_2}
&&\text{s. t.}\quad \int f_{\scriptscriptstyle X}(\mathbf{x}) d\mathbf{x} =1,\\
\label{EF_eq30_add}
&&\hspace{10mm} \int \x f_{\scriptscriptstyle X}(\mathbf{x}) d\mathbf{x}=\boldsymbol{\mu}_{\scriptscriptstyle X}\\
\label{EF_eq30_3}
&&\hspace{10mm} \int \mathbf{x}\mathbf{x}^{\scriptscriptstyle T} f_{\scriptscriptstyle X}(\mathbf{x}) d\mathbf{x}=\boldsymbol{\Omega}_{\scriptscriptstyle X}.
\end{eqnarray}
Using Theorem \ref{CAL_thm3}, the functional  in (\ref{EF_eq30_1}) is expressed as
\begin{eqnarray}
\label{EF_eq31_1}
\min_{f_{\scriptscriptstyle X}} \quad U[f_{\scriptscriptstyle X}],
\end{eqnarray}
where $U[f_{\scriptscriptstyle X}]=$ $\int K(\mathbf{x},f_{\scriptscriptstyle X}) d\mathbf{x} =$ $ \int f_{\scriptscriptstyle X}(\mathbf{x})$ $ \left(\log f_{\scriptscriptstyle X}(\mathbf{x}) +\alpha + \sum_{i=1}^n \zeta_i x_i + \sum_{i=1}^n \sum_{j=1}^n \lambda_{ij} x_i x_j \right)d\mathbf{x}$,  $\alpha$ is the Lagrange multiplier
associated with the constraint (\ref{EF_eq30_2}), and $\zeta_i$ and $\lambda_{ij}$ stand for the  Lagrange multipliers corresponding to  the constraints (\ref{EF_eq30_add}) and (\ref{EF_eq30_3}), respectively.

Based on Theorem \ref{CAL_thm1}, by checking the first-order variation condition, we can find the optimal solution $f_{\scriptscriptstyle X^*}(\mathbf{x})$ as follows:
\begin{equation}
\label{EF_eq32_1}
K'_{f_{\scriptscriptstyle X}} \Big|_{f_{\scriptscriptstyle X}=f_{\scriptscriptstyle X^*}} = 1+\log f_{\scriptscriptstyle X^*}(\mathbf{x}) + \alpha + \boldsymbol{\zeta}^T\x + \mathbf{x}^{\scriptscriptstyle T} \boldsymbol{\Lambda}\mathbf{x}=0
\end{equation}
with $\boldsymbol{\zeta} = [\zeta_1, \cdots, \zeta_n]^T$ and the matrix $\boldsymbol{\Lambda} =[\lambda_{ij}]$, $i,j=1,\ldots,n$.
%\begin{eqnarray}
%\boldsymbol{\Lambda} = \left[ \begin{array}{cccc}
%\lambda_{11} & \lambda_{12} & \cdots & \lambda_{1n}\\
%\vdots & \ddots & \vdots & \vdots\\
%\lambda_{n1} & \lambda_{n2} & \cdots & \lambda_{nn}
%\end{array}\right].\nonumber
%\end{eqnarray}
Considering the constraints in (\ref{EF_eq30_2}) - (\ref{EF_eq30_3}), from (\ref{EF_eq32_1}) it turns out that
\begin{eqnarray}
\label{EF_eq33_1}
f_{\scriptscriptstyle X^*}(\mathbf{x}) \hspace*{-2mm}& = &\hspace*{-2mm} \exp\left\{-\mathbf{x}^{\scriptscriptstyle T} \boldsymbol{\Lambda}\mathbf{x} - \boldsymbol{\zeta}^T\x -\alpha -1\right\} \nonumber\\
\hspace*{-2mm}& = &\hspace*{-2mm}\left(2\pi\right)^{-\frac{n}{2}}\left|\frac{1}{2}\boldsymbol{\Lambda}^{-1}\right|^{-\frac{1}{2}}
\exp\left\{-\frac{1}{2}( \mathbf{x} - \boldsymbol{\mu})^T \left(\frac{1}{2}\boldsymbol{\Lambda}^{-1}\right)^{-1}( \mathbf{x} - \boldsymbol{\mu}) \right\}  \left(2\pi\right)^{\frac{n}{2}}\left|\frac{1}{2}\boldsymbol{\Lambda}^{-1}
\right|^{\frac{1}{2}}\exp\left\{-1-\alpha + \boldsymbol{\mu}^T \mathbf{\Lambda} \boldsymbol{\mu}\right\}\nonumber\\
\hspace*{-2mm}& = &\hspace*{-2mm}\left(2\pi\right)^{-\frac{n}{2}}\left|\boldsymbol{\Omega}_{\scriptscriptstyle X}\right|^{-\frac{1}{2}}\exp\left\{-\frac{1}{2}( \mathbf{x} - \boldsymbol{\mu})^T \boldsymbol{\Omega}_{\scriptscriptstyle X}^{-1}(\mathbf{x}-\boldsymbol{\mu})\right\},
\end{eqnarray}
where
\begin{eqnarray}
\label{EF_eq34_1}
\alpha & = & -1 + \boldsymbol{\mu}^T \mathbf{\Lambda} \boldsymbol{\mu} + \frac{1}{2}\log \left(2\pi\right)^n \left|\boldsymbol{\Omega}_{\scriptscriptstyle X}\right|,\nonumber\\
\boldsymbol{\Lambda} & = & \frac{1}{2}\boldsymbol{\Omega}_{\scriptscriptstyle X}^{-1}, \nonumber\\
\boldsymbol{\zeta} & = & -2\mathbf{\Lambda}\boldsymbol{\mu}.
\end{eqnarray}

Two remarks are  now in order. First, the correlation matrix $\boldsymbol{\Omega}_{\scriptscriptstyle X}$ is assumed to be invertible. When the correlation matrix is non-invertible, similar to the method shown in \cite{Extremal:Liu}, we can equivalently re-write the functional problem in (\ref{EF_eq30_1}) as
\begin{eqnarray}\label{EF_add_eq1}
&& \min_{f_{\scriptscriptstyle X} (\mathbf{x})} -h(\mathbf{X}) \quad \Leftrightarrow \quad \min_{f_{\scriptscriptstyle \bar{X}} (\mathbf{x})} -h(\bar{\mathbf{X}}),
\end{eqnarray}
where $\mathbf{X}=\mathbf{Q}_{\scriptscriptstyle \Omega} \bar{\mathbf{X}}$,  and in the spectral factorization $\boldsymbol{\Omega}_{\scriptscriptstyle X} =\mathbf{Q}_{\scriptscriptstyle \Omega} \boldsymbol{\Lambda}_{\scriptscriptstyle \Omega} \mathbf{Q}^{\scriptscriptstyle T}_{\scriptscriptstyle \Omega}$, $\mathbf{Q}_{\scriptscriptstyle \Omega}$ is an orthogonal matrix, $\boldsymbol{\Lambda}_{\scriptscriptstyle \Omega} = diag\left(\Lambda_1, \ldots, \right.$ $\left. \Lambda_m, 0, \ldots, 0 \right)$  and $diag$ denotes a diagonal matrix.

Let $\bar{\mathbf{X}}=\left[\bar{\mathbf{X}}_{a}^{\scriptscriptstyle T}, \bar{\mathbf{X}}_{b}^{\scriptscriptstyle T} \right]^{\scriptscriptstyle T}$, where the dimensions of $\bar{\mathbf{X}}_{a}$ and $\bar{\mathbf{X}}_{b}$ are $m$ and $n-m$, respectively. It can be observed that the correlation matrix (or covariance matrix) of $\bar{\mathbf{X}}$, $\boldsymbol{\Omega}_{\bar{X}}$, is equal to the diagonal matrix $\boldsymbol{\Lambda}_{\scriptscriptstyle \Omega}$. Furthermore, the correlation of $\bar{\mathbf{X}}_b$ is a zero matrix and $\bar{\mathbf{X}}_b$ can be considered as a deterministic vector. Thus, $\bar{\mathbf{X}}_{a}$ and $\bar{\mathbf{X}}_{b}$ are statistically independent and the equation in (\ref{EF_add_eq1}) and constraints in (\ref{EF_eq30_2})-(\ref{EF_eq30_3}) are equivalently re-written as
\begin{eqnarray}
&&\min_{f_{\scriptscriptstyle \bar{X}_a}(\mathbf{x})} - h(\bar{\mathbf{X}}_a),\nonumber\\
&&\text{s.t.} \quad \int f_{\scriptscriptstyle \bar{X}_a}(\mathbf{x}) d\mathbf{x} = 1,\nonumber\\
&&\hspace{8mm} \int \x f_{\scriptscriptstyle \bar{X}_a}(\mathbf{x}) d\mathbf{x} = \boldsymbol{\mu},\nonumber\\
&&\hspace{8mm} \int \mathbf{x}\mathbf{x}^{\scriptscriptstyle T} f_{\scriptscriptstyle \bar{X}_a}(\mathbf{x}) d\mathbf{x} = \boldsymbol{\Lambda}_{\scriptscriptstyle \Omega_a},\nonumber
\end{eqnarray}
where $\boldsymbol{\Lambda}_{\scriptscriptstyle \Omega_a}=diag(\Lambda_1, \ldots, \Lambda_m) \succ \boldsymbol{0}$  is  a positive-definite matrix. Therefore, without loss of generality, we can assume  that the correlation matrix $\boldsymbol{\Omega}_{\scriptscriptstyle X}$ is invertible. %Second, if an additional constraint, related to the vector mean of $\mathbf{X}$, $\boldsymbol{\mu}_{\scriptscriptstyle X}$, is enforced, the optimal solution is a multi-variate Gaussian density function with mean  $\boldsymbol{\mu}_{\scriptscriptstyle X}$, instead of the multi-variate Gaussian density function with zero mean  (\ref{EF_eq33_1}).

Based on Theorem \ref{CAL_thm2},  since
\begin{eqnarray}
\label{EF_eq35_1}
K''_{f_{\scriptscriptstyle X} f_{\scriptscriptstyle X}}\Big|_{f_{\scriptscriptstyle X}=f_{\scriptscriptstyle X^*}} = \frac{1}{f_{\scriptscriptstyle X^*}(\mathbf{x})} >0,\nonumber
\end{eqnarray}
the second-order variation $\delta^2 U\left[f_{\scriptscriptstyle X^*}\right]$ is positive, and the optimal solution $f_{\scriptscriptstyle X^*}$ is a minimal solution for the variational problem in (\ref{EF_eq30_1}).

Therefore, the negative of  differential entropy $-h(\mathbf{X})$ is minimized (or equivalently $h(\mathbf{X})$ is maximized) when $\mathbf{X}$ is a multi-variate Gaussian random vector. Even though Theorems \ref{CAL_thm1}, \ref{CAL_thm2} are necessary conditions for the minimum, in this case, a multi-variate Gaussian density function is the actual solution since there is only one solution, namely the  multi-variate Gaussian density function, in the feasible set. An alternative justification of global optimality of multi-variate Gaussian pdf can be achieved by exploiting the convexity of $K(\x,\fx)$  wrt $\fx$.
\begin{rem}
The proof in \cite{inf:cover} relies on calculus of variations to find the first-order necessary condition, which only represents a necessary 
(and    not   sufficient) condition for optimality. Therefore, an additional technique, referred to as the Kullback-Leibler divergence, was used to prove  that the necessary solution globally maximizes the differential entropy. Unlike this proof, by confirming the convexity of the variational problem, we show that Gaussian distribution is indeed the global optimal solution solely based on calculus of variations arguments.
\end{rem}
%\begin{rem}
%Depending on the existence of the constraint related to the vector mean, the mean of the optimal Gaussian density function will change. However, the constraint on the vector mean is not necessarily required. Details of  the proof are presented in Appendix \ref{EF_thm2_proof}.
%\end{rem}
\end{proof}
\end{thm}

%%%%%%%%%%%%%%%%%%%%%
The maximum entropy result can be extended in various ways. A simple variation of the maximum entropy  considers only non-negative random variables. Then it turns out that Gaussian random variables are no longer the optimal solution that maximizes the differential entropy. The following theorem can be easily established and states that a half-normal random variable maximizes the differential entropy over the set of non-negative random variables.
% Maximizing entropy for non-negative random
\begin{thm}\label{EF_thm3}
Within the class of  non-negative random variables with given second-order moment $m_{\scriptscriptstyle X}^2$, a half-normal random variable $X_{\scriptscriptstyle HN}$  maximizes the  differential entropy, i.e.,
\begin{eqnarray}
\label{EF_eq9_1}
h(X) & \leq & h(X_{\scriptscriptstyle HN}),
\end{eqnarray}
where $X$ is an arbitrary (but fixed)  non-negative random variable with the second-order moment $m_{\scriptscriptstyle X}^2$, and $h(\cdot)$ denotes differential entropy.
\begin{proof}
The proof is omitted since it can be established following similar steps to the proof of Theorem \ref{EF_thm2}.
\end{proof}
\end{thm}

%Similar to Theorems \ref{EF_thm1}, \ref{EF_thm2}, and \ref{EF_thm3}, we can find a probability density function, which minimizes the Fisher information.

%%%%%%%%%%%%%%%%%%%%%%%%%%

%Theorem \ref{EF_thm4} can be generalized to random vectors as shown in the following theorem.
Adopting a similar variational approach to the one in Theorem \ref{EF_thm2}, we can also determine the probability density function that  minimizes the Fisher information matrix as shown by the following theorem.
%%%%%%%%%%%%%%%%%%%%%%%%%% Minimizing Fisher information Matrix
\begin{thm}[Cram\'{e}r-Rao Inequality (a vector version)]\label{EF_thm5}
Given a vector mean $\boldsymbol{\mu}_{\scriptscriptstyle X}$ and a correlation matrix $\boldsymbol{\Omega}_{\scriptscriptstyle X}$, the Gaussian density function with the vector mean $\boldsymbol{\mu}_{\scriptscriptstyle X}$ and the correlation matrix $\boldsymbol{\Omega}_{\scriptscriptstyle X}$ minimizes the Fisher information matrix, i.e.,
\begin{eqnarray}
\label{EF_eq25_1}
\mathbb{J}(\mathbf{X}) & \succeq & \mathbb{J}(\mathbf{X}_{\scriptscriptstyle G}),
\end{eqnarray}
where $\mathbf{X}$ and $\mathbf{X}_{\scriptscriptstyle G}$  stand for an arbitrary (but fixed) random vector and Gaussian random vector, respectively, with   given mean   $\boldsymbol{\mu}_{\scriptscriptstyle X}$ and correlation matrix  $\boldsymbol{\Omega}_{\scriptscriptstyle X}$, and  $\mathbb{J}(\cdot)$ denotes the Fisher information matrix:
\begin{eqnarray}
\label{EF_eq26_1}
\mathbb{J}(\mathbf{X}) = \left[\begin{array}{ccc}
s_{11} & \cdots & s_{1n} \\
\vdots & \ddots & \vdots \\
s_{n1} & \cdots & s_{nn}
\end{array}\right],
\end{eqnarray}
\begin{eqnarray}
\label{EF_eq27_1}
s_{ij} = \int \left(\frac{\frac{d}{d x_i} f_{\scriptscriptstyle X}(\mathbf{x})}{f_{\scriptscriptstyle X}(\mathbf{x})} \right)\left(\frac{\frac{d}{d x_j} f_{\scriptscriptstyle X}(\mathbf{x})}{f_{\scriptscriptstyle X}(\mathbf{x})}\right) f_{\scriptscriptstyle X}(\mathbf{x})d\mathbf{x}.\nonumber
\end{eqnarray}
%The following regularity condition is assumed to be satisfied:
%\begin{eqnarray}
%\int \nabla f_{\scriptscriptstyle X} (\mathbf{x}) d\mathbf{x} = 0,\nonumber
%\end{eqnarray}
%where $\nabla$ denotes the gradient, and the covariance matrix is supposed to be non-singular.
\begin{proof}
We first  represent the inequality in (\ref{EF_eq25_1}) as a functional with the required constraints as follows:
\begin{eqnarray}
\label{EF_eq36_1}
&&\min_{f_{\scriptscriptstyle X}} \quad \int \boldsymbol{\xi}^{\scriptscriptstyle T} \nabla f_{\scriptscriptstyle X}(\mathbf{x}) \nabla f_{\scriptscriptstyle X}(\mathbf{x})^{\scriptscriptstyle T} \boldsymbol{\xi} \frac{1}{f_{\scriptscriptstyle X}(\mathbf{x})}d\mathbf{x},\\
\label{EF_eq36_2}
&&\text{s. t.}\quad \int f_{\scriptscriptstyle X}(\mathbf{x}) d\mathbf{x} =1,\nonumber\\
&&\hspace{10mm} \int \mathbf{x} f_{\scriptscriptstyle X}(\mathbf{x}) d\mathbf{x} = \boldsymbol{\mu}_{\scriptscriptstyle X},\nonumber\\
&&\hspace{10mm} \int \mathbf{x}\mathbf{x}^T f_{\scriptscriptstyle X}(\mathbf{x}) d\mathbf{x} = \boldsymbol{\Omega}_{\scriptscriptstyle X},
\end{eqnarray}
where $\boldsymbol{\xi}$ is an arbitrary but fixed non-zero vector, defined as $\boldsymbol{\xi}=[\xi_1, \ldots, \xi_n]^{\scriptscriptstyle T}$.

Using Theorem \ref{CAL_thm3}, the functional problem in (\ref{EF_eq36_1}) is expressed as
\begin{eqnarray}
\label{EF_eq37_1}
\min_{f_{\scriptscriptstyle X}} \quad U[f_{\scriptscriptstyle X}],
\end{eqnarray}
where $U[f_{\scriptscriptstyle X}]= \int K(\mathbf{x},f_{\scriptscriptstyle X}, \nabla f_{\scriptscriptstyle X}) d\mathbf{x}$, $K(\mathbf{x},f_{\scriptscriptstyle X}, \nabla f_{\scriptscriptstyle X}) = \left(\boldsymbol{\xi}^{\scriptscriptstyle T} \nabla f_{\scriptscriptstyle X}(\mathbf{x}) \nabla f_{\scriptscriptstyle X}(\mathbf{x})^{\scriptscriptstyle T} \boldsymbol{\xi} / f_{\scriptscriptstyle X}(\mathbf{x})\right)  + \alpha f_{\scriptscriptstyle X}(\mathbf{x}) +f_{\scriptscriptstyle X}(\mathbf{x}) \sum_{i=1}^n \zeta_i x_i + f_{\scriptscriptstyle X}(\mathbf{x}) \sum_{i=1}^n \sum_{j=1}^n \lambda_{ij}x_i x_j$, and $\alpha$, $\zeta_i$, and $\lambda_{ij}$ are the Lagrange multipliers
 corresponding to the three constraints in (\ref{EF_eq36_2}).

Based on Theorem \ref{CAL_thm1}, by confirming the first-order variation condition, i.e., $\delta U[f_{\scriptscriptstyle X^*}]=0$, we can find the optimal solution $f_{\scriptscriptstyle X^*}(x)$ as follows:
\begin{eqnarray}
\label{EF_eq38_1}
K'_{f_{\scriptscriptstyle X}} - \sum_{i=1}^n \frac{\partial}{\partial x_i} K'_{f'_{\scriptscriptstyle X_i}} \Bigg|_{f_{\scriptscriptstyle X}=f_{\scriptscriptstyle X^*}} & = & 0,
\end{eqnarray}
where
\begin{eqnarray}
\label{EF_eq39_1}
K'_{f_{\scriptscriptstyle X}} \hspace*{-2mm} & = & \hspace*{-2mm}- \frac{\boldsymbol{\xi}^{\scriptscriptstyle T} \nabla f_{\scriptscriptstyle X}(\mathbf{x}) \nabla f_{\scriptscriptstyle X}(\mathbf{x})^{\scriptscriptstyle T} \boldsymbol{\xi}}{f_{\scriptscriptstyle X}(\mathbf{x})^{2}} + \alpha +  \boldsymbol{\zeta}^{\scriptscriptstyle T} \mathbf{x}+ \mathbf{x}^{\scriptscriptstyle T} \boldsymbol{\Lambda} \mathbf{x},\nonumber\\
\frac{\partial}{\partial x_i} K'_{f'_{\scriptscriptstyle X_i}} \hspace*{-2mm}& = &\hspace*{-2mm} \frac{\partial}{\partial x_i}\left(\frac{2\sum\limits_{j=1}^n \frac{\partial}{\partial x_j}f_{\scriptscriptstyle X}(\mathbf{x}) \xi_i \xi_j}{f_{\scriptscriptstyle X}(\mathbf{x})}\right)\nonumber\\
\hspace*{-2mm}& = & \hspace*{-2mm} \frac{2\sum\limits_{j=1}^n\frac{\partial}{\partial x_i} \frac{\partial}{\partial x_j}f_{\scriptscriptstyle X}(\mathbf{x}) \xi_i \xi_j}{f_{\scriptscriptstyle X}(\mathbf{x})}-\frac{2\sum\limits_{j=1}^n \frac{\partial}{\partial x_j}f_{\scriptscriptstyle X}(\mathbf{x}) \xi_i \xi_j \frac{\partial}{\partial x_i}f_{\scriptscriptstyle X}(\mathbf{x})}{f_{\scriptscriptstyle X}(\mathbf{x})^2}.
\end{eqnarray}

Therefore, the left-hand side of the equation in (\ref{EF_eq38_1}) is expressed as
\begin{eqnarray}
\label{EF_eq40_1}
K'_{f_{\scriptscriptstyle X}}-\sum_{i=1}^n \frac{\partial}{\partial x_i} K'_{f'_{\scriptscriptstyle X_i}}
\hspace*{-2mm}&=&\hspace*{-2mm} \frac{\sum\limits_{i=1}^n \sum\limits_{j=1}^n \frac{\partial}{\partial x_i}f_{\scriptscriptstyle X}(\mathbf{x}) \frac{\partial}{\partial x_j}f_{\scriptscriptstyle X}(\mathbf{x}) \xi_i \xi_j }{f_{\scriptscriptstyle X}(\mathbf{x})^2} -\frac{2\sum\limits_{i=1}^n\sum\limits_{j=1}^n\frac{\partial}{\partial x_i} \frac{\partial}{\partial x_j}f_{\scriptscriptstyle X}(\mathbf{x}) \xi_i \xi_j}{f_{\scriptscriptstyle X}(\mathbf{x})} +\alpha + \sum_{i=1}^n \zeta_i x_i +  \sum_{i=1}^n \sum_{j=1}^n \lambda_{ij} x_i x_j\nonumber\\
\label{EF_eq40_2}
\hspace*{-2mm}& = &\hspace*{-2mm} 0.
\end{eqnarray}

Unlike Theorem \ref{EF_thm2}, we cannot directly calculate $f_{\scriptscriptstyle X^*}(\mathbf{x})$ from   (\ref{EF_eq38_1}). Fortunately, the first two parts in equation (\ref{EF_eq40_1}) are  expressed as quadratic forms when $f_{\scriptscriptstyle X^*}(\mathbf{x})$ is a multi-variate Gaussian density function, and therefore, the multi-variate Gaussian density function satisfies the equality in (\ref{EF_eq40_2}). When $f_{\scriptscriptstyle X^*}(\mathbf{x})$ is a multi-variate Gaussian density function:
\begin{eqnarray}
\label{EF_eq41_1}
f_{\scriptscriptstyle X^*}(\mathbf{x})=\left(2\pi\right)^{-\frac{n}{2}} \left|\boldsymbol{\Sigma}_{\scriptscriptstyle X}\right|^{-\frac{1}{2}} e^{-\frac{\left(\mathbf{x}-\boldsymbol{\mu}_{\scriptscriptstyle X}\right)^{\scriptscriptstyle T} \boldsymbol{\Sigma}_{\scriptscriptstyle X}^{-1}\left(\mathbf{x}-\boldsymbol{\mu}_{\scriptscriptstyle X}\right)}{2}} \nonumber
\end{eqnarray}
with  $\boldsymbol{\Sigma}_{\scriptscriptstyle X}=\boldsymbol{\Omega}_{\scriptscriptstyle X}-\boldsymbol{\mu}_{\scriptscriptstyle X}\boldsymbol{\mu}_{\scriptscriptstyle X}^{\scriptscriptstyle T}$  and
\begin{eqnarray}
\label{EF_eq42_1}
\boldsymbol{\Sigma}_{\scriptscriptstyle X}^{-1} =\left[
\begin{array}{ccc}
\sigma_{\scriptscriptstyle X_{11}}^2 & \cdots & \sigma_{\scriptscriptstyle X_{1n}}^2\\
\vdots & \ddots & \vdots\\
\sigma_{\scriptscriptstyle X_{n1}}^2 & \cdots & \sigma_{\scriptscriptstyle X_{nn}}^2
\end{array}\right],
\end{eqnarray}
its partial derivatives  can be  expressed as  follows:
\begin{eqnarray}
\label{EF_eq43_1}
\frac{\partial}{\partial x_i}f_{\scriptscriptstyle X^*}(\mathbf{x}) \hspace*{-2mm}&=&\hspace*{2mm}-\frac{1}{2} (\sum_{l=1}^n \sigma_{\scriptscriptstyle X_{il}}^2 \left(x_l-\mu_{\scriptscriptstyle X_l} \right)+\sum_{m=1}^n \sigma_{\scriptscriptstyle X_{mi}}^2 \left(x_m-\mu_{\scriptscriptstyle X_m}\right))f_{\scriptscriptstyle X^*}(\mathbf{x}) , \nonumber  \\
\frac{\partial}{\partial x_j}\frac{\partial}{\partial x_i}f_{\scriptscriptstyle X^*}(\mathbf{x})\hspace*{-2mm}&=&\hspace*{-2mm}-\frac{1}{2}\left(\sigma_{\scriptscriptstyle X_{ij}}^2+\sigma_{\scriptscriptstyle X_{ji}}^2\right)f_{\scriptscriptstyle X^*}(\mathbf{x})+\frac{1}{4} \left(\sum_{l=1}^n \hspace*{-1mm} \sigma_{\scriptscriptstyle X_{il}}^2 \left(x_l-\mu_{\scriptscriptstyle X_l}\right) + \hspace*{-1mm} \sum_{m=1}^n \hspace*{-1mm} \sigma_{\scriptscriptstyle X_{mi}}^2 \left(x_m-\mu_{\scriptscriptstyle X_m}\right)\right)\nonumber\\
\label{EF_eq43_2}
\hspace*{-2mm}&&\hspace*{-2mm} \cdot \left(\sum_{l=1}^n \hspace*{-1mm} \sigma_{\scriptscriptstyle X_{jl}}^2  \hspace*{-.1cm} \left(x_l-\mu_{\scriptscriptstyle X_l}\right) \hspace*{-.1cm}+ \hspace*{-1mm} \sum_{m=1}^n \hspace*{-1mm} \sigma_{\scriptscriptstyle X_{mj}}^2 \hspace*{-.1cm} \left(x_m-\mu_{\scriptscriptstyle X_m}\right)\right)\hspace*{-1mm}f_{\scriptscriptstyle X^*}(\mathbf{x})
\end{eqnarray}
%Without loss of generality,  the covariance matrix $\boldsymbol{\Sigma}_{\scriptscriptstyle X}$ is assumed to be invertible due to the same reason mentioned in the proof of Theorem \ref{EF_thm2}.

By substituting (\ref{EF_eq43_2})  into  (\ref{EF_eq40_1}), it turns out that
\begin{eqnarray}
\label{EF_eq44_1}
K'_{f_{\scriptscriptstyle X^*}}-\sum_{i=1}^n \frac{\partial}{\partial x_i} K'_{f'_{\scriptscriptstyle X_i^*}} \hspace*{-2mm}&=&\hspace*{-2mm} \frac{1}{4} \sum_{i=1}^n \sum_{j=1}^n \xi_i \xi_j \left(\sum_{l=1}^n \left(\sigma_{\scriptscriptstyle X_{il}}^2+\sigma_{\scriptscriptstyle X_{li}}^2\right) \left(x_l-\mu_{\scriptscriptstyle X_l}\right) \right)  \left(\sum_{m=1}^n  (\sigma_{\scriptscriptstyle X_{jm}}^2+\sigma_{\scriptscriptstyle X_{mj}}^2) \left(x_m-\mu_{\scriptscriptstyle X_m}\right) \right)\nonumber\\
\hspace*{-2mm}& + &\hspace*{-2mm} \sum_{i=1}^n \sum_{j=1}^n  (\sigma_{\scriptscriptstyle X_{ij}}^2 + \sigma_{\scriptscriptstyle X_{ji}}^2) \xi_i \xi_j +\alpha + \sum_{i=1}^n  \zeta_i x_i + \sum_{i=1}^n \sum_{j=1}^n  \lambda_{ij}x_i x_j\nonumber\\
%& = & \sum_{l=1}^n \sum_{m=1}^n \left[\left(x_l-\mu_{\scriptscriptstyle X_l}\right)\left(x_m-\mu_{\scriptscriptstyle X_m}\right)\left(\frac{1}{4}\sum_{i=1}^n \sum_{j=1}^n \xi_i \xi_j \left(\sigma_{\scriptscriptstyle X_{il}}^2+\sigma_{\scriptscriptstyle X_{li}}^2\right)\left(\sigma_{\scriptscriptstyle X_{jm}}^2+\sigma_{\scriptscriptstyle X_{mj}}^2\right)\right)\right]\nonumber\\
%&&  + \sum_{i=1}^n \sum_{j=1}^n \left(\sigma_{\scriptscriptstyle X_{ij}}^2 + \sigma_{\scriptscriptstyle X_{ji}}^2\right) \xi_i \xi_j  +\alpha + \sum_{i=1}^n \zeta_i x_i + \sum_{i=1}^n \sum_{j=1}^n \lambda_{ij}x_i x_j\nonumber\\
%& = & \sum_{l=1}^n \sum_{m=1}^n \left[\left(x_l-\mu_{\scriptscriptstyle X_l}\right)\left(x_m - \mu_{\scriptscriptstyle X_m}\right)\boldsymbol{\xi}^{\scriptscriptstyle T}\boldsymbol{\Sigma}_{\scriptscriptstyle X_{lm}} \boldsymbol{\xi} \right]+ \sum_{i=1}^n \sum_{j=1}^n \left(\sigma_{\scriptscriptstyle X_{ij}}^2 + \sigma_{\scriptscriptstyle X_{ji}}^2\right) \xi_i \xi_j \nonumber\\
%&&+\alpha + \sum_{i=1}^n \zeta_i x_i + \sum_{i=1}^n \sum_{j=1}^n \lambda_{ij}x_i x_j\nonumber\\
\hspace*{-2mm}& = &\hspace*{-2mm} \sum_{l=1}^n \sum_{m=1}^n \omega_{lm}\left(x_l-\mu_{\scriptscriptstyle X_l}\right)\left(x_m-\mu_{\scriptscriptstyle X_m}\right) + \sum_{i=1}^n \sum_{j=1}^n \left(\sigma_{\scriptscriptstyle X_{ij}}^2\sigma_{\scriptscriptstyle X_{ji}}^2\right) \xi_i \xi_j+\alpha + \sum_{i=1}^n  \zeta_i x_i +\sum_{i=1}^n \sum_{j=1}^n  \lambda_{ij}x_i x_j\nonumber\\
%& = & \left(\mathbf{x} - \boldsymbol{\mu}_{\scriptscriptstyle X}\right)^{\scriptscriptstyle T} \boldsymbol{\Omega} \left(\mathbf{x}- \boldsymbol{\mu}_{\scriptscriptstyle X}\right)+ \boldsymbol{\xi}^{\scriptscriptstyle T} \boldsymbol{\Psi} \boldsymbol{\xi} +\alpha + \boldsymbol{\zeta}^{\scriptscriptstyle T} \mathbf{x} + \mathbf{x}^{\scriptscriptstyle T}\boldsymbol{\Lambda} \mathbf{x}\nonumber\\
\hspace*{-2mm}& = &\hspace*{-2mm} \mathbf{x}^{\scriptscriptstyle T} \boldsymbol{\Omega} \mathbf{x} + \mathbf{x}^{\scriptscriptstyle T}\boldsymbol{\Lambda} \mathbf{x}+\boldsymbol{\zeta}^{\scriptscriptstyle T} \mathbf{x}- 2\boldsymbol{\mu}_{\scriptscriptstyle X}^{\scriptscriptstyle T} \boldsymbol{\Omega} \mathbf{x}+ \boldsymbol{\mu}_{\scriptscriptstyle X}^{\scriptscriptstyle T} \boldsymbol{\Omega} \boldsymbol{\mu}_{\scriptscriptstyle X} + \boldsymbol{\xi}^{\scriptscriptstyle T} \boldsymbol{\Psi} \boldsymbol{\xi} +\alpha  \nonumber \\
\hspace*{-2mm}& = & \hspace*{-2mm} 0,
\end{eqnarray}
where
\begin{eqnarray}
\label{EF_eq45_1}
\boldsymbol{\Sigma}_{\scriptscriptstyle X_{lm}} \hspace*{-4mm}&=&\hspace*{-4mm} \left[\begin{array}{ccc}
\Sigma_{\scriptscriptstyle X_{11}}^{\scriptscriptstyle lm} & \cdots & \Sigma_{\scriptscriptstyle X_{1n}}^{\scriptscriptstyle lm}\\
\vdots & \ddots & \vdots \\
\Sigma_{\scriptscriptstyle X_{n1}}^{\scriptscriptstyle lm} & \cdots & \Sigma_{\scriptscriptstyle X_{nn}}^{\scriptscriptstyle lm}
\end{array}\right], \; \boldsymbol{\Lambda} \hspace*{-1mm}=\hspace*{-1mm}\left[\begin{array}{ccc}
\lambda_{11} & \cdots & \lambda_{1n}\\
\vdots & \ddots & \vdots \\
\lambda_{n1} & \cdots & \lambda_{nn}
\end{array}\right], \nonumber\\
\boldsymbol{\Psi} \hspace*{-4mm}&=&\hspace*{-4mm} \left[\begin{array}{ccc}
\psi_{11} & \cdots & \psi_{1n}\\
\vdots & \ddots & \vdots \\
\psi_{n1} & \cdots & \psi_{nn}
\end{array}\right], \; \boldsymbol{\Omega} \hspace*{-1mm}=\hspace*{-1mm} \left[\begin{array}{ccc}
\omega_{11} & \cdots & \omega_{1n}\\
\vdots & \ddots & \vdots \\
\omega_{n1} & \cdots & \omega_{nn}
\end{array}\right]\nonumber\\
\Sigma_{\scriptscriptstyle X_{ij}}^{\scriptscriptstyle lm} \hspace*{-3mm}& =&\hspace*{-3mm} \frac{1}{4} \left(\sigma_{\scriptscriptstyle X_{il}}^2+\sigma_{\scriptscriptstyle X_{li}}^2\right)\left(\sigma_{\scriptscriptstyle X_{jm}}^2+\sigma_{\scriptscriptstyle X_{mj}}^2\right)\nonumber\\
\hspace*{-3mm}& = &\hspace*{-3mm} \sigma_{\scriptscriptstyle X_{li}}^2 \sigma_{\scriptscriptstyle X_{jm}}^2,\; i, j=1,\ldots,n,\quad l, m=1,\ldots,n \nonumber\\
\psi_{ij} \hspace*{-3mm}& = &\hspace*{-3mm} 2\sigma_{\scriptscriptstyle X_{ij}}^2, \; i, j=1,\ldots,n \nonumber\\
\omega_{lm} \hspace*{-3mm}& = &\hspace*{-3mm} \boldsymbol{\xi}^{\scriptscriptstyle T}\boldsymbol{\Sigma}_{\scriptscriptstyle X_{lm}} \boldsymbol{\xi}, \;  l, m=1,\ldots,n.
\end{eqnarray}
Therefore, the Lagrange multipliers $\alpha$ and $\lambda_{ij}$ must be selected as
\begin{eqnarray}
\label{EF_eq46_1}
\alpha & = & - \boldsymbol{\mu}_{\scriptscriptstyle X}^{\scriptscriptstyle T} \boldsymbol{\Omega} \boldsymbol{\mu}_{\scriptscriptstyle X} - \boldsymbol{\xi}^{\scriptscriptstyle T} \boldsymbol{\Psi} \boldsymbol{\xi}, \nonumber\\
\boldsymbol{\zeta} & = & 2\boldsymbol{\Omega}\boldsymbol{\mu}_{\scriptscriptstyle X},\nonumber\\
\boldsymbol{\Lambda} & = & -\boldsymbol{\Omega}.
\end{eqnarray}
Since the second-order variation is positive:
\begin{eqnarray}
\label{EF_eq47_1}
K''_{\nabla\fx \nabla\fx} \Big|_{\fx=\fxstar} & = & 2 \frac{\boldsymbol{\xi} \boldsymbol{\xi}^{\subt}}{f_{\scriptscriptstyle X^*}(\mathbf{x})} \succeq 0,
\end{eqnarray}
based on Theorem \ref{CAL_thm2}, the Gaussian distribution $f_{\scriptscriptstyle X^*}(\mathbf{x})$ is necessary optimal for the variational problem in (\ref{EF_eq36_1}).
Even though Theorems \ref{CAL_thm1} and  \ref{CAL_thm2} are necessary conditions for the minimum, in this case, the multi-variate Gaussian density function is sufficiently the global minimum solution since this is a convex optimization problem (the objective function is strictly convex and its constraint set is convex).
\end{proof}
\end{thm}

%%%%%%%%%%%%%%%%%%%%%%%%%%
Using similar variational arguments, one can show that a half-normal and a chi density function minimize the Fisher information over the set of non-negative random variables as shown by the following two theorems.
% Minimizing Fisher information among non-negative random variables (no regularity)
\begin{thm}\label{EF_thm6}
 Within the class of  non-negative continuous random variables with fixed second-order moment $m_{\scriptscriptstyle X}^2$, the Fisher information is minimized by a half-normal random variable $X_{\scriptscriptstyle HN}$:
\begin{eqnarray}
\label{EF_eq28_1}
J(X) & \succeq & J(X_{\scriptscriptstyle HN}),
\end{eqnarray}
where $X$ is an arbitrary (but fixed) non-negative random variable with the second-order moment $m_{\scriptscriptstyle X}^2$, and $J(\cdot)$ denotes the Fisher information.
\end{thm}
\begin{rem}
Theorem \ref{EF_thm6} does not assume the following regularity condition:
\begin{eqnarray}
\label{EF_eq28_1_1}
\int_0^{\infty} \nabla f_{\scriptscriptstyle } (x) dx = 0.
\end{eqnarray}
 for the Fisher information.
\end{rem}
% Minimizing Fisher information among non-negative random variables (with regularity conditions)
The following result establishes the counterpart of Theorem \ref{EF_thm6} for the class of non-negative random variables with fixed second order moment and whose distribution satisfies the  regularity condition in (\ref{EF_eq28_1_1}).
\begin{thm}[\cite{MinFisher:Bercher}]\label{EF_thm7}
Within the class of non-negative continuous random variables $X$ with fixed second-order moment and whose distributions satisfy  the regularity condition in (\ref{EF_eq28_1_1}),  the Fisher information is minimized by a chi-distributed random variable $X_{\scriptscriptstyle C}$:
\begin{eqnarray}
\label{EF_eq29_1}
J(X) & \succeq & J(X_{\scriptscriptstyle C}),
\end{eqnarray}
where $J(\cdot)$ stands for the Fisher information.
\begin{proof}
Unlike the proof in \cite{MinFisher:Bercher}, by considering the first-order and the second-order moments instead of variance, %we obtain the convex constraint sets. Since the Fisher information is a strictly convex functional with respect to a probability density function, the variational problem is convex, and hence has a unique solution.
we construct a variational problem and address the problem using the first-order and second-order necessary conditions, as well as the convexity property of the problem. The details of the proof are omitted because of the similar steps to those encountered in the proof
 of Theorem   \ref{EF_thm5}.  %The details of the proof are deferred to  Appendix \ref{EF_thm7_proof}.
\end{proof}
\end{thm}

\section{Worst Additive Noise Lemma}\label{WA}

Worst additive noise lemma was introduced and exploited in several references \cite{WANoise:Cover}, \cite{EPI:Rioul}, \cite{WAN:Ihara}, and it has been widely used in  numerous other applications. One of the main applications of the worst additive noise lemma pertains to the capacity calculation of a wireless
communication channel  subject to different constraints such as   Gaussian MIMO broadcasting,  Gaussian MIMO wire-tap, etc. In this section, the worst additive noise lemma for random vectors will be proved solely based on variational arguments.

%Similarly, Theorem \ref{WA_thm1} can be generalized to random vectors as shown in the following theorem.
\begin{thm}\label{WA_thm2}
Assume $\mathbf{X}$ is an arbitrary but fixed random vector and $\mathbf{X}_{\scriptscriptstyle G}$ is a Gaussian random vector, whose mean and correlation matrix are  identical to those of $\mathbf{X}$, denoted as $\boldsymbol{\mu}_{\scriptscriptstyle X}$ and $\boldsymbol{\Omega}_{\scriptscriptstyle X}$, respectively. Given a Gaussian random vector $\mathbf{W}_{\scriptscriptstyle G}$, assumed  independent of both $\mathbf{X}$ and $\mathbf{X}_{\scriptscriptstyle G}$ and  with zero mean and the correlation matrix $\boldsymbol{\Omega}_{\scriptscriptstyle W}$, then the following relation holds:
\begin{eqnarray}
\label{WA_eq14_1}
I(\mathbf{X}+\mathbf{W}_{\scriptscriptstyle G};\mathbf{W}_{\scriptscriptstyle G}) \geq I(\mathbf{X}_{\scriptscriptstyle G}+\mathbf{W}_{\scriptscriptstyle G}; \mathbf{W}_{\scriptscriptstyle G}).
\end{eqnarray}
\begin{proof}
Our  proof is entirely  anchored in the variational  calculus framework. A summary of our proof runs as follows. First, we construct a variational problem, which represents the inequality in (\ref{WA_eq14_1}) and required constraints in a functional form. Second, using the first-order variation condition, we find the necessary optimal solutions, which satisfy Euler's equation. Third, using the second-order variation condition, we show that the optimal solutions are necessarily local minima. Finally, we justify that the local  minimum is also global.

By setting $\mathbf{Y}=\mathbf{X}+\mathbf{W}_{\subg}$, where $\mathbf{X}$ and $\mathbf{W}_{\scriptscriptstyle G}$ are independent of each other, in (\ref{WA_eq14_1}), the mutual information $I(\mathbf{X}+\mathbf{W}_{\subg};\mathbf{W}_{\subg})$ can be expressed as
$$I(\mathbf{X}+\mathbf{W}_{\subg};\mathbf{W}_{\subg}) = h(\mathbf{Y}) - h(\mathbf{Y}|\mathbf{W}_{\subg}) = h(\mathbf{Y})- h(\mathbf{X}).$$
Then, we consider the functional:
\begin{eqnarray}
\hspace*{-3mm}&&\hspace*{-3mm}\min_{f_{\scriptscriptstyle X}} -\int \hspace*{-2mm}\int \hspace*{-2mm}f_{\scriptscriptstyle X}(\mathbf{x})f_{\scriptscriptstyle Y|X}(\mathbf{y}|\mathbf{x}) \log\left( \int \hspace*{-2mm}f_{\scriptscriptstyle X}(\mathbf{x})f_{\scriptscriptstyle Y|X}(\mathbf{y}|\mathbf{x}) d\mathbf{x}\right) d\mathbf{x} d\mathbf{y}
+\int \hspace*{-2mm} \int f_{\scriptscriptstyle X}(\mathbf{x})f_{\scriptscriptstyle Y|X}(\mathbf{y}|\mathbf{x}) \log f_{\scriptscriptstyle X}(\mathbf{x}) d\mathbf{x} d\mathbf{y} \label{WA_eq15_1} \\
\hspace*{-3mm}&&\hspace*{-3mm} \text{s. t.}  \int f_{\scriptscriptstyle X}(\mathbf{x}) d\mathbf{x} =1,\nonumber\\
\hspace*{-3mm}&&\hspace*{+2mm} \int \mathbf{x} f_{\scriptscriptstyle X}(\mathbf{x})d\mathbf{x} = \boldsymbol{\mu}_{\scriptscriptstyle X},\nonumber\\
\label{WA_eq15_2}
\hspace*{-3mm}&&\hspace*{+2mm} \int \mathbf{x}\mathbf{x}^{\scriptscriptstyle T}f_{\scriptscriptstyle X}(\mathbf{x}) d\mathbf{x} = \boldsymbol{\Omega}_{\scriptscriptstyle X}.
\end{eqnarray}
The density function $f_{\scriptscriptstyle Y}(\mathbf{y})$ and conditional density function $f_{\scriptscriptstyle Y|X}(\mathbf{y}|\mathbf{x})$ are expressed as
\begin{eqnarray}
\label{WA_eq16_1}
f_{\scriptscriptstyle Y}(\mathbf{y}) & = & \int f_{\scriptscriptstyle X}(\mathbf{x}) f_{\scriptscriptstyle Y|X}(\mathbf{y}|\mathbf{x}) d\mathbf{x},\\
\label{WA_eq16_2}
f_{\scriptscriptstyle Y|X}(\mathbf{y}|\mathbf{x}) & = & f_{\scriptscriptstyle W}(\mathbf{y}-\mathbf{x}),
\end{eqnarray}
respectively.
Therefore, by substituting $f_{\scriptscriptstyle Y}(\mathbf{y})$  for $\int f_{\scriptscriptstyle X}(\mathbf{x}) f_{\scriptscriptstyle Y|X}(\mathbf{y}|\mathbf{x}) d\mathbf{x}$ and  $f_{\scriptscriptstyle W}(\mathbf{y}-\mathbf{x})$  for  $f_{\scriptscriptstyle Y|X}(\mathbf{y}|\mathbf{x})$, respectively, and appropriately changing the constrains in (\ref{WA_eq15_2}), the variational problem in (\ref{WA_eq15_1})  is  expressed as
\begin{eqnarray}
\label{WA_eq17_1}
\hspace*{-4mm}&&\hspace*{-4mm}\min_{f_{\scriptscriptstyle X}, f_{\scriptscriptstyle Y}} \hspace*{-1mm} \int \hspace*{-2mm} \int \hspace*{-2mm} f_{\scriptscriptstyle X}(\mathbf{x})f_{\scriptscriptstyle W}(\mathbf{y}-\mathbf{x}) \left[-\log f_{\scriptscriptstyle Y}(\mathbf{y})+\log f_{\scriptscriptstyle X}(\mathbf{x}) \right] d\mathbf{x}d\mathbf{y}     \\
\label{WA_eq17_3_1}
\hspace*{-4mm}&&\hspace*{-4mm}\text{s. t.} \int \hspace*{-2mm} \int \hspace*{-2mm} f_{\scriptscriptstyle X}(\mathbf{x})f_{\scriptscriptstyle W}(\mathbf{y}-\mathbf{x}) d\mathbf{x} d\mathbf{y} =1,\\
\label{WA_eq17_3_2}
\hspace*{-3mm}&& \hspace*{+2mm}\int \hspace*{-2mm} \int \hspace*{-1mm}\mathbf{x} f_{\scriptscriptstyle X}(\mathbf{x})f_{\scriptscriptstyle W}(\mathbf{y}-\mathbf{x}) d\mathbf{x} d\mathbf{y} =\boldsymbol{\mu}_{\scriptscriptstyle X},\\
\label{WA_eq17_3_3}
\hspace*{-3mm}&& \hspace*{+2mm}\int  \hspace*{-2mm} \int \hspace*{-1mm} \mathbf{x}\mathbf{x}^{\scriptscriptstyle T} f_{\scriptscriptstyle X}(\mathbf{x})f_{\scriptscriptstyle W}(\mathbf{y}-\mathbf{x}) d\mathbf{x} d\mathbf{y} =\boldsymbol{\Omega}_{\scriptscriptstyle X},\\
%\label{WA_eq17_3_4}
%\hspace*{-3mm}&& \hspace*{+2mm} \int f_{\scriptscriptstyle Y}(\mathbf{y}) d\mathbf{y} =1,\\
\label{WA_eq17_3_5}
\hspace*{-3mm}&& \hspace*{+2mm} \iint \mathbf{y} \fx(\x)\fw(\y-\x) d\x d\mathbf{y} =\boldsymbol{\mu}_{\scriptscriptstyle Y},\\
\label{WA_eq17_3_6}
\hspace*{-3mm}&& \hspace*{+2mm}\iint \mathbf{y}\mathbf{y}^{\scriptscriptstyle T} \fx(\x)\fw(\y-\x) d\x d\mathbf{y} = \boldsymbol{\Omega}_{\scriptscriptstyle Y},\\
\label{WA_eq17_3_7}
\hspace*{-3mm}&& \hspace*{+2mm}f_{\scriptscriptstyle Y}(\mathbf{y}) = \int f_{\scriptscriptstyle X}(\mathbf{x})f_{\scriptscriptstyle W}(\mathbf{y}-\mathbf{x})d\mathbf{x}.
\end{eqnarray}

Based on Corollary \ref{CAL_cor1}, the functional problem in (\ref{WA_eq17_1}) can be re-cast into the following equivalent form:
\begin{eqnarray}
\label{WA_eq18_1}
\hspace*{-4mm}& &\hspace*{-4mm}\min_{f_{\scriptscriptstyle X}, f_{\scriptscriptstyle Y}}\quad \hspace*{-4mm} \int  ( \int f_{\scriptscriptstyle X}(\mathbf{x})f_{\scriptscriptstyle W}(\mathbf{y}-\mathbf{x}) [- \log f_{\scriptscriptstyle Y}(\mathbf{y}) + \log f_{\scriptscriptstyle X}(\mathbf{x})
+\alpha_0+\sum_{i=1}^n \zeta_i x_i + \sum_{i=1}^n \sum_{j=1}^n \gamma_{ij}x_i x_j + \sum_{i=1}^n \eta_{i} y_i \nonumber\\
\hspace*{-4mm}&+&\hspace*{-4mm} \sum_{i=1}^n \sum_{j=1}^n \theta_{ij} y_i y_j - \lambda(\mathbf{y})] d\mathbf{x}) +
 f_{\scriptscriptstyle Y}(\mathbf{y})\lambda(\mathbf{y})d\mathbf{y},
\end{eqnarray}
where $\mathbf{x}^{\scriptscriptstyle T} = [x_1, \ldots, x_n]$, $\mathbf{y}^{\scriptscriptstyle T} = [y_1, \ldots, y_n]$, and $\alpha_0$, $\zeta_{i}$, $\gamma_{ij}$, $\eta_{i}$, $\theta_{ij}$, and $\lambda(\mathbf{y})$ stand for the Lagrange multipliers corresponding to the constraints (\ref{WA_eq17_3_1}), (\ref{WA_eq17_3_2}), (\ref{WA_eq17_3_3}), (\ref{WA_eq17_3_5}), (\ref{WA_eq17_3_6}), and (\ref{WA_eq17_3_7}), respectively.

Define now the functional $U$ as
\begin{eqnarray}
\label{WA_eq19_1}
U[f_{\scriptscriptstyle X},f_{\scriptscriptstyle Y}] \hspace*{-2mm}& = & \hspace*{-2mm} \int \left(\int K(\mathbf{x},\mathbf{y},f_{\scriptscriptstyle X},f_{\scriptscriptstyle Y}) d\mathbf{x}\right) + \tilde{K}(\mathbf{y},f_{\scriptscriptstyle Y}) d\mathbf{y},\nonumber
\end{eqnarray}
where
\begin{eqnarray}
\label{WA_eq20_1}
\hspace*{-4mm}& &\hspace*{-4mm}K(\mathbf{x},\mathbf{y},f_{\scriptscriptstyle X},f_{\scriptscriptstyle Y}) =f_{\scriptscriptstyle X}(\mathbf{x})f_{\scriptscriptstyle W}(\mathbf{y}-\mathbf{x}) [- \log f_{\scriptscriptstyle Y}(\mathbf{y})
+ \log f_{\scriptscriptstyle X}(\mathbf{x}) + \alpha_0 + \sum_{i=1}^n \zeta_i x_i  + \sum_{i=1}^n \sum_{j=1}^n \gamma_{ij}x_i x_j \nonumber\\
\hspace*{-4mm}& &\hspace*{-4mm}~~~~~~~~~~~~~~~~~~~~~ +\hspace*{-1mm}\sum_{i=1}^n \hspace*{-1mm}\eta_{i} y_i+\hspace*{-1mm}\sum_{i=1}^n \sum_{j=1}^n \hspace*{-1mm}\theta_{ij}y_i y_j -\lambda(\mathbf{y})],\nonumber\\
\hspace*{-4mm}&&\hspace*{-4mm}\tilde{K}(\mathbf{y},f_{\scriptscriptstyle Y})=\lambda(\y) f_{\scriptscriptstyle Y}(\mathbf{y}).
\end{eqnarray}
Based on Corollary 2, we can find the optimal solution $f_{\scriptscriptstyle X^*}$ and $f_{\scriptscriptstyle Y^*}$ as follows:
\begin{eqnarray}
\label{WA_eq21_1}
\int K'_{f_{\scriptscriptstyle X}}\Big|_{f_{\scriptscriptstyle X}=  f_{\scriptscriptstyle X^*}, f_{\scriptscriptstyle Y}=f_{\scriptscriptstyle Y^*}} d\y%\nonumber\\
% & = & f_{\scriptscriptstyle W}(\mathbf{y}-\mathbf{x}) \Bigg(- \log f_{\scriptscriptstyle Y^*}(\mathbf{y}) + \log f_{\scriptscriptstyle X^*}(\mathbf{x}) + \alpha_0 + \sum_{i=1}^n \zeta_i x_i\nonumber\\
%&&\hspace{30mm} + \sum_{i=1}^n \sum_{j=1}^n \gamma_{ij}x_i x_j + 1 - \lambda(\mathbf{y})\Bigg)\nonumber\\
\hspace*{-2mm}& = &\hspace*{-2mm} \int f_{\scriptscriptstyle W}(\mathbf{y}-\mathbf{x})(- \log f_{\scriptscriptstyle Y^*}(\mathbf{y}) +\log f_{\scriptscriptstyle X^*}(\mathbf{x})+\alpha_0 +\boldsymbol{\zeta} \mathbf{x}^{\scriptscriptstyle T} + \mathbf{x}^{\scriptscriptstyle T}\boldsymbol{\Gamma} \mathbf{x} + \boldsymbol{\eta}^T\y \nonumber\\
\hspace*{-2mm} & & \hspace*{-2mm}~~+ \y^T\boldsymbol{\Theta} \mathbf{y}+1-\lambda(\mathbf{y})) d\y \nonumber\\
\label{WA_eq21_2} \hspace*{-2mm} & = & \hspace*{-2mm}  0\\
\int K'_{f_{\scriptscriptstyle Y}} d\mathbf{x}+ \tilde{K}'_{f_{\scriptscriptstyle Y}}\Bigg|_{f_{\scriptscriptstyle X}=f_{\scriptscriptstyle X^*}, f_{\scriptscriptstyle Y}=f_{\scriptscriptstyle Y^*}} %\nonumber\\
%& = & - \int f_{\scriptscriptstyle X^*}(\mathbf{x})f_{\scriptscriptstyle W}(\mathbf{y}-\mathbf{x})d\mathbf{x} \frac{1}{f_{\scriptscriptstyle Y^*}(\mathbf{y})} + \alpha_1 + \sum_{i=1}^n \eta_{i} y_i + \sum_{i=1}^n \sum_{j=1}^n \theta_{ij} y_i y_j +\lambda(\mathbf{y})\nonumber\\
 \hspace*{-2mm}&=&\hspace*{-2mm} -\int  \frac{f_{\scriptscriptstyle X^*}(\mathbf{x})f_{\scriptscriptstyle W}(\mathbf{y}-\mathbf{x})d\mathbf{x}}{f_{\scriptscriptstyle Y^*}(\mathbf{y})} + \lambda(\mathbf{y})\nonumber\\
\label{WA_eq21_3}
\hspace*{-2mm}& = & \hspace*{-2mm}  0,
\end{eqnarray}
where
\begin{eqnarray}
\label{WA_eq22_1}
\boldsymbol{\Gamma} = \left[\begin{array}{ccc}
\gamma_{11}  & \cdots & \gamma_{1n}\\
\vdots & \ddots & \vdots\\
\gamma_{n1} & \cdots & \gamma_{nn}
\end{array}\right], \; \boldsymbol{\Theta} = \left[\begin{array}{ccc}
\theta_{11}  & \cdots & \theta_{1n}\\
\vdots & \ddots & \vdots\\
\theta_{n1} & \cdots & \theta_{nn}
\end{array}\right]
\end{eqnarray}
$\boldsymbol{\zeta} = [\zeta_1, \ldots, \zeta_n]^{\scriptscriptstyle T}$  and $\boldsymbol{\eta} = [\eta_1, \ldots, \eta_n]^{\scriptscriptstyle T}$.

%Since the equalities in (\ref{WA_eq21_2}) and (\ref{WA_eq21_3}) must be satisfied for any $\mathbf{x}$ and $\mathbf{y}$, it follows that
The following relationships satisfy  the necessary conditions (\ref{WA_eq21_2}) and (\ref{WA_eq21_3}):
\begin{eqnarray}
\label{WA_eq23_1}
0&=&- \log f_{\scriptscriptstyle Y^*}(\mathbf{y}) +\log f_{\scriptscriptstyle X^*}(\mathbf{x})+\alpha_0 +\boldsymbol{\zeta} \mathbf{x}^{\scriptscriptstyle T} + \mathbf{x}^{\scriptscriptstyle T}\boldsymbol{\Gamma} \mathbf{x} + \boldsymbol{\eta}^T\y + \y^T\boldsymbol{\Theta} \mathbf{y}+1-\lambda(\mathbf{y}), \nonumber\\
0 & = & -1 + \lambda(\mathbf{y}).
\end{eqnarray}
%%and  hence:
%%\begin{eqnarray}
%%\label{WA_eq24_1}
%%f_{\scriptscriptstyle X^*}(\mathbf{x}) & = & \exp\left(-\alpha_0 - \boldsymbol{\zeta}^{\scriptscriptstyle T} \mathbf{x} - \mathbf{x}^{\scriptscriptstyle T} \boldsymbol{\Gamma} \mathbf{x}  - 1 \right),\nonumber\\
%%f_{\scriptscriptstyle Y^*}(\mathbf{y}) & = & \exp\left(  - 1 + \alpha_1 +\boldsymbol{\eta}^{\scriptscriptstyle T} \mathbf{y} + \mathbf{y}^{\scriptscriptstyle T} \boldsymbol{\Theta} \mathbf{y} \right).
%\end{eqnarray}

Considering the constraints in (\ref{WA_eq17_3_1})-(\ref{WA_eq17_3_7}), $f_{\scriptscriptstyle X^*}(\mathbf{x})$ and $f_{\scriptscriptstyle Y^*}(\mathbf{y})$ in (\ref{WA_eq23_1}) can be expressed as
\begin{eqnarray}
\label{WA_eq25_1}
f_{\scriptscriptstyle X^*}(\mathbf{x}) \hspace*{-3mm}& = & \hspace*{-3mm}\left(2\pi\right)^{-\frac{n}{2}} \left|\boldsymbol{\Sigma}_{\scriptscriptstyle X}\right|^{-\frac{1}{2}} e^{-\frac{ \left(\mathbf{x}-\boldsymbol{\mu}_{\scriptscriptstyle X}\right)^{\scriptscriptstyle T}\boldsymbol{\Sigma}_{\scriptscriptstyle X}^{-1} \left(\mathbf{x}-\boldsymbol{\mu}_{\scriptscriptstyle X}\right)}{2}}\nonumber\\
%& = &  \exp \left\{-\frac{1}{2}\log \left(2\pi\right)^n \left|\boldsymbol{\Sigma}_{\scriptscriptstyle X}\right| -\frac{1}{2} \mathbf{x}^{\scriptscriptstyle T} \boldsymbol{\Sigma}_{\scriptscriptstyle X}^{-1} \mathbf{x} + \boldsymbol{\mu}_{\scriptscriptstyle X}^{\scriptscriptstyle T} \boldsymbol{\Sigma}_{\scriptscriptstyle X}^{-1}\mathbf{x}-\frac{1}{2} \boldsymbol{\mu}_{\scriptscriptstyle X}^{\scriptscriptstyle T} \boldsymbol{\Sigma}_{\scriptscriptstyle X}^{-1} \boldsymbol{\mu}_{\scriptscriptstyle X}\right\}\nonumber\\
%\hspace*{-3mm}& = &\hspace*{-3mm} \exp\left(-\alpha_0 - \boldsymbol{\zeta}^{\scriptscriptstyle T} \mathbf{x} - \mathbf{x}^{\scriptscriptstyle T} \boldsymbol{\Gamma} \mathbf{x}  - 1\right),\nonumber\\
%%%%%%%%%%%%%%%%%%%%%%%%%%
f_{\scriptscriptstyle Y^*}(\mathbf{y}) \hspace*{-3mm}& = &\hspace*{-3mm} \left(2\pi\right)^{-\frac{n}{2}} \left|\boldsymbol{\Sigma}_{\scriptscriptstyle Y}\right|^{-\frac{1}{2}} e^{-\frac{ \left(\mathbf{y}-\boldsymbol{\mu}_{\scriptscriptstyle Y}\right)^{\scriptscriptstyle T}\boldsymbol{\Sigma}_{\scriptscriptstyle Y}^{-1} \left(\mathbf{y}-\boldsymbol{\mu}_{\scriptscriptstyle Y}\right)}{2}}\nonumber
%& = &  \exp \left\{-\frac{1}{2}\log \left(2\pi\right)^n \left|\boldsymbol{\Sigma}_{\scriptscriptstyle Y}\right| -\frac{1}{2} \mathbf{y}^{\scriptscriptstyle T} \boldsymbol{\Sigma}_{\scriptscriptstyle Y}^{-1} \mathbf{y} + \boldsymbol{\mu}_{\scriptscriptstyle Y}^{\scriptscriptstyle T} \boldsymbol{\Sigma}_{\scriptscriptstyle Y}^{-1}\mathbf{y}-\frac{1}{2} \boldsymbol{\mu}_{\scriptscriptstyle Y}^{\scriptscriptstyle T} \boldsymbol{\Sigma}_{\scriptscriptstyle Y}^{-1} \boldsymbol{\mu}_{\scriptscriptstyle Y}\right\}\nonumber\\
%\hspace*{-3mm}& = &\hspace*{-3mm} \exp\left( - 1 + \alpha_1 +\boldsymbol{\eta}^{\scriptscriptstyle T} \mathbf{y} + \mathbf{y}^{\scriptscriptstyle T} \boldsymbol{\Theta} \mathbf{y} \right),
\end{eqnarray}
where $\boldsymbol{\Sigma}_{\scriptscriptstyle X} = \boldsymbol{\Omega}_{\scriptscriptstyle X}-\boldsymbol{\mu}_{\scriptscriptstyle X}\boldsymbol{\mu}_{\scriptscriptstyle X}^{\scriptscriptstyle T}$, $\boldsymbol{\Sigma}_{\scriptscriptstyle Y} = \boldsymbol{\Sigma}_{\scriptscriptstyle X}+\boldsymbol{\Sigma}_{\scriptscriptstyle W}$,
and $\boldsymbol{\Sigma}_{\scriptscriptstyle W}$ is the  covariance matrix of $\mathbf{W}_{\scriptscriptstyle G}$.
Based on the equations in (\ref{WA_eq25_1}), it turns out that
\begin{eqnarray}
\label{WA_eq26_1}
\alpha_0 & = &\frac{1}{2}\log \left(2\pi\right)^n \left|\boldsymbol{\Sigma}_{\scriptscriptstyle X}\right| + \frac{1}{2} \boldsymbol{\mu}_{\scriptscriptstyle X}^{\scriptscriptstyle T} \boldsymbol{\Sigma}_{\scriptscriptstyle X}^{-1} \boldsymbol{\mu}_{\scriptscriptstyle X} -\frac{1}{2}\log \left(2\pi\right)^n \left|\boldsymbol{\Sigma}_{\scriptscriptstyle Y}\right|-\frac{1}{2} \boldsymbol{\mu}_{\scriptscriptstyle Y}^{\scriptscriptstyle T} \boldsymbol{\Sigma}_{\scriptscriptstyle Y}^{-1} \boldsymbol{\mu}_{\scriptscriptstyle Y},\nonumber\\
\boldsymbol{\Gamma} & = & \frac{1}{2}\boldsymbol{\Sigma}_{\scriptscriptstyle X}^{-1},\nonumber\\
\boldsymbol{\zeta} & = & -\boldsymbol{\Sigma}_{\scriptscriptstyle X}^{-1} \boldsymbol{\mu}_{\scriptscriptstyle X},\nonumber\\
\boldsymbol{\Theta} & = & -\frac{1}{2}\boldsymbol{\Sigma}_{\scriptscriptstyle Y}^{-1},\nonumber\\
\boldsymbol{\eta} & = & -\boldsymbol{\Sigma}_{\scriptscriptstyle Y}^{-1} \boldsymbol{\mu}_{\scriptscriptstyle Y}.
\end{eqnarray}
Therefore, $f_{\scriptscriptstyle X^*}$ and $f_{\scriptscriptstyle Y^*}$ are multi-variate Gaussian density functions (without loss of generality, and we can assume that the covariance matrix $\boldsymbol{\Sigma}_{\scriptscriptstyle X}$ is invertible due to the arguments mentioned in Appendix \ref{NON}).

Now, by confirming the second-order variation condition, we will show that the optimal solutions  $f_{\scriptscriptstyle X^*}$ and $f_{\scriptscriptstyle Y^*}$ are necessarily local minima. Using Corollary \ref{CAL_cor2}, we will show that the following matrix is positive semi-definite:
\begin{eqnarray}
\label{WA_eq27_1}
\left[\begin{array}{cc}
K''_{f_{\scriptscriptstyle X} f_{\scriptscriptstyle X}} & K''_{f_{\scriptscriptstyle X} f_{\scriptscriptstyle Y}}\\
K''_{f_{\scriptscriptstyle Y} f_{\scriptscriptstyle X}} & K''_{f_{\scriptscriptstyle Y} f_{\scriptscriptstyle Y}}
\end{array}\right] \bigg|_{\fx=\fxstar,\fy=\fystar} \succeq {\bf  0}.
\end{eqnarray}
Since the elements of the matrix in (\ref{WA_eq27_1}) are defined as
\begin{eqnarray}
\label{WA_eq28_1}
K''_{f_{\scriptscriptstyle X} f_{\scriptscriptstyle X}}\Big|_{f_{\scriptscriptstyle X}=f_{\scriptscriptstyle X^*}, f_{\scriptscriptstyle Y}=f_{\scriptscriptstyle Y^*}} & = & \frac{f_{\scriptscriptstyle W}(\mathbf{y}-\mathbf{x})}{f_{\scriptscriptstyle X^*}(\mathbf{x})},\nonumber\\
K''_{f_{\scriptscriptstyle Y} f_{\scriptscriptstyle Y}}\Big|_{f_{\scriptscriptstyle X}=f_{\scriptscriptstyle X^*}, f_{\scriptscriptstyle Y}=f_{\scriptscriptstyle Y^*}} & = & \frac{f_{\scriptscriptstyle X^*}(\mathbf{x})f_{\scriptscriptstyle W}(\mathbf{y}-\mathbf{x})}{f_{\scriptscriptstyle Y^*}(\mathbf{y})^2},\nonumber\\
K''_{f_{\scriptscriptstyle X} f_{\scriptscriptstyle Y}}\Big|_{f_{\scriptscriptstyle X}=f_{\scriptscriptstyle X^*}, f_{\scriptscriptstyle Y}=f_{\scriptscriptstyle Y^*}} & = & -\frac{f_{\scriptscriptstyle W}(\mathbf{y}-\mathbf{x})}{f_{\scriptscriptstyle Y^*}(\mathbf{y})},\nonumber\\
K''_{f_{\scriptscriptstyle Y} f_{\scriptscriptstyle X}}\Big|_{f_{\scriptscriptstyle X}=f_{\scriptscriptstyle X^*}, f_{\scriptscriptstyle Y}=f_{\scriptscriptstyle Y^*}} & = & -\frac{f_{\scriptscriptstyle W}(\mathbf{y}-\mathbf{x})}{f_{\scriptscriptstyle Y^*}(\mathbf{y})},
\end{eqnarray}
the matrix is a positive semi-definite matrix, and therefore $\delta^2 U \geq 0$.
Because of the convexity of functional
$K(\mathbf{x},\mathbf{y},f_{\scriptscriptstyle X},f_{\scriptscriptstyle Y})$  wrt variables $f_{\scriptscriptstyle X}$ and $f_{\scriptscriptstyle Y}$, the optimal solutions  $f_{\scriptscriptstyle X^*}$ and $f_{\scriptscriptstyle Y^*}$ actually globally minimize the variational functional in (\ref{WA_eq17_1}). Even though these optimal solutions are necessarily optimal, there exists only one solution, which is the multi-variate Gaussian density function and it satisfies Euler's equation in (\ref{WA_eq21_2}) and (\ref{WA_eq21_3}). Therefore,  $f_{\scriptscriptstyle X^*}$ and $f_{\scriptscriptstyle Y^*}$ are also sufficient in this case.

An alternative more detailed proof of the fact that $f_{\scriptscriptstyle X^*}$ and $f_{\scriptscriptstyle Y^*}$ represent global optimal solutions is to show  that $U[\fxhat,\fyhat] \geq U[\fxstar,\fystar]$, where  $\fxhat, \fyhat$ denote any arbitrary functions satisfying the boundary conditions and the constraints. First, the following functionals are defined:
\begin{eqnarray*}
F(\x,\y,\fx,\fy) &=& \fx(\x)\fw(\y-\x)[-\log{\fy(\y)}+\log{\fx(\x)}],  \\
F_0(\x,\y,\fx) &=& \fx(\x)\fw(\y-\x),  \\
 F_1^{(i)}(\x,\y,\fx) &=& x_i \fx(\x)\fw(\y-\x),  \\
F_2^{(i,j)}(\x,\y,\fx) &=& x_i x_j \fx(\x)\fw(\y-\x), \\
F_3^{(i)}(\x,\y,\fx) &=& y_i \fx(\x)\fw(\y-\x), \\
F_4^{(i,j)}(\x,\y,\fx) &=& y_i y_j \fx(\x)\fw(\y-\x),
\end{eqnarray*}
and thus $K(\x,\y,\fx,\fy)$ can be expressed as
\begin{equation}\notag
\begin{split}
& K(\x,\y,\fx,\fy) = F(\x,\y,\fx,\fy) + \alpha_0 F_0(\x,\y,\fx) + \sum\limits_{i=1}^n \zeta_i F_1^{(i)}(\x,\y,\fx) + \sum\limits_{i=1}^n \sum\limits_{j=1}^n \gamma_{ij} F_2^{(i,j)}(\x,\y,\fx) \\
& +\sum\limits_{i=1}^n \eta_i F_3^{(i)}(\x,\y,\fx) + \sum\limits_{i=1}^n \sum\limits_{j=1}^n \theta_{ij}F_4^{(i,j)}(\x,\y,\fx)-\lambda(\y) \fx(\x)\fw(\y-\x).
\end{split}
\end{equation}
Since the Hessian matrix of $K(\x,\y,\fx,\fy)$ wrt $\fx$ and $\fy$ is given by
$$\left[
  \begin{array}{cc}
    \fw(\y-\x) / \fx(\x) & -\fw(\y-\x) / \fy(\y) \\
    -\fw(\y-\x) / \fy(\y) & \fx(\x)\fw(\y-\x)/ \fy(\y)^2 \\
  \end{array}
\right],$$
which is positive semi-definite, $K(\x,\y,\fx,\fy)$ is convex wrt $\fx$ and $\fy$, and the following inequality holds
\begin{equation}\label{eq:Wanl-tangent}
 K(\x,\y,\fxhat,\fyhat) - K(\x,\y,\fxstar,\fystar)
\geq  \left[ (\fxhat-\fxstar) K'_{\fx} + (\fyhat-\fystar) K'_{\fy} \right]\Big|_{\fx=\fxstar,\fy=\fystar},
\end{equation}
due to the fact that the convex function lies above its tangents.
Therefore, it follows that
\begin{eqnarray}\label{eq:Wanl-suff}
 \hspace{-2mm} &  & \hspace{-2mm} U[\fxhat,\fyhat] - U[\fxstar,\fystar]  \nonumber \\
\hspace{-3mm} &=& \hspace{-3mm} \iint \hspace{-2mm} F(\x,\y,\fxhat,\fyhat) - F(\x,\y,\fxstar,\fystar) d\x d\y  \nonumber \\
\hspace{-3mm} &=& \hspace{-3mm} \iint \hspace{-2mm} F(\x,\y,\fxhat,\fyhat) - F(\x,\y,\fxstar,\fystar) d\x d\y
+ \alpha_0 \iint \hspace{-2mm} F_0(\x,\y,\fxhat) - F_0(\x,\y,\fxstar) d\x d\y \nonumber \\
\hspace{-3mm} &+& \hspace{-3mm} \sum\limits_{i=1}^n \hspace{-1mm}\zeta_i \hspace{-1mm} \iint \hspace{-2mm} F_1^{(i)}(\x,\y,\fxhat) \hspace{-1mm}- \hspace{-1mm} F_1^{(i)}(\x,\y,\fxstar) d\x d\y + \sum\limits_{i=1}^n \sum\limits_{j=1}^n \hspace{-1mm}\gamma_{ij} \hspace{-1mm} \iint \hspace{-2mm} F_2^{(i,j)}(\x,\y,\fxhat)\hspace{-1mm} - \hspace{-1mm}F_2^{(i,j)}(\x,\y,\fxstar) d\x d\y \nonumber \\
\hspace{-3mm} &+& \hspace{-3mm} \sum\limits_{i=1}^n \hspace{-1mm}\eta_i \hspace{-1mm} \iint \hspace{-2mm} F_3^{(i)}(\x,\y,\fxhat) \hspace{-1mm}- \hspace{-1mm} F_3^{(i)}(\x,\y,\fxstar) d\x d\y + \sum\limits_{i=1}^n \sum\limits_{j=1}^n \hspace{-1mm}\theta_{ij} \hspace{-1mm} \iint \hspace{-2mm} F_4^{(i,j)}(\x,\y,\fxhat)\hspace{-1mm} - \hspace{-1mm}F_4^{(i,j)}(\x,\y,\fxstar) d\x d\y \nonumber \\
\hspace{-3mm} &+& \hspace{-3mm} \int \hspace{-2mm} \lambda(\y) \bigg[ \fyhat(\y) - \int \hspace{-2mm} \fxhat(\x)\fw(\y-\x) d\x \bigg] d\y
- \int \hspace{-2mm} \lambda(\y) \bigg[ \fystar(\y) - \int \fxstar(\x)\fw(\y-\x) d\x \bigg] d\y \nonumber \\
\hspace{-3mm} &=& \hspace{-3mm} \iint \hspace{-1mm} K(\x,\y,\fxhat,\fyhat) - K(\x,\y,\fxstar,\fystar) d\x d\y
+ \int \hspace{-1mm} (\fyhat - \fystar) \lambda(\y) d\y  \nonumber \\
\end{eqnarray}

Based on (\ref{eq:Wanl-tangent}),  the righthand side of (\ref{eq:Wanl-suff}) can be lower bounded as follows:
\begin{equation}\label{eq:Wanl-suff-con}
\begin{split}
 U[\fxhat,\fyhat] - U[\fxstar,\fystar] & \\
\geq & \hspace*{-.1cm} \iint \hspace*{-.1cm} \left[ (\fxhat \hspace*{-.1cm} - \hspace*{-.1cm} \fxstar) K'_{\fx} \hspace*{-.12cm} + \hspace*{-.1cm} (\fyhat \hspace*{-.1cm} - \hspace*{-.1cm} \fystar) K'_{\fy} \right]\Big|_{\fx=\fxstar,\fy=\fystar} \hspace*{-.15cm} d\x d\y \\
&+ \int (\fyhat - \fystar) \lambda(\y) d\y \\
\overset{(a)}{=}& \int (\fxhat-\fxstar) \left[ \int K'_{\fx} \Big|_{\fx=\fxstar,\fy=\fystar} d\y \right] d\x \\
&+ \int (\fyhat - \fystar) \left[ \int K'_{\fy} d\x + \tilde{K}'_{\fy}  \right] \Big|_{\fx=\fxstar,\fy=\fystar} d\y \\
\overset{(b)}{=}& 0,
\end{split}
\end{equation}
where (a) follows from the fact that
$$\tilde{K}'_{\fy}\Big|_{\fx=\fxstar,\fy=\fystar} = \lambda(\y),$$
and (b) is due to (\ref{WA_eq21_2}) and (\ref{WA_eq21_3}). This proves the sufficiency of the Gaussian distributions, and therefore, $f_{\scriptscriptstyle X^*}$ and $f_{\scriptscriptstyle Y^*}$ minimize the variational problem.
 % there exists only one solution, which is the multi-variate Gaussian density function and it satisfies Euler's equation in (\ref{WA_eq21_2}) and (\ref{WA_eq21_3}). Therefore,  $f_{\scriptscriptstyle X^*}$ and $f_{\scriptscriptstyle Y^*}$ are also sufficient in this case.

\begin{rem}
The constraints related to the vector means in (\ref{WA_eq17_3_2}) and (\ref{WA_eq17_3_5}) are unnecessary. Without these constraints, the optimal solutions are still multi-variate Gaussian density functions but the vector  means are  equal to zero.
\end{rem}
\end{proof}
\end{thm}

%%%%%%%%%%

%%%%%% The section of EPI is removed herein the paper.

%%%%%%%%%%%%%%%%%%
\section{Extremal Entropy Inequality}\label{EI}

Extremal entropy inequality, proposed by Liu and Viswanath \cite{Extremal:Liu}, was motivated by multi-terminal information theoretic problems such as the vector Gaussian
broadcast channel and the distributed source coding with a single quadratic distortion constraint.  EEI  is an entropy power
inequality which includes a covariance constraint. Because of the covariance constraint, the extremal entropy inequality could not be proved directly by using the classical Entropy Power Inequality (EPI). Therefore, new techniques (\cite{CapBroad:Shamai}, \cite{Extrem:SPark}) were adopted in the proofs reported in \cite{Extremal:Liu}, \cite{Extrem:SPark}. In this section, the extremal entropy inequality will be proved using a variational approach.

%%%%%%%%%%%%%%%%%%%%

%Theorem \ref{EI_thm1} can be generalized for random vectors as shown in the following two theorems.
%%%%%%%%%%%%%%%%%%
%Theorem 3
\begin{thm}\label{EI_thm2}
Assume that $\mu \geq 1$ is an arbitrary but fixed constant  and $\boldsymbol{\Sigma}$ is a positive semi-definite matrix. A Gaussian random vector $\mathbf{W}_{\scriptscriptstyle G}$ with  positive definite covariance matrix $\boldsymbol{\Sigma}_{\scriptscriptstyle W}$  is assumed to be independent of an arbitrary random vector $\mathbf{X}$ whose  covariance matrix $\boldsymbol{\Sigma}_{\scriptscriptstyle X}$ satisfies $\boldsymbol{\Sigma}_{\scriptscriptstyle X} \preceq \boldsymbol{\Sigma}$. Then, there exists a Gaussian random vector $\mathbf{X}_{\scriptscriptstyle G}^{*}$ with covariance matrix $\boldsymbol{\Sigma}_{\scriptscriptstyle X^{*}}$ which satisfies the following inequality:
\begin{eqnarray}
\label{EI_eq13_1}
    h(\mathbf{X}) - \mu h(\mathbf{X}+\mathbf{W}_{\scriptscriptstyle G}) \leq h(\mathbf{X}_{\scriptscriptstyle G}^{*}) - \mu h(\mathbf{X}_{\scriptscriptstyle G}^{*}+\mathbf{W}_{\scriptscriptstyle G}),
\end{eqnarray}
where $\boldsymbol{\Sigma}_{\scriptscriptstyle X^{*}} \preceq \boldsymbol{\Sigma}$.
\begin{proof}
By setting $\mathbf{Y}=\mathbf{X}+\mathbf{W_{\subg}}$, we first consider the following variational problem (without loss of generality, we assume that $\mathbf{X}$, $\mathbf{W}_{\subg}$, and $\mathbf{Y}$ have zero mean):
\begin{eqnarray}
\label{EI_eq20_1}
\hspace{-10mm}&&\min_{f_{\scriptscriptstyle X}, f_{\scriptscriptstyle Y}} \quad \int \int f_{\scriptscriptstyle X}(\mathbf{x}) f_{\scriptscriptstyle W}(\mathbf{y}-\mathbf{x}) ( -\mu\log f_{\scriptscriptstyle Y}(\mathbf{y}) + \log f_{\scriptscriptstyle X}(\mathbf{x})
+ \mu\left(\mu-1\right) \log f_{\scriptscriptstyle W}(\mathbf{y}-\mathbf{x}) ) d\mathbf{x} d\mathbf{y}\\
\hspace{-10mm}&&\text{s.t.}\quad\quad \int\int f_{\scriptscriptstyle X}(\mathbf{x}) f_{\scriptscriptstyle W}(\mathbf{y}-\mathbf{x})d\mathbf{x}d\mathbf{y} = 1, \nonumber\\
\hspace{-10mm}&&\hspace{12mm} \int\int \mathbf{y}\mathbf{y}^{\scriptscriptstyle T} f_{\scriptscriptstyle X}(\mathbf{x}) f_{\scriptscriptstyle W}(\mathbf{y}-\mathbf{x})d\mathbf{x} d\mathbf{y} =
\int\int \mathbf{x}\mathbf{x}^{\scriptscriptstyle T} f_{\scriptscriptstyle X}(\mathbf{x}) f_{\scriptscriptstyle W}(\mathbf{y}-\mathbf{x})d\mathbf{x}d\mathbf{y}\nonumber\\
\hspace{-10mm}&&\hspace{15mm} + \int\int \left(\mathbf{y}-\mathbf{x}\right)\left(\mathbf{y}-\mathbf{x}\right)^{\scriptscriptstyle T} f_{\scriptscriptstyle X}(\mathbf{x}) f_{\scriptscriptstyle W}(\mathbf{y}-\mathbf{x})d\mathbf{x}d\mathbf{y},\nonumber\\
\hspace{-10mm}&&\hspace{12mm} \int\int \mathbf{x}\mathbf{x}^{\scriptscriptstyle T} f_{\scriptscriptstyle X}(\mathbf{x}) f_{\scriptscriptstyle W}(\mathbf{y}-\mathbf{x})d\mathbf{x}d\mathbf{y}  \preceq  \boldsymbol{\Sigma},\nonumber\\
\hspace{-10mm}&&\hspace{12mm} \int\int \mathbf{y}\mathbf{y}^{\scriptscriptstyle T} f_{\scriptscriptstyle X}(\mathbf{x}) f_{\scriptscriptstyle W}(\mathbf{y}-\mathbf{x})d\mathbf{x}d\mathbf{y} = \boldsymbol{\Sigma}_{\scriptscriptstyle Y^*}, \nonumber\\
\hspace{-10mm}&&\hspace{12mm} -\int \int f_{\scriptscriptstyle X}(\mathbf{x}) f_{\scriptscriptstyle W}(\mathbf{y}-\mathbf{x}) \log f_{\scriptscriptstyle X}(\mathbf{x}) d\mathbf{x} d\mathbf{y} \geq p_{\scriptscriptstyle X}, \nonumber\\
\label{EI_eq20_2}
\hspace{-10mm}&&\hspace{12mm} f_{\scriptscriptstyle Y}(\mathbf{y}) = \int f_{\scriptscriptstyle X}(\mathbf{x}) f_{\scriptscriptstyle W}(\mathbf{y}-\mathbf{x})d\mathbf{x},
\end{eqnarray}
where $p_{\scriptscriptstyle X}$ is a constant, and $\boldsymbol{\Sigma}_{\scriptscriptstyle Y^*}$ stands for  the covariance matrix of the optimal solution  $\mathbf{Y}$.  The constraint $-\iint \fx(\x) \fw(\y-\x) \log\fx(\x)d\x d\y \geq p_x$   means that the differential entropy of $\mathbf{X}$ is greater than a constant $p_x$, i.e., $H(\mathbf{X})\geq p_x$, and it is introduced because it helps to convexify the problem by enforcing the semi-positive definiteness of the resulting functional second-order variation.  This is due to the fact that this constraint introduces an additional Lagrange multiplier $\alpha_1$, which can be selected appropriately to ensure the non-negative definiteness of the second-order variation.  Since $p_x$ can be any arbitrary small number, we believe that adding this additional constraint is reasonable.  In addition, the term $\mu(\mu-1)\iint \fx(\x)\fw(\y-\x)\log{\fw(\y-\x)}d\x d\y = \mu(\mu-1) h(\mathbf{W}_{\subg})$ is added to the objective functional (\ref{EI_eq20_1}), and being a constant, it does not affect the optimization problem.  Without loss of generality, the matrix $\boldsymbol{\Sigma}$ is assumed to be a positive definite matrix due to the same arguments mentioned in \cite{Extremal:Liu}.

The optimization problem  (\ref{EI_eq20_1})  is re-cast  as follows:
\begin{eqnarray}
\label{EI_eq21_1}
\hspace*{-4mm}&&\hspace*{-4mm}\min_{f_{\scriptscriptstyle X}, f_{\scriptscriptstyle Y}} \;  \int \hspace*{-2mm} \int f_{\scriptscriptstyle X}(\mathbf{x}) f_{\scriptscriptstyle W}(\mathbf{y}-\mathbf{x}) [ -\mu\log f_{\scriptscriptstyle Y}(\mathbf{y}) + \log f_{\scriptscriptstyle X}(\mathbf{x})
+ \mu\left(\mu-1\right) \log f_{\scriptscriptstyle W}(\mathbf{y}-\mathbf{x})] d\mathbf{x} d\mathbf{y}\\
\label{EI_eq21_5_1}
\hspace*{-4mm}&&\hspace*{-4mm}\text{s.t.}\; \int \hspace*{-2mm} \int f_{\scriptscriptstyle X}(\mathbf{x}) f_{\scriptscriptstyle W}(\mathbf{y}-\mathbf{x})d\mathbf{x}d\mathbf{y} = 1,\\
\label{EI_eq21_5_2}
\hspace*{-4mm}&&\hspace*{-1mm} \int \hspace*{-2mm} \int \left(y_i y_j -x_i x_j - \left(y-x\right)_i \left(y-x\right)_j \right) f_{\scriptscriptstyle X}(\mathbf{x}) f_{\scriptscriptstyle W}(\mathbf{y}-\mathbf{x})d\mathbf{x} d\mathbf{y} = 0,\\
\label{EI_eq21_5_3}
\hspace*{-4mm}&&\hspace*{-1mm}\sum_{i=1}^{n}\sum_{j=1}^n\left( \int\int x_i x_j \xi_i \xi_j f_{\scriptscriptstyle X}(\mathbf{x}) f_{\scriptscriptstyle W}(\mathbf{y}-\mathbf{x})d\mathbf{x}d\mathbf{y} \right)  \leq  \sum_{i=1}^{n}\sum_{j=1}^n\sigma^2_{ij} \xi_i \xi_j,\\
\label{EI_eq21_5_4}
\hspace*{-4mm}&&\hspace*{-1mm} \int \hspace*{-2mm}\int y_iy_j f_{\scriptscriptstyle X}(\mathbf{x}) f_{\scriptscriptstyle W}(\mathbf{y}-\mathbf{x})d\mathbf{x}d\mathbf{y} = {\sigma}^2_{\scriptscriptstyle Y^*_{ij}}, \\
\label{EI_eq21_5_5}
\hspace*{-4mm}&&\hspace*{-1mm} -\int \hspace*{-2mm}\int f_{\scriptscriptstyle X}(\mathbf{x}) f_{\scriptscriptstyle W}(\mathbf{y}-\mathbf{x}) \log f_{\scriptscriptstyle X}(\mathbf{x}) d\mathbf{x} d\mathbf{y} \geq  p_{\scriptscriptstyle X},\\
\label{EI_eq21_5_6}
\hspace*{-4mm}&&\hspace*{-1mm} f_{\scriptscriptstyle Y}(\mathbf{y}) = \int f_{\scriptscriptstyle X}(\mathbf{x}) f_{\scriptscriptstyle W}(\mathbf{y}-\mathbf{x})d\mathbf{x},
\end{eqnarray}
where the  arbitrary deterministic non-zero vector $\boldsymbol{\xi}$ is defined as $[\xi_1, \ldots, \xi_n]^T$, $\sigma^2_{ij}$ and ${\sigma}^2_{\scriptscriptstyle Y^*_{ij}}$ denote the $i^{\text{th}}$ row and  $j^{\text{th}}$ column entry of $\boldsymbol{\Sigma}$ and $\boldsymbol{\Sigma}_{\scriptscriptstyle Y^*}$ ($i=1,\ldots,n$, and $j=1,\ldots,n$), respectively.

Using Lagrange multipliers, as shown in Corollary \ref{CAL_cor1}, the functional problem in (\ref{EI_eq21_1}) and the constraints in (\ref{EI_eq21_5_1})-(\ref{EI_eq21_5_6}) can be  expressed in terms of the Lagrangian:
\begin{eqnarray}
\label{EI_eq22_1}
\min_{f_{\scriptscriptstyle X}, f_{\scriptscriptstyle Y}} \quad  \int \left(\int K(\mathbf{x},\mathbf{y},f_{\scriptscriptstyle X}, f_{\scriptscriptstyle Y}) d\mathbf{x}\right) + \tilde{K}(\mathbf{y},f_{\scriptscriptstyle Y})d\mathbf{y},\nonumber\\
\end{eqnarray}
where
\begin{eqnarray}
\label{EI_eq23_1}
K(\mathbf{x},\mathbf{y},f_{\scriptscriptstyle X},f_{\scriptscriptstyle Y}) \hspace*{-2mm}&= &\hspace*{-2mm} f_{\scriptscriptstyle X}(\mathbf{x})f_{\scriptscriptstyle W}(\mathbf{y}-\mathbf{x}) [ -\mu\log f_{\scriptscriptstyle Y}(\mathbf{y}) +\log f_{\scriptscriptstyle X}(\mathbf{x}) + \mu\left(\mu-1\right) \log f_{\scriptscriptstyle W}(\mathbf{y}-\mathbf{x}) + \alpha_0\nonumber\\
\hspace*{-2mm}& + &\hspace*{-2mm} \sum_{i=1}^{n} \sum_{j=1}^{n} (\gamma_{ij} y_i y_j-\gamma_{ij} x_i x_j - \gamma_{ij} \left(y-x\right)_i \left(y-x\right)_j +  \theta x_i x_j \xi_i \xi_j + \phi_{ij} y_i y_j )-\alpha_1 \log f_{\scriptscriptstyle X}(\mathbf{x}) - \lambda(\mathbf{y})],\nonumber\\
\tilde{K}(\mathbf{y},f_{\scriptscriptstyle Y}) \hspace*{-2mm} &= & \hspace*{-2mm} \lambda(\mathbf{y})f_{\scriptscriptstyle Y}(\mathbf{y}).
\end{eqnarray}
The Lagrange multipliers $\alpha_0$, $\gamma_{ij}$, $\theta$, $\phi_{ij}$, $\alpha_1$, and $\lambda(\mathbf{y})$ correspond to the constraints in (\ref{EI_eq21_5_1}), (\ref{EI_eq21_5_2}), (\ref{EI_eq21_5_3}), (\ref{EI_eq21_5_4}), (\ref{EI_eq21_5_5}), and (\ref{EI_eq21_5_6}), respectively.

To find the optimal solutions, based on Corollary \ref{CAL_cor2}, the first-order variation condition is checked as follows:
\begin{eqnarray}
\label{EI_eq24_1}
\hspace*{-2mm}&&\hspace*{-2mm} \int K'_{f_{\scriptscriptstyle X}}\Big|_{f_{\scriptscriptstyle X}=f_{\scriptscriptstyle X^*}, f_{\scriptscriptstyle Y}=f_{\scriptscriptstyle Y^*}} d\y = \int f_{\scriptscriptstyle W}(\mathbf{y}-\mathbf{x}) [ -\mu\log f_{\scriptscriptstyle Y^*}(\mathbf{y}) +(1-\alpha_1)\log f_{\scriptscriptstyle X^*}(\mathbf{x})
+\mu\left(\mu-1\right) \log f_{\scriptscriptstyle W}(\mathbf{y}-\mathbf{x})+\alpha_0
 \nonumber \\
\hspace*{-3mm} &+& \hspace*{-3mm}\sum_{i=1}^{n} \sum_{j=1}^{n} (  \gamma_{ij} y_i y_j -  \gamma_{ij} x_i x_j - \gamma_{ij} \left(y-x\right)_i \left(y-x\right)_j + \theta x_i x_j \xi_i \xi_j +\phi_{ij} y_i y_j)- \lambda(\mathbf{y})+1-\alpha_1] d\y \nonumber \\
\hspace*{-2mm}& = & \hspace*{-2mm}  0.\\
\label{EI_eq24_2}
\hspace*{-2mm}&&\hspace*{-2mm} \int K'_{f_{\scriptscriptstyle Y}} d\x + \tilde{K}'_{\fy} \Big|_{f_{\scriptscriptstyle X}=f_{\scriptscriptstyle X^*}, f_{\scriptscriptstyle Y}=f_{\scriptscriptstyle Y^*}}
=  -\frac{\mu \int f_{\scriptscriptstyle X}(\mathbf{x})f_{\scriptscriptstyle W}(\mathbf{y}-\mathbf{x}) d\mathbf{x}}{f_{\scriptscriptstyle Y}(\mathbf{y})} + \lambda(\mathbf{y}) =  0.
\end{eqnarray}
The following expressions  satisfy the equalities in (\ref{EI_eq24_1}) and (\ref{EI_eq24_2}):
\begin{eqnarray}
\label{EI_eq25_1}
\lambda(\mathbf{y}) \hspace*{-2mm}&=&\hspace*{-2mm}\mu,\nonumber\\
f_{\scriptscriptstyle Y^*}(\mathbf{y}) %& = & \exp\left\{\frac{1}{\mu} \left(\mathbf{y}^{\scriptscriptstyle T}\left(\boldsymbol{\Gamma}+\boldsymbol{\Phi}\right) \mathbf{y} + c_{\scriptscriptstyle Y}\right)\right\}\nonumber\\
\hspace*{-2mm}&=&\hspace*{-2mm}  \left(2\pi\right)^{-\frac{n}{2}} \left|-\frac{\mu}{2} \left(\boldsymbol{\Gamma}+\boldsymbol{\Phi}\right)^{-1} \right|^{-\frac{1}{2}} \exp\left\{-\frac{1}{2}\mathbf{y}^{\scriptscriptstyle T} \left(-\frac{\mu}{2}\left(\boldsymbol{\Gamma}+\boldsymbol{\Phi}\right)^{-1}\right)^{-1}\mathbf{y}\right\}
\left(2\pi\right)^{\frac{n}{2}} \left|-\frac{\mu}{2} \left(\boldsymbol{\Gamma}+\boldsymbol{\Phi}\right)^{-1} \right|^{\frac{1}{2}} \exp\left\{\frac{c_{\scriptscriptstyle Y}}{\mu}\right\}\nonumber\\
%%%%%%%%%%%
f_{\scriptscriptstyle W}(\mathbf{y}-\mathbf{x}) %& = & \exp\left\{\frac{1}{\mu\left(\mu-1\right)}\left( \left(\mathbf{y}-\mathbf{x}\right)^{\scriptscriptstyle T} \boldsymbol{\Gamma}\left(\mathbf{y}-\mathbf{x}\right)-c_{\scriptscriptstyle W}\right)\right\}\nonumber\\
\hspace*{-2mm}&=&\hspace*{-2mm}  \left(2\pi \right)^{-\frac{n}{2}} \left|-\frac{\mu\left(\mu-1\right)}{2}  \boldsymbol{\Gamma}^{-1}\right|^{-\frac{1}{2}}
\exp\left\{-\frac{1}{2}\left(\mathbf{y}-\mathbf{x}\right)^{\scriptscriptstyle T}\left(-\frac{\mu\left(\mu-1\right)}{2} \boldsymbol{\Gamma}^{-1}\right)^{-1}\left(\mathbf{y}-\mathbf{x}\right)\right\}\nonumber\\
\hspace*{-2mm}&  & \hspace*{-2mm} \cdot \left(2\pi \right)^{\frac{n}{2}} \left|-\frac{\mu\left(\mu-1\right)}{2} \boldsymbol{\Gamma}^{-1}\right|^{\frac{1}{2}} \exp\left\{-\frac{c_{\scriptscriptstyle W}}{\mu\left(\mu-1\right)}\right\},\nonumber\\
%%%%%%%%%%%
\label{EI_eq25_3}
f_{\scriptscriptstyle X^*}(\mathbf{x}) %& = & \exp\left\{\frac{1}{1-\alpha_1} \left( \mathbf{x}^{\scriptscriptstyle T} \left(\boldsymbol{\Gamma}-\theta\boldsymbol{\Xi}\right) \mathbf{x} - \alpha_0 + \mu - 1 +\alpha_1 +c_{\scriptscriptstyle W} + c_{\scriptscriptstyle Y}\right)\right\}\nonumber\\
\hspace*{-2mm}&=&\hspace*{-2mm} \left(2\pi \right)^{-\frac{n}{2}}\left|-\frac{1-\alpha_1}{2} \left(\boldsymbol{\Gamma}-\theta\boldsymbol{\Xi}\right)^{-1}\right|^{-\frac{1}{2}}
\exp\left\{-\frac{1}{2}\mathbf{x}^{\scriptscriptstyle T} \left(-\frac{1-\alpha_1}{2} \left(\boldsymbol{\Gamma}-\theta\boldsymbol{\Xi}\right)^{-1}\right)^{-1}\mathbf{x}\right\}\nonumber\\
\hspace*{-2mm} &  & \hspace*{-2mm} \cdot \left(2\pi \right)^{\frac{n}{2}}\left|-\frac{1-\alpha_1}{2} \left(\boldsymbol{\Gamma}-\theta\boldsymbol{\Xi}\right)^{-1}\right|^{\frac{1}{2}}
\exp\left\{\frac{-\alpha_0+\mu-1+\alpha_1+c_{\scriptscriptstyle W}+c_{\scriptscriptstyle Y}}{1-\alpha_1}\right\},
\end{eqnarray}
where
\begin{eqnarray}
\label{EI_eq26_1}
\boldsymbol{\Phi} \hspace*{-3mm}& = & \hspace*{-3mm} \left[\begin{array}{ccc}
\phi_{11} & \cdots & \phi_{1n}\\
\vdots & \ddots & \vdots \\
\phi_{n1} & \cdots & \phi_{nn} \end{array}\right],  \quad \boldsymbol{\Gamma}  =  \left[\begin{array}{ccc}
\gamma_{11} & \cdots & \gamma_{1n}\\
\vdots & \ddots & \vdots \\
\gamma_{n1} & \cdots & \gamma_{nn} \end{array}\right] \nonumber \\
\boldsymbol{\Xi} \hspace*{-3mm} &=& \hspace*{-3mm}  \left[\begin{array}{ccc}
\xi_1 \xi_1 & \cdots & \xi_1 \xi_n\\
\vdots & \ddots & \vdots \\
\xi_n \xi_1& \cdots & \xi_n \xi_n \end{array}\right], \nonumber\\
\mathbf{x} \hspace*{-3mm} & = & \hspace*{-3mm} \left[x_1, \cdots, x_n\right]^T,\nonumber\\
\mathbf{y} \hspace*{-3mm} & = & \hspace*{-3mm} \left[y_1, \cdots, y_n\right]^T.\nonumber
%\theta \hspace*{-3mm} & \geq & \hspace*{-3mm} 0.
\end{eqnarray}
Now considering the constraints in (\ref{EI_eq21_5_1})-(\ref{EI_eq21_5_6}), the equations in (\ref{EI_eq25_3}) are further processed as follows:
\begin{eqnarray}
\label{EI_eq27_1}
f_{\scriptscriptstyle Y^*}(y) %& = & \left(2\pi\right)^{-\frac{n}{2}} \left|-\frac{\mu}{2} \left(\boldsymbol{\Gamma}+\boldsymbol{\Phi}\right)^{-1} \right|^{-\frac{1}{2}}\exp\left\{-\frac{1}{2}\mathbf{y}^{\scriptscriptstyle T} \left(-\frac{\mu}{2}\left(\boldsymbol{\Gamma}+\boldsymbol{\Phi}\right)^{-1}\right)^{-1}\mathbf{y}\right\}\nonumber\\
%&&\times \left(2\pi\right)^{\frac{n}{2}} \left|-\frac{\mu}{2} \left(\boldsymbol{\Gamma}+\boldsymbol{\Phi}\right)^{-1} \right|^{\frac{1}{2}} \exp\left\{\frac{c_{\scriptscriptstyle Y}}{\mu}\right\}\nonumber\\
\hspace*{-2mm}&=&\hspace*{-2mm} \left(2\pi\right)^{-\frac{n}{2}} \left|\boldsymbol{\Sigma}_{\scriptscriptstyle Y^*}\right|^{-\frac{1}{2}}\exp\left\{-\frac{1}{2}\mathbf{y}^{\scriptscriptstyle T} \boldsymbol{\Sigma}_{\scriptscriptstyle Y^*}^{-1}\mathbf{y}\right\}\nonumber\\
%%%%%%%%%%%%%
f_{\scriptscriptstyle W}(y-x) %& = & \left(2\pi \right)^{-\frac{n}{2}} \left|-\frac{\mu\left(\mu-1\right)}{2}\boldsymbol{\Gamma}^{-1}\right|^{-\frac{1}{2}} \exp\left\{-\frac{1}{2}\left(\mathbf{y}-\mathbf{x}\right)^{\scriptscriptstyle T}\left(-\frac{\mu\left(\mu-1\right)}{2} \boldsymbol{\Gamma}^{-1}\right)^{-1}\left(\mathbf{y}-\mathbf{x}\right)\right\}\nonumber\\
%&&\times \left(2\pi \right)^{\frac{n}{2}} \left|-\frac{\mu\left(\mu-1\right)}{2}\boldsymbol{\Gamma}^{-1}\right|^{\frac{1}{2}} \exp\left\{-\frac{c_{\scriptscriptstyle W}}{\mu\left(\mu-1\right)}\right\}\nonumber\\
\hspace*{-2mm}&=&\hspace*{-2mm}
\left(2\pi \right)^{-\frac{n}{2}} \left|\boldsymbol{\Sigma}_{\scriptscriptstyle W} \right|^{-\frac{1}{2}} \exp\left\{-\frac{1}{2}\left(\mathbf{y}-\mathbf{x}\right)^{\scriptscriptstyle T}\boldsymbol{\Sigma}_{\scriptscriptstyle W}^{-1} \left(\mathbf{y}-\mathbf{x}\right)\right\}\nonumber\\
f_{\scriptscriptstyle X^*}(x) %& = & \left(2\pi \right)^{-\frac{n}{2}}\left|-\frac{1-\alpha_1}{2} \left(\boldsymbol{\Gamma}-\theta\boldsymbol{\Xi}\right)^{-1}\right|^{-\frac{1}{2}}\exp\left\{-\frac{1}{2}\mathbf{x}^{\scriptscriptstyle T} \left(-\frac{1-\alpha_1}{2} \left(\boldsymbol{\Gamma}-\theta\boldsymbol{\Xi}\right)^{-1}\right)^{-1}\mathbf{x}\right\}\nonumber\\
%&&\times \left(2\pi \right)^{\frac{n}{2}}\left|-\frac{1-\alpha_1}{2} \left(\boldsymbol{\Gamma}-\theta\boldsymbol{\Xi}\right)^{-1}\right|^{\frac{1}{2}}\exp\left\{\frac{-\alpha_0+\mu-1+\alpha_1+c_{\scriptscriptstyle W}+c_{\scriptscriptstyle Y}}{1-\alpha_1}\right\}\nonumber\\
\hspace*{-2mm}&=&\hspace*{-2mm}\left(2\pi \right)^{-\frac{n}{2}}\left|\boldsymbol{\Sigma}_{\scriptscriptstyle X^*}\right|^{-\frac{1}{2}}\exp\left\{-\frac{1}{2}\mathbf{x}^{\scriptscriptstyle T} \boldsymbol{\Sigma}_{\scriptscriptstyle X^*}^{-1}\mathbf{x}\right\}
\end{eqnarray}
where
\begin{eqnarray}
\alpha_0 & = &  \mu-\left(1-\alpha_1\right)+\frac{\mu\left(\mu-1\right)}{2}\log \left(2\pi\right)^n \left|\boldsymbol{\Sigma}_{\scriptscriptstyle W}\right| \nonumber \\
&& -\frac{\mu}{2}\log \left(2\pi\right)^n\left|\boldsymbol{\Sigma}_{\scriptscriptstyle Y^*}\right|+\frac{1-\alpha_1}{2} \log\left(2\pi \right)^n \left|\boldsymbol{\Sigma}_{\scriptscriptstyle X^*}\right|,\nonumber\\
%%%
\boldsymbol{\Gamma} & = & -\frac{\mu\left(\mu-1\right)}{2} \boldsymbol{\Sigma}_{\scriptscriptstyle W}^{-1},\nonumber\\
\boldsymbol{\Phi} & = & -\boldsymbol{\Gamma} - \frac{\mu}{2} \boldsymbol{\Sigma}_{\scriptscriptstyle Y^*}^{-1}\nonumber\\
& = & \frac{\mu\left(\mu-1\right)}{2}\boldsymbol{\Sigma}_{\scriptscriptstyle W}^{-1} - \frac{\mu}{2} \left(\boldsymbol{\Sigma}_{\scriptscriptstyle X^*}+\boldsymbol{\Sigma}_{\scriptscriptstyle W}\right)^{-1} ,\nonumber\\
\boldsymbol{\Sigma}_{\scriptscriptstyle X^*}  & = & -\frac{1-\alpha_1}{2}\left(\boldsymbol{\Gamma}-\theta \boldsymbol{\Xi}   \right)^{-1}\nonumber\\
%\label{EI_eq28_4}
& = & \frac{1-\alpha_1}{2}\left(\frac{\mu\left(\mu-1\right)}{2} \boldsymbol{\Sigma}_{\scriptscriptstyle W}^{-1} +\theta \boldsymbol{\Xi} \right)^{-1} \nonumber\\
%\label{EI_eq28_5}
%& \succeq & \mathbf{0},\\
\label{EI_eq28_X}
\theta & \geq & 0,\\
\label{EI_eq28_6}
\alpha_1 & \leq & 1-\mu,\\
c_{\scriptscriptstyle W} & = & \frac{\mu\left(\mu-1\right)}{2}\log \left(2\pi \right)^n \left|\boldsymbol{\Sigma}_{\scriptscriptstyle W}\right|,\nonumber\\
c_{\scriptscriptstyle Y} & = & -\frac{\mu }{2}\log \left(2\pi\right)^n \left|\boldsymbol{\Sigma}_{\scriptscriptstyle Y^*}\right|,\nonumber\\
%\label{EI_eq28_7}
\left|\boldsymbol{\Sigma}_{\scriptscriptstyle X^*}\right| & = & \left(\frac{1}{2\pi e} \exp \left\{\frac{2}{n}p_{\scriptscriptstyle X}\right\}\right)^n. \nonumber %\leq \left|\boldsymbol{\Sigma}\right|.
\end{eqnarray}
%%%%%%%%%%
The inequality in (\ref{EI_eq28_6}) is due to the second-order variation condition, which will be presented later in this proof. The inequality (\ref{EI_eq28_X}) is based on the theory of KKT conditions since the multiplier associated with the inequality constraint is nonnegative. Moreover, the complementary slackness condition  in the KKT conditions leads to the following relationship:
\begin{equation}\label{eq:EEI-vec-comple}
\theta \Bigg[ \iint \Bigg(\sum\limits_{i=1}^n \sum\limits_{j=1}^n x_i x_j \xi_i \xi_j\Bigg) \fxstar(\x) \fw(\y-\x) d\x d\y  - \sum\limits_{i=1}^n \sum\limits_{j=1}^n \sigma_{ij}^2 \xi_i \xi_j \Bigg] = 0.
\end{equation}
%The inequality in (\ref{EI_eq28_5}) is always satisfied since the matrix $\boldsymbol{\Xi}$ is   positive semi-definite and $\theta$ is non-negative. The inequality in (\ref{EI_eq28_6}) will be proved later in this proof. The constant $p_{\scriptscriptstyle X}$ must be chosen to satisfy the inequality in (\ref{EI_eq28_6}). Then, the Lagrange multipliers always exist, and the necessary optimal solutions exist.

%Interestingly, similar to the complementary slackness in KKT conditions, when $\theta=0$ in (\ref{EI_eq28_4}), $\boldsymbol{\Sigma}_{\scriptscriptstyle X^*}=\left(1-\alpha_1\right)\mu^{-1}(\mu-1)^{-1} \boldsymbol{\Sigma}_{\scriptscriptstyle W}$, and it requires $\left(1-\alpha_1\right)\mu^{-1}(\mu-1)^{-1}\boldsymbol{\Sigma}_{\scriptscriptstyle W} \preceq \boldsymbol{\Sigma}$. When $\theta$ is non-zero, the equation in (\ref{EI_eq28_4}) is positive semi-definite, and it means  $\boldsymbol{\Sigma}_{\scriptscriptstyle X^*} = \left(1-\alpha_1\right)\mu^{-1}\left(\mu-1\right)^{-1} \boldsymbol{\Sigma}_{\scriptscriptstyle \tilde{W}}$, where $\boldsymbol{\Sigma}_{\scriptscriptstyle \tilde{W}} = \boldsymbol{\Sigma}_{\scriptscriptstyle W}-\boldsymbol{\Sigma}_{\scriptscriptstyle \hat{W}}$, where  $\boldsymbol{\Sigma}_{\scriptscriptstyle \hat{W}}$ and $\boldsymbol{\Sigma}_{\scriptscriptstyle \tilde{W}} $ are positive semi-definite matrices. When $1-\alpha_1=\mu$, then  $\boldsymbol{\Sigma}_{\scriptscriptstyle X^*} = \left(\mu-1\right)^{-1} \boldsymbol{\Sigma}_{\scriptscriptstyle \tilde{W}}$, which is exactly the same as the one in \cite{Extremal:Liu} and \cite{Extrem:SPark}.

Based on Corollary \ref{CAL_cor2}, to make the second variation nonnegative, the positive semi-definiteness of the following matrix is required:
\begin{eqnarray}
\label{EI_eq34_1}
\left[\begin{array}{cc}
K''_{f_{\scriptscriptstyle X^*} f_{\scriptscriptstyle X^*}} &  K''_{f_{\scriptscriptstyle X^*} f_{\scriptscriptstyle Y^*}}\\
K''_{f_{\scriptscriptstyle Y^*} f_{\scriptscriptstyle X^*}} &  K''_{f_{\scriptscriptstyle Y^*} f_{\scriptscriptstyle Y^*}}
\end{array}\right],
\end{eqnarray}
which further reduces to  the following condition:
\begin{eqnarray}
\label{EI_eq35_1}
\hspace*{-4mm}&&\hspace*{-4mm}\left[\begin{array}{cc} h_{\scriptscriptstyle X} & h_{\scriptscriptstyle Y}\end{array} \right] \left[\begin{array}{cc}
K''_{f_{\scriptscriptstyle X^*} f_{\scriptscriptstyle X^*}} &  K''_{f_{\scriptscriptstyle X^*} f_{\scriptscriptstyle Y^*}}\\
K''_{f_{\scriptscriptstyle Y^*} f_{\scriptscriptstyle X^*}} &  K''_{f_{\scriptscriptstyle Y^*} f_{\scriptscriptstyle Y^*}}
\end{array}\right]
\left[\begin{array}{c} h_{\scriptscriptstyle X} \\ h_{\scriptscriptstyle Y}\end{array} \right] \nonumber\\
\hspace*{-4mm}&=&\hspace*{-4mm}K''_{f_{\scriptscriptstyle X^*} f_{\scriptscriptstyle X^*}} h_{\scriptscriptstyle X}^2 + K''_{f_{\scriptscriptstyle Y^*} f_{\scriptscriptstyle Y^*}} h_{\scriptscriptstyle Y}^2 +  (K''_{f_{\scriptscriptstyle X^*} f_{\scriptscriptstyle Y^*}}+K''_{f_{\scriptscriptstyle Y^*} f_{\scriptscriptstyle X^*}}) h_{\scriptscriptstyle Y} h_{\scriptscriptstyle X}\nonumber\\
\hspace*{-4mm}&\geq & \hspace*{-2mm} 0,
\end{eqnarray}
where $h_{\scriptscriptstyle X}$ and $h_{\scriptscriptstyle Y}$ are arbitrary admissible functions.
Since $K''_{f_{\scriptscriptstyle X^*} f_{\scriptscriptstyle X^*}}$, $K''_{f_{\scriptscriptstyle X^*} f_{\scriptscriptstyle Y^*}}$, $K''_{f_{\scriptscriptstyle Y^*} f_{\scriptscriptstyle X^*}}$, and $K''_{f_{\scriptscriptstyle Y^*} f_{\scriptscriptstyle Y^*}}$ are defined as
\begin{eqnarray}
\label{EI_eq36_1}
K''_{f_{\scriptscriptstyle X^*} f_{\scriptscriptstyle X^*}} & = & \frac{(1-\alpha_1)f_{\scriptscriptstyle W}(\mathbf{y}-\mathbf{x})}{f_{\scriptscriptstyle X^*}(\mathbf{x})},\nonumber \\
K''_{f_{\scriptscriptstyle X^*} f_{\scriptscriptstyle Y^*}}  &= & -\frac{\mu f_{\scriptscriptstyle W}(\mathbf{y}-\mathbf{x})}{f_{\scriptscriptstyle Y^*}(\mathbf{y})},\nonumber\\
K''_{f_{\scriptscriptstyle Y^*} f_{\scriptscriptstyle X^*}} & = & -\frac{\mu f_{\scriptscriptstyle W}(\mathbf{y}-\mathbf{x})}{f_{\scriptscriptstyle Y^*}(\mathbf{y})}, \nonumber \\
K''_{f_{\scriptscriptstyle Y^*} f_{\scriptscriptstyle Y^*}}  & = &  \frac{\mu f_{\scriptscriptstyle X^*}(\mathbf{x}) f_{\scriptscriptstyle W}(\mathbf{y}-\mathbf{x})}{f_{\scriptscriptstyle Y^*}(\mathbf{y})^2},
\end{eqnarray}
the condition in (\ref{EI_eq35_1}) requires
\begin{eqnarray}
\label{EI_eq37_1}
\hspace*{-4mm}&&\hspace*{-4mm} \frac{(1-\alpha_1) f_{\scriptscriptstyle W} (\mathbf{y}-\mathbf{x})}{f_{\scriptscriptstyle X^*}(\mathbf{x})}h_{\scriptscriptstyle X}(\mathbf{x})^2 - 2\frac{\mu f_{\scriptscriptstyle W}(\mathbf{y}-\mathbf{x})}{f_{\scriptscriptstyle Y^*}(\mathbf{y})}h_{\scriptscriptstyle X}(\mathbf{x})h_{\scriptscriptstyle Y}(\mathbf{y}) +\frac{\mu f_{\scriptscriptstyle X^*}(\mathbf{x})f_{\scriptscriptstyle W}(\mathbf{y}-\mathbf{x})}{f_{\scriptscriptstyle Y^*}(\mathbf{y})^2}h_{\scriptscriptstyle Y}(\mathbf{y})^2\nonumber\\
& \geq & \frac{\mu f_{\scriptscriptstyle W}(\mathbf{y}-\mathbf{x})}{f_{\scriptscriptstyle X^*}(\mathbf{x})} \left( h_{\scriptscriptstyle X}(\mathbf{x}) - \frac{f_{\scriptscriptstyle X^*}(\mathbf{x})}{f_{\scriptscriptstyle Y^*}(\mathbf{y})}h_{\scriptscriptstyle Y}(\mathbf{y}) \right)^2,
\end{eqnarray}
which holds true if $1-\alpha_1 \geq \mu$ (i.e., $\alpha_1 \leq 1-\mu \leq 0$). Condition $\alpha_1\leq 0$ is also imposed by the KKT complementary slackness condition corresponding to the constraint (\ref{EI_eq21_5_5}).  Therefore, the optimal solutions $f_{\scriptscriptstyle X^*}$ and $f_{\scriptscriptstyle Y^*}$ minimize the functional problem in (\ref{EI_eq21_1}), and the proof is completed because of convexity of the functional $K(\mathbf{x},\mathbf{y},f_{\scriptscriptstyle X},f_{\scriptscriptstyle Y})$ wrt variables $f_{\scriptscriptstyle X}$  and $f_{\scriptscriptstyle Y}$.

A more detailed alternative justification of the fact the Gaussian distributions $f_{\scriptscriptstyle X^*}$ and $f_{\scriptscriptstyle Y^*}$ are global  minima is next presented. We will prove the sufficiency of the Gaussian distributions by showing $U[\fxhat,\fyhat] \geq U[\fxstar,\fystar]$, where $U[\cdot,\cdot]$ represents the objective functional in the problem and $\fxhat,\fyhat$ denote any arbitrary functions satisfying the boundary conditions and the constraints. First, the following functionals are defined:
\begin{eqnarray*}
F(\x,\y,\fx,\fy) &=& \fx(\x)\fw(\y-\x)(-\mu\log{\fy(\y)} + \log{\fx(\x)} + \mu(\mu-1)\log{\fw(\y-\x)}), \\
F_0(\x,\y,\fx) &=& \fx(\x)\fw(\y-\x),  \\
F_1^{(i,j)}(\x,\y,\fx) &=& \left(y_i y_j -x_i x_j - \left(y-x\right)_i \left(y-x\right)_j \right) \fx(\x)\fw(\y-\x), \\
F_2(\x,\y,\fx) &=& \left( \sum_{i=1}^{n}\sum_{j=1}^n x_i x_j \xi_i \xi_j \right) \fx(\x)\fw(\y-\x), \\
F_3^{(i,j)}(\x,\y,\fx) &=& y_i y_j \fx(\x)\fw(\y-\x),  \\
F_4(\x,\y,\fx) &=& -\fx(\x)\fw(\y-\x)\log{\fx(\x)},
\end{eqnarray*}
and thus
\begin{eqnarray*}
K(\x,\y,\fx,\fy)
&=& F(\x,\y,\fx,\fy) + \alpha_0 F_0(\x,\y,\fx)  + \sum\limits_{i=1}^n \sum\limits_{j=1}^n \gamma_{ij} F_1^{(i,j)}(\x,\y,\fx) + \theta F_2(\x,\y,\fx) \\
&+& \sum\limits_{i=1}^n \sum\limits_{j=1}^n \phi_{ij} F_3^{(i,j)}(\x,\y,\fx) + \alpha_1 F_4(\x,\y,\fx)
 - \lambda(\y) \fx(\x) \fw(\y-\x).
\end{eqnarray*}
It can be verified that the Hessian matrix of $K(\x,\y,\fx,\fy)$ w.r.t $\fx$ and $\fy$ is given by
$$\left[
    \begin{array}{cc}
      (1-\alpha_1)\fw(\y-\x)/\fx(\x) & -\mu\fw(\y-\x)/\fy(\y) \\
      -\mu\fw(\y-\x)/\fy(\y) & \mu\fx(\x)\fw(\y-\x)/\fy(\y)^2 \\
    \end{array}
  \right],
$$
which is positive semi-definite due to (\ref{EI_eq28_6}). The convexity property of $K(\x,\y,\fx,\fy)$ yields that
\begin{equation}\label{eq:EEI-vec-tan}
 K(\x,\y,\fxhat,\fyhat) - K(\x,\y,\fxstar,\fystar)
\geq  \left[ (\fxhat-\fxstar) K'_{\fx} + (\fyhat-\fystar) K'_{\fy} \right] \Big|_{\fx=\fxstar,\fy=\fystar},
\end{equation}
and it follows that
\begin{eqnarray}\label{eq:EEI-vec-suff}
 \hspace{-2mm} &  & \hspace{-2mm}  U[\fxhat,\fyhat] \hspace{-1mm}- \hspace{-1mm}U[\fxstar,\fystar]    \nonumber  \\
\hspace*{-4mm} &=& \hspace*{-4mm} \iint \hspace*{-1mm}F(\x,\y,\fxhat,\fyhat) - F(\x,\y,\fxstar,\fystar) d\x d\y \nonumber \\
\hspace*{-4mm} & \overset{(a)}{\geq} & \hspace*{-4mm} \iint \hspace*{-1mm}F(\x,\y,\fxhat,\fyhat) - F(\x,\y,\fxstar,\fystar) d\x d\y
 + \alpha_0 \left[ \iint \hspace*{-1mm} F_0(\x,\y,\fxhat) - F_0(\x,\y,\fxstar) d\x d\y \right] \nonumber \\
\hspace*{-4mm} &+& \hspace*{-4mm} \sum\limits_{i=1}^n \sum\limits_{j=1}^n \gamma_{ij} \left[ \iint \hspace*{-1mm} F_1^{(i,j)}(\x,\y,\fxhat) - F_1^{(i,j)}(\x,\y,\fxstar) d\x d\y \right]
 + \theta \left[ \iint \hspace*{-1mm} F_2(\x,\y,\fxhat) - F_2(\x,\y,\fxstar) d\x d\y \right] \nonumber \\
\hspace*{-4mm} &+& \hspace*{-4mm} \sum\limits_{i=1}^n \sum\limits_{j=1}^n \phi_{ij} \left[ \iint \hspace*{-1mm} F_3^{(i,j)}(\x,\y,\fxhat) - F_3^{(i,j)}(\x,\y,\fxstar) d\x d\y \right]
+ \alpha_1 \hspace*{-.1cm} \left[ \iint \hspace*{-1mm} F_4(\x,\y,\fxhat) - F_4(\x,\y,\fxstar) d\x d\y  \right] \nonumber \\
\hspace*{-4mm} &+& \hspace*{-4mm} \int \hspace*{-1mm} \lambda(\y) \left[ \fyhat(\y) - \int \hspace*{-1mm} \fxhat(\x)\fw(\y-\x) d\x \right] d\y - \int \hspace*{-1mm} \lambda(\y) \left[ \fystar(\y) - \int \hspace*{-1mm} \fxstar(\x)\fw(\y-\x) d\x \right] d\y \\
\hspace*{-4mm} &=& \hspace*{-4mm} \iint \hspace*{-1mm} K(\x,\y,\fxhat,\fyhat) - K(\x,\y,\fxstar,\fystar) d\x d\y
 + \int \hspace*{-1mm} \lambda(\y) \left( \fyhat(\y) - \fystar(\y) \right) d\y \nonumber \\
\hspace*{-4mm} & \overset{(b)}{\geq}& \hspace*{-4mm} \iint \hspace*{-1mm} \left[ (\fxhat - \fxstar) K'_{\fx}  +  (\fyhat \hspace*{-.1cm}-\hspace*{-.1cm} \fystar) K'_{\fy} \right] \Big|_{\fx=\fxstar,\fy=\fystar}  d\x d\y
 + \int \hspace*{-1mm} \lambda(\y) \left( \fyhat(\y) - \fystar(\y) \right) d\y \nonumber \\
\hspace*{-4mm} &=& \hspace*{-4mm} \int \hspace*{-1mm} (\fxhat-\fxstar) \left[\int \hspace*{-1mm} K'_{\fx}\Big|_{\fx=\fxstar,\fy=\fystar} d\y \right] d\x
 + \int (\fyhat-\fystar) \left[ \int \hspace*{-1mm} K'_{\fy}\Big|_{\fx=\fxstar,\fy=\fystar} d\x + \lambda(\y) \right] d\y \nonumber \\
\hspace*{-4mm} & \overset{(c)}{=}& \hspace*{-4mm} 0,
\end{eqnarray}
where the inequality (a) follows from the complementary slackness condition in the KKT conditions  (\ref{eq:EEI-vec-comple}). Indeed, since $\fxhat$ only represents an arbitrary feasible solution and $\theta \geq 0$, it follows that
$$\theta \left[ \iint F_2(\x,\y,\fxstar) d\x d\y - \sum\limits_{i=1}^n \sum\limits_{j=1}^n \sigma_{ij}^2 \xi_i \xi_j \right] = 0,$$
and
$$\theta \left[ \iint F_2(\x,\y,\fxhat) d\x d\y - \sum\limits_{i=1}^n \sum\limits_{j=1}^n \sigma_{ij}^2 \xi_i \xi_j \right] \leq 0,$$
and therefore, $\theta \left[ \iint F_2(\x,\y,\fxhat) - F_2(\x,\y,\fxstar) d\x d\y \right] \leq 0.$
Similarly, the complementary slackness condition associated with (\ref{EI_eq21_5_5}) leads to $\alpha_1  \left[ \iint F_4(\x,\y,\fxhat) - F_4(\x,\y,\fxstar) d\x d\y  \right]\leq 0 $. In addition, (b) is due to (\ref{eq:EEI-vec-tan}), and (c) follows from (\ref{EI_eq24_1}) and (\ref{EI_eq24_2}). This proves the sufficiency of Gaussian distributions. %Therefore, the optimal solutions $f_{\scriptscriptstyle X^*}$ and $f_{\scriptscriptstyle Y^*}$ minimize the functional problem in (\ref{EI_eq21_1}), and the proof is completed.
\begin{rem}
%In \cite{Extremal:Liu} and \cite{Extrem:SPark}, it was proved that a Gaussian density function is necessarily optimal for the extremal entropy inequality. Likewise, the current proof shows that the necessary optimal solution of the extremal entropy inequality is Gaussian.
The proposed proof only exploits calculus of variations tools. Unlike the previous proofs, this proof does not adopt neither the channel enhancement technique and EPI as in \cite{Extremal:Liu} nor the EPI and data processing inequality as in \cite{Extrem:SPark}.
\end{rem}
\end{proof}
\end{thm}
%%%%%%%%%%%%%%%%%%%%

%Theorem 3
\begin{thm}\label{EI_thm3}
Assume that $\mu \geq 1$ is an arbitrary but fixed constant and $\boldsymbol{\Sigma}$ is a positive semi-definite matrix. Independent Gaussian random vectors $\mathbf{W}_{\scriptscriptstyle G}$ with covariance matrix $\boldsymbol{\Sigma}_{\scriptscriptstyle W}$ and $\mathbf{V}_{\scriptscriptstyle G}$ with  covariance matrix $\boldsymbol{\Sigma}_{\scriptscriptstyle V}$  are assumed to be independent of an arbitrary random vector $\mathbf{X}$ with covariance matrix $\boldsymbol{\Sigma}_{\scriptscriptstyle X} \preceq \boldsymbol{\Sigma}$. Both  covariance matrices  $\boldsymbol{\Sigma}_{\scriptscriptstyle W}$  and  $\boldsymbol{\Sigma}_{\scriptscriptstyle V}$ are assumed to be positive definite. Then, there exists a Gaussian random vector $\mathbf{X}_{\scriptscriptstyle G}^{*}$ with covariance matrix $\boldsymbol{\Sigma}_{X^{*}}$ which satisfies the following inequality:
\begin{eqnarray}
\label{EI_eq50_1}
    h(\mathbf{X}+\mathbf{W}_{\scriptscriptstyle G}) - \mu h(\mathbf{X}+\mathbf{V}_{\scriptscriptstyle G}) \hspace*{-3mm}&\leq &\hspace*{-3mm} h(\mathbf{X}_{\scriptscriptstyle G}^{*}+\mathbf{W}_{\scriptscriptstyle G}) - \mu h(\mathbf{X}_{\scriptscriptstyle G}^{*}+\mathbf{V}_{\scriptscriptstyle G}),
\end{eqnarray}
where $\boldsymbol{\Sigma}_{X^{*}} \preceq \boldsymbol{\Sigma}$.
\begin{proof}
See Appendix \ref{EI_thm3_proof}.
\begin{rem}
The proposed  proof does not borrow any techniques from \cite{Extremal:Liu}.  Even though the proposed proof adopts the equality condition for the data processing inequality,  a result which  was also  exploited in  \cite{Extrem:SPark}, the proposed proof is different from the one in \cite{Extrem:SPark} due to the following features.  First, the proposed proof uses the equality condition of the data processing inequality only once while the proof in \cite{Extrem:SPark} uses it twice.  The proof in \cite{Extremal:Liu} exploited the channel enhancement technique twice, which is equivalent to using the equality condition in the data processing inequality. Second,  the proposed proof does not use the moment generating function technique unlike the proof proposed in \cite{Extrem:SPark}; instead the current proof directly exploits a property of the conditional mutual information pertaining to a Markov chain.
\end{rem}
\end{proof}
\end{thm}

\section{Applications}\label{APP}

Because of the easiness to incorporate a broad class of constraints,  the proposed variational framework  finds usage  in a large number of applications. Herein section, we will briefly illustrate some potential applications in this regard and state several open research problems which might  be also addressed within the considered functional framework.

\subsection{Gaussian Wire-tap Channel}
The secrecy capacity of Gaussian wire-tap channel has been studied by many researchers \cite{SecrecyCap:Liu}, \cite{Cheong}. We will approach the Gaussian wire-tap problem from the estimation viewpoint, rather than considering the secrecy capacity from an information theoretic perspective.

The following scalar Gaussian wire-tap channel is considered:
\begin{eqnarray}\label{wire_eq1_1}
Y_1 & = & a X + W_{\scriptscriptstyle G},\nonumber\\
Y_2 & = & a X + W_{\scriptscriptstyle G} + Z_{\scriptscriptstyle G},
\end{eqnarray}
where $X$ is an arbitrary but fixed random variable with zero mean and unit variance, $a$ is a constant, and $W_{\scriptscriptstyle G}$ and $Z_{\scriptscriptstyle G}$ are Gaussian random variables with variances $\sigma_{\scriptscriptstyle W}^2$ and $\sigma_{\scriptscriptstyle Z}^2$, respectively. The random variables $W_{\scriptscriptstyle G}$  and $Z_{\scriptscriptstyle G}$  are independent of each other, and they have zero mean. In the channel model (\ref{wire_eq1_1}), $Y_1$ and $Y_2$ are considered as a legitimate receiver and as an eavesdropper, respectively.  The goal of this problem is the following.  Assume that both receivers use  minimum mean square error (MMSE) estimators.  Given the value of the mean square error (MSE), which allows to correctly decode  the legitimate receiver, what is the optimal distribution which maximizes the difference between the MSE in the legitimate receiver and the MSE in the eavesdropper?

The above mentioned problem  adopts both  practical and reasonable assumptions due to the following reasons.  First, the MMSE estimator is an optimal estimator in the sense that it minimizes the MSE. Therefore, it is reasonable to use such an optimal estimator. Second, to prevent from eavesdropping, finding the signal distribution that maximizes the difference between  the MSEs corresponding to the legitimate receiver and the eavesdropper, respectively, represents a legitimate design objective. To find the optimal distribution, the following functional problem is constructed:
\begin{eqnarray}\label{wire_eq2_1}
&& \max_{f_{\scriptscriptstyle X}(x)} \quad Var(X|Y_2) - Var(X|Y_1),\nonumber\\
&& \text{s.t.}\quad Var(X|Y_1) = R,
\end{eqnarray}
where $Var(X|Y)=\mathbb{E}\left[\left(X-\mathbb{E}\left[X|Y\right]\right)^2\right]$, $\mathbb{E}[\cdot]$ denotes the expectation operator, and $R$ is a constant.

The optimization problem in (\ref{wire_eq2_1}) is expressed as
\begin{eqnarray}
\label{wire_eq3_1}
&& \max_{f_{\scriptscriptstyle X}(x)} \quad Var(\mathbb{E}\left[X|Y_1\right]|Y_2),\\
\label{wire_eq3_2}
&& \text{s.t.}\quad \mathbb{E}\left[ \mathbb{E} \left[X|Y_1\right]^2\right] = 1- R.
\end{eqnarray}
The equation in (\ref{wire_eq3_1}) is due to the total law of variance and the Markov chain $X \rightarrow Y_1 \rightarrow Y_2$. Since $\mathbb{E}[X^2]=1$, the equation (\ref{wire_eq3_2}) follows from the constraint in (\ref{wire_eq2_1}).

The objective function in (\ref{wire_eq3_1}) is further expressed as
\begin{eqnarray}
\label{wire_eq4_1}
\hspace*{-2mm}Var\left(\mathbb{E}\left[X|Y_1\right]|Y_2\right)=\mathbb{E} \left[\mathbb{E}\left[X|Y_1\right]^2\right]-\mathbb{E} \left[\mathbb{E} \left[X|Y_2\right]^2\right]
\end{eqnarray}
and using the equations (\ref{wire_eq3_2}), (\ref{wire_eq4_1}), the optimization problem in (\ref{wire_eq3_1}) is re-formulated in terms of the following variational problem:
\begin{eqnarray}
\label{wire_eq5_1}
&&\min_{f_{\scriptscriptstyle Y_2}, g} \quad \int \frac{1}{f_{\scriptscriptstyle Y_2} (y)} g(y)^2 dy,\\
\label{wire_eq5_3}
&&\hspace{10mm} \int y^2 f_{\scriptscriptstyle Y_2}(y) dy = m_{\scriptscriptstyle Y_2}^2,\\
\label{wire_eq5_4}
&&\hspace{10mm}g(y) = \int x f_{\scriptscriptstyle Y_2|X}(y|x) f_{\scriptscriptstyle X}(x) dx,
\end{eqnarray}
where $f_{\scriptscriptstyle X}(x)$ and $f_{\scriptscriptstyle Y_2}(y)$ are the probability density functions of $X$ and $Y_2$, respectively, and $m_{\scriptscriptstyle Y_2}^2$ stands for  the second-order moment of $Y_2$.

Since the first term in (\ref{wire_eq4_1}) is given and
\begin{eqnarray}
\mathbb{E} \left[\mathbb{E} \left[X|Y_2\right]^2\right] = \int f_{\scriptscriptstyle Y_2}(y)\left(\int x \frac{f_{\scriptscriptstyle Y_2|X}(y|x) f_{\scriptscriptstyle X}(x)}{f_{\scriptscriptstyle Y_2}(y)} dx\right)^2dy,\nonumber
\end{eqnarray}
the objective function in (\ref{wire_eq5_1}) is derived from the equation (\ref{wire_eq3_1}). Also, the additional constraint in (\ref{wire_eq5_3}) is required to solve this variational problem.

Considering the Lagrange multipliers $\lambda_1$  and $\lambda(y)$ to account for the constraints in (\ref{wire_eq5_3})  and (\ref{wire_eq5_4}), respectively,  the following variational problem is constructed:
\begin{eqnarray}
\int K(y, f_{\scriptscriptstyle Y_2}, g) dy,\nonumber
\end{eqnarray}
where
\begin{eqnarray}
\label{wire_eq6_1}
\hspace*{-4mm}&&\hspace*{-4mm}K(y, f_{\scriptscriptstyle Y_2}, g)=\frac{g(y)^2}{f_{\scriptscriptstyle Y_2}(y)}  + \lambda_1 y^2 f_{\scriptscriptstyle Y_2}(y) +\lambda(y)\left(g(y)-\int \hspace*{-2mm}x f_{\scriptscriptstyle Y_2|X}(y|x) f_{\scriptscriptstyle X}(x) dx \right).
\end{eqnarray}
In accordance with Theorem \ref{CAL_thm1}, we can determine $g^*$ and $f_{\scriptscriptstyle Y_2}^*$  to  enforce  the first-order variation to be zero:
\begin{eqnarray}
\label{wire_eq7_1}
K_{f_{\scriptscriptstyle Y_2}^*} & = &-\frac{g^*(y)^2}{f_{\scriptscriptstyle Y_2}^*(y)^2} +\lambda_1 y^2 =0,\\
K_{g^*} & = & \frac{2g^*(y)}{f_{\scriptscriptstyle Y_2}^*(y)} + \lambda(y) = 0,\nonumber
\end{eqnarray}
Taking into account (\ref{wire_eq7_1}),  it follows further that
\begin{eqnarray}
\label{wire_eq8_1}
\mathbb{E}\left[X^*|Y_2^*\right] & = & \frac{g^*(y)}{f_{\scriptscriptstyle Y_2}^*(y)} = \sqrt{\lambda_1} y.
\end{eqnarray}
Since $\mathbb{E}[X^*|Y_2^*]$, the MMSE estimator,  is a linear function of $y$ and the channel is corrupted with additive Gaussian noise,  it  is necessary  that $X^*$ is a Gaussian random variable. 
%It can be observed that given $Y^*_2$, a Gaussian random variable $X^*$, i.e.,
%$$f_{X^*|Y_2^*}(x|y) = \frac{1}{\sqrt{2\pi}} \exp{-\frac{(x-\frac{y}{a})^2}{2}},$$
%satisfies the equation defined in \eqref{wire_eq8_1}.
 Based on Theorem \ref{CAL_thm2}, it can be verified that the second-order variation is nonnegative. Moreover, due to the convexity of $K(y, f_{Y_2} , g)$ wrt $f_{Y_2}$ and $g$, we can
confirm that the Gaussian solution is optimal, and the proof is completed.

\subsection{Additional Applications}

The importance of the variational framework in establishing some fundamental  information theoretic inequalities was  already illustrated herein paper. At their turn, these information theoretic inequalities played a fundamental role in establishing other important results and applications. For example, the minimum Fisher information theorem (Cram\'{e}r-Rao inequality) and maximum entropy theorem were used for developing min-max robust estimation techniques  \cite{LargeCRB:Stoica},   results which were recently further extended to the more general framework of noise with arbitrary distribution (and correlation) in \cite{park:ieeespm} and used to explain why the MIMO channel estimation scheme proposed in \cite{TrainingSeq:Stoica} exhibits a min-max robustness property.   Along the same line of potential applications, the extensions of the maximum entropy and minimum Fisher information results to positive random variables, as stated in Theorems \ref{EF_thm3}, \ref{EF_thm6} and \ref{EF_thm7},  play a fundamental role in developing robust clock synchronization algorithms for wireless sensor networks  and other wireless networks that rely on message exchanges to acquire the timing information. A large class of clock synchronization protocols (see e.g., TPSN, Internet, PBS \cite{esbook})  rely on the two-way message exchange mechanism and for which the
timing synchronization approach reduces to estimating a linear regression model for  which the distribution of
additive noise has positive support but it is otherwise arbitrary \cite{esbook}. Designing robust timing
synchronization algorithms for such protocols is difficult,  because of the variability
of delay distributions caused by the variable network traffic. However,  this problem can now be resolved at the light of the results brought by Theorems
  \ref{EF_thm3}, \ref{EF_thm6} and \ref{EF_thm7}. By optimizing the design of timing messages for the scenario of a chi or log-normal distributed delay, then  min-max robust time synchronization algorithms could be developed.

%The entropy power inequality was first adapted to prove a lower bound on the capacity of additive noise channels by Shannon \cite{Infor:Shannon}, and received huge interest recently \cite{EPI:Rioul}, \cite{guo}. Furthermore, EPI was exploited for the scalar Gaussian broadcast channel \cite{CapBroad:Bergmans} and  the scalar quadratic Gaussian CEO problem \cite{Rate:Oohama}, among numerous other applications. Also,
The extremal entropy inequality was used in  the vector Gaussian broadcast channel  \cite{Extremal:Liu}, the distributed source coding with a single quadratic distortion constraint problem \cite{Extremal:Liu},  the Gaussian wire-tap channel \cite{Extrem:SPark}, and many other problems. Even though these applications were traditionally addressed using the information theoretic inequalities,  one  can directly approach these applications by means of the proposed variational calculus techniques. One of the benefits of such a variational approach is the fact that it can cope with many types of constraints as opposed to the EEI which is still quite rigid in its formulation. As Prof. Max Costa suggested the authors of this paper in a private communication, in the context of Z Gaussian interference channels, such a variational approach might be helpful to develop novel entropy-power-like inequalities, where the limiting variables are Gaussian and independent but not anymore identically distributed, and to assess the capacity of the Z-Gaussian interference channel.

Additional important extensions of maximum entropy theorem, minimum Fisher information theorem, additive worst noise lemma, and extremal entropy inequality might be envisioned
within the proposed variational  framework by imposing various restrictions on the range of values assumed by random variables/vectors (e.g., random variables whose support is limited to a finite length interval or finite set of values) or on their  second or higher-order moments and correlations. For example, the problem of finding  the  worst  additive noise under a covariance constraint \cite{WANoise:Cover} as well as establishing multivariate extensions of Costa's entropy power inequality \cite{costa} along the lines mentioned by Liu et al. \cite{tieliu1} and Palomar \cite{palomar3}, \cite{palomar4}  might be also  addressed  within the proposed variational framework. However, all  these challenges together with finding a variational proof of EPI remain open research problems for future study.

\section{Conclusions}\label{CON}

In this paper, we derived several fundamental information theoretic inequalities using a functional analysis framework.  
 The main benefit for employing calculus of variations is due to the fact for any information theoretic inequality as long as it can be expressed in terms of a convex functional, the global optimal solution can be obtained from the necessary conditions. 
 A brief summary of this paper contributions is the following.  First, the entropy maximizing theorem and Fisher information minimizing theorem were derived under different assumptions.  Second, the worst additive noise lemma was proved from the perspective of a functional problem. Third,  the extremal entropy inequality was derived using calculus of variations techniques. Finally, applications and possible extensions that could be addressed within the proposed variational framework were briefly presented.
Many open research problems were also formulated.

\appendices

\section{Proof of Corollaries  \ref{CAL_cor1} and \ref{CAL_cor2}}\label{sec:appendix_a}

Even though the functionals in Corollary \ref{CAL_cor1} involve double integrations, they can be regarded as a special case of the functionals in Theorem \ref{CAL_thm3}. For example, the functional $U[\fx,\fy]$ in \eqref{CAL_eq12_1} can be considered as $\int_a^b G(y,\fy)dy$ where
$G(y,\fy) = \int_a^b K(x,y,\fx,\fy)dx$. In this way, the augmented functional is given by
\begin{equation}\notag
\begin{split}
J[\fx,\fy] =& \int_a^b \left[ \int_a^b K(x,y,\fx,\fy)dx + \sum\limits_{i=1}^n \int_a^b \tilde{L}_i(x,y,\fx,\fy)dx + \lambda(y)\left( g(y,\fy) - \int_a^b \tilde{k}(x,y,\fx)dx\right) \right]dy \\
=& \int_{a}^{b} \Big\{ \Big[\int_{a}^{b}  (K(x,y,\fx , \fy )+ \sum\limits_{i=1}^n \lambda_i \tilde{L}_i(x,y, \fx , \fy )
 - \lambda(y) \tilde{k}(x,y, \fx ))dx \Big] +\lambda(y)g(y,\fy) \Big\}dy.
\end{split}
\end{equation}
This completes the proof of Corollary \ref{CAL_cor1}.

Based on the definitions in Section \ref{CAL}, the first-order variation of the above augmented functional can be calculated as
\begin{equation}\label{app-1st}
\begin{split}
\delta J[\fx,\fy] =& \int_a^b\int_a^b \bigg\{ \frac{\partial K(x,y,\fx,\fy)}{\partial \fx}\eta(x) + \frac{\partial K(x,y,\fx,\fy)}{\partial \fy}\xi(y)
+ \sum\limits_{i=1}^n \bigg[ \frac{\partial \tilde{L}_i(x,y,\fx,\fy)}{\partial \fx}\eta(x) + \\
& \frac{\partial \tilde{L}_i(x,y,\fx,\fy)}{\partial \fy}\xi(y)\bigg] -\lambda(y) \frac{\partial \tilde{k}(x,y,\fx)}{\partial \fx}\eta(x) \bigg\} dxdy
+ \int_a^b \lambda(y) \frac{\partial g(y,\fy)}{\partial \fy}\xi(y) dy \\
=& \int_a^b \bigg\{ \int_a^b \frac{\partial K(x,y,\fx,\fy)}{\partial \fx} + \sum\limits_{i=1}^n \frac{\partial \tilde{L}_i(x,y,\fx,\fy)}{\partial \fx} -\lambda(y) \frac{\partial \tilde{k}(x,y,\fx)}{\partial \fx} dy\bigg\} \eta(x) dx \\
&+ \int_a^b \bigg\{ \int_a^b  \frac{\partial K(x,y,\fx,\fy)}{\partial \fy} + \sum\limits_{i=1}^n \frac{\partial \tilde{L}_i(x,y,\fx,\fy)}{\partial \fy}dx + \lambda(y)\frac{\partial g(y,\fy)}{\partial \fy}  \bigg\} \xi(y) dy ,
\end{split}
\end{equation}
where $\eta(x)$  and $\xi(y)$ represent any admissible increments for $\fx$ and $\fy$, respectively.  
Due to Theorem \ref{CAL_thm1}, a necessary condition for the function $J[\fx,\fy]$ to have an extremum for given functions $\fxstar$ and $\fystar$ is that $\delta J[\fx,\fy]$ vanishes at $\fxstar$ and $\fystar$ for any admissible $\eta(x)$ and $\xi(y)$. This leads to
\begin{eqnarray}\notag
\int  K'_{\fxstar}(x,y,\fxstar , \fystar ) + \sum\limits_{i=1}^n \lambda_i \tilde{L_i}'_{\fxstar }(x,y, \fxstar , \fystar )  -\lambda(y) \tilde{k}'_{\fxstar}(x,y, \fxstar ) dy \hspace*{-2mm}&=&\hspace*{-2mm} 0,  \\
\notag
\int K'_{\fystar }(x,y,\fxstar,\fystar) + \sum\limits_{i=1}^n \lambda_i \tilde{L_i}'_{\fystar }(x,y, \fxstar , \fystar )dx
 + \lambda(y)g'_{\fystar }(y,\fystar ) \hspace*{-2mm}&=&\hspace*{-2mm} 0,
\end{eqnarray}
which are exactly \eqref{CAL_eq16_1} and \eqref{CAL_eq16_2}.

In order to calculate the second-order variation of $J[\fx,\fy]$ from the first-order variation \eqref{app-1st}, we rewrite the term $\lambda(y)\frac{\partial g(y,\fy)}{\partial \fy}$ in \eqref{app-1st} as $\int_a^b q(x) \lambda(y)\frac{\partial g(y,\fy)}{\partial \fy} dx$, where $q(x)$ is an arbitrary but fixed function satisfying $\int_a^b q(x)dx = 1$. Thus, the first-order variation \eqref{app-1st} can be rewritten as
\begin{equation}\label{app-2nd}
\begin{split}
& \int_a^b \bigg\{ \int_a^b \frac{\partial K(x,y,\fx,\fy)}{\partial \fx} + \sum\limits_{i=1}^n \frac{\partial \tilde{L}_i(x,y,\fx,\fy)}{\partial \fx} -\lambda(y) \frac{\partial \tilde{k}(x,y,\fx)}{\partial \fx} dy\bigg\} \eta(x) dx \\
&+ \int_a^b \bigg\{ \int_a^b  \frac{\partial K(x,y,\fx,\fy)}{\partial \fy} + \sum\limits_{i=1}^n \frac{\partial \tilde{L}_i(x,y,\fx,\fy)}{\partial \fy} + q(x)\lambda(y)\frac{\partial g(y,\fy)}{\partial \fy} dx  \bigg\} \xi(y) dy
\end{split}
\end{equation}
Based on \eqref{app-2nd}, the second-order variation of $J[\fx,\fy]$ is derived as
$$\delta^2 J[\fx,\fy] = \int_a^b\int_a^b
\begin{bmatrix}
\eta(x) & \xi(y)
\end{bmatrix}
\begin{bmatrix}
G''_{\fx  \fx } & G''_{\fx  \fy }\\
G''_{\fy  \fx } & G''_{\fy  \fy }
\end{bmatrix}
\begin{bmatrix}
\eta(x) \\
\xi(y)
\end{bmatrix}dxdy,
$$
where
\begin{eqnarray}
G(x,y,\fxstar , \fystar ) = K(x,y,\fxstar , \fystar ) + \sum\limits_{i=1}^N \lambda_i \tilde{L}_i(x,y, \fxstar , \fystar )
-\lambda(y)\tilde{k}(x,y, \fxstar )+ \lambda(y)g(y,\fystar ) q(x),\nonumber
\end{eqnarray}
Since a necessary condition for the functional $J[\fx,\fy]$ to have a minimum for given functions $\fxstar$ and $\fystar$ is that $\delta^2 J[\fx,\fy] \geq 0$, this leads to the positive semi-definiteness of
$$\begin{bmatrix}
G''_{\fx  \fx } & G''_{\fx  \fy }\\
G''_{\fy  \fx } & G''_{\fy  \fy }
\end{bmatrix}$$
and completes the proof of Corollary \ref{CAL_cor2}.

\section{Non-invertible Correlation (or Covariance) Matrix} \label{NON}
%%%%%%%%%%%%%%%
Let $\boldsymbol{\Omega}_{\scriptscriptstyle X} =  \mathbf{Q}_{\scriptscriptstyle \Omega} \boldsymbol{\Lambda}_{\scriptscriptstyle \Omega} \mathbf{Q}^{\scriptscriptstyle T}_{\scriptscriptstyle \Omega}$ and $\bar{\mathbf{X}} =  \mathbf{Q}_{\scriptscriptstyle \Omega}^{\scriptscriptstyle T} \mathbf{X} = [ \bar{\mathbf{X}}_{a}^{\scriptscriptstyle T}, \bar{\mathbf{X}}_b^{\scriptscriptstyle T} ]$, where $\boldsymbol{\Lambda}_{\scriptscriptstyle \Omega} = diag(\Lambda_1, \ldots, \Lambda_m, 0, \ldots, 0)$, $\boldsymbol{\Omega}_{\scriptscriptstyle X}$ is a singular matrix,  $ \mathbf{Q}_{\scriptscriptstyle \Omega} $ is an orthogonal matrix, and $diag(\cdot)$ denotes a diagonal matrix. The correlation matrix of $ \bar{\mathbf{X}}_b$ is the zero matrix, and therefore, it is considered as a deterministic vector. Without loss of generality, we can assume $ \bar{\mathbf{X}}_b=\mathbf{0}$. The following matrices are also considered:
\begin{eqnarray}
\label{IM_eq0_0}
\mathbf{Q}_{\scriptscriptstyle \Omega}^{\scriptscriptstyle T} \boldsymbol{\Omega}_{\scriptscriptstyle W} \mathbf{Q}_{\scriptscriptstyle \Omega} & = & \left[\begin{array}{cc} \mathbf{A} & \mathbf{B}^{\scriptscriptstyle T} \\ \mathbf{B} & \mathbf{C} \end{array} \right],\nonumber\\
\mathbf{D} & = & \left[\begin{array}{cc} \mathbf{I} & -\mathbf{B}^{\scriptscriptstyle T} \mathbf{C}^{-1} \\ \mathbf{0} & \mathbf{I} \end{array}\right],
\end{eqnarray}
where the dimensions of $\mathbf{A}$, $\mathbf{B}$, and $\mathbf{C}$ are $m \times m$, $(n-m) \times m$, and $(n-m) \times (n-m)$, respectively. Then,
\begin{eqnarray}
\hspace*{-4mm}& & \hspace*{-4mm}\mathbf{D}\mathbf{Q}^{\scriptscriptstyle T}_{\scriptscriptstyle \Omega} \mathbf{X} =  \left[\begin{array}{cc} \mathbf{I} & -\mathbf{B}^{\scriptscriptstyle T}\mathbf{C}^{-1} \\ \mathbf{0} & \mathbf{I} \end{array}\right]\left[\begin{array}{c} \bar{\mathbf{X}}_a \\ \mathbf{0} \end{array}\right] = \left[\begin{array}{c}\bar{\mathbf{X}}_a \\ \mathbf{0} \end{array}\right],\nonumber\\
\hspace*{-4mm} & & \hspace*{-4mm} \mathbf{D}\mathbf{Q}^{\scriptscriptstyle T}_{\scriptscriptstyle \Omega} \mathbf{W}_{\scriptscriptstyle G}  =  \left[\begin{array}{c} \bar{\mathbf{W}}_{\scriptscriptstyle G_a} \\ \bar{\mathbf{W}}_{\scriptscriptstyle G_b} \end{array}\right],\nonumber\\
\label{IM_eq0_1}
\hspace*{-4mm} & & \hspace*{-4mm} \mathbb{E} \left[\mathbf{D} \mathbf{Q}_{\scriptscriptstyle \Omega}^{\scriptscriptstyle T} \mathbf{W}_{\scriptscriptstyle G}  \mathbf{W}_{\scriptscriptstyle G}^{\scriptscriptstyle T} \mathbf{Q}_{\scriptscriptstyle \Omega} \mathbf{D}^{\scriptscriptstyle T} \right]  =  \left[\begin{array}{cc} \mathbf{A} - \mathbf{B}^{\scriptscriptstyle T} \mathbf{C}^{-1}\mathbf{B} & \mathbf{0} \\ \mathbf{0} & \mathbf{C} \end{array}\right].
\end{eqnarray}
Due to  (\ref{IM_eq0_1}), the random vectors $\bar{\mathbf{W}}_{\scriptscriptstyle G_a}$ and $\bar{\mathbf{W}}_{\scriptscriptstyle G_b}$ are statistically independent of each other.

The left-hand side of the equation in (\ref{WA_eq14_1}) can be re-expressed as
\begin{eqnarray}
\label{IM_eq1_1}
\hspace*{-4mm}&&\hspace*{-4mm}h(\mathbf{X}+\mathbf{W}_{\scriptscriptstyle G}) - h(\mathbf{X}) \hspace*{-0.5mm}=\hspace*{-0.5mm} h(\mathbf{D}\mathbf{Q}^{\scriptscriptstyle T}_{\scriptscriptstyle \Omega}\mathbf{X}+\mathbf{D}\mathbf{Q}^{\scriptscriptstyle T}_{\scriptscriptstyle \Omega}\mathbf{W}_{\scriptscriptstyle G}) - h(\mathbf{D}\mathbf{Q}^{\scriptscriptstyle T}_{\scriptscriptstyle \Omega}\mathbf{X})\nonumber\\
\hspace*{-4mm}&&\hspace*{-4mm} =h(\bar{\mathbf{X}}_{a} + \bar{\mathbf{W}}_{\scriptscriptstyle G_a}, \bar{\mathbf{X}}_{b} + \bar{\mathbf{W}}_{\scriptscriptstyle G_b})-h(\bar{\mathbf{X}}_{a}, \bar{\mathbf{X}}_{b})\nonumber\\
\hspace*{-4mm}&&\hspace*{-4mm} =h(\bar{\mathbf{X}}_{a} + \bar{\mathbf{W}}_{\scriptscriptstyle G_a})-h(\bar{\mathbf{X}}_{a}) + \underbrace{h(\bar{\mathbf{X}}_{b} + \bar{\mathbf{W}}_{\scriptscriptstyle G_b})-h(\bar{\mathbf{X}}_{b})}_{(a)}.
\end{eqnarray}
In (\ref{IM_eq1_1}), $\bar{\mathbf{X}}_{b}$ is considered as a deterministic variable, $\bar{\mathbf{W}}_{\scriptscriptstyle G_b}$ is given, the term $(a)$ can be ignored in the optimization, and the correlation matrix of $\bar{\mathbf{X}}_a$ is non-singular. Therefore, we can always assume the correlation matrix to be  invertible.

%%%%%%%%%%%%%
%%%%%%%%%%%%%%%%%%%%%%%%%%%%%%%
%%% EPI random variables

%%%%%%%%%%%%%%%%%%%%%%%%%%%%%%%%
% EPI vector versions
%\section{Proof of Theorem \ref{EPI_thm2}}\label{EPI_thm2_proof}

%%%%%%%%%%%%%%%%%%%%%%%%%%%%%%
%%%% EI simple random variables

%%%%%%%%%%%%%%%%%%%%%%%%%%%%%%%%%%%%%%%%%%%%%%%%%%%%%%%
%%%% EI simple vector version
%\section{Proof of Theorem \ref{EI_thm2}}\label{EI_thm2_proof}

%%%%%%%%%%%%%%%%%%%%%%%%%%%%%%%%%%%%%%%%%%%%%%%%%%%%
%%% EI general vector version
\section{Proof of Theorem \ref{EI_thm3}}\label{EI_thm3_proof}
\begin{proof}
First, choose a Gaussian random vector $\tilde{\mathbf{W}}_{\scriptscriptstyle G}$ whose covariance matrix $\boldsymbol{\Sigma}_{\scriptscriptstyle \tilde{W}}$ satisfies $\boldsymbol{\Sigma}_{\scriptscriptstyle \tilde{W}}\preceq \boldsymbol{\Sigma}_{\scriptscriptstyle W}$ and $\boldsymbol{\Sigma}_{\scriptscriptstyle \tilde{W}}\preceq \boldsymbol{\Sigma}_{\scriptscriptstyle V}$. Since the Gaussian random vectors $\mathbf{V}_{\scriptscriptstyle G}$ and $\mathbf{W}_{\scriptscriptstyle G}$ can be represented as the summation of two independent random vectors $\tilde{\mathbf{W}}_{\scriptscriptstyle G}$ and $\hat{\mathbf{V}}_{\scriptscriptstyle G}$, and the summation of  two independent random vectors $\tilde{\mathbf{W}}_{\scriptscriptstyle G}$ and $\hat{\mathbf{W}}_{\scriptscriptstyle G}$, respectively, the left-hand side of the equation in (\ref{EI_eq50_1}) is written as follows:
\begin{eqnarray}
\label{EI_eq51_1}
\hspace*{-3mm}&&\hspace*{-3mm} \mu h(\mathbf{X}+\mathbf{V}_{\scriptscriptstyle G}) - h(\mathbf{X}+\mathbf{W}_{\scriptscriptstyle G}) \nonumber\\
\hspace*{-3mm}& \geq &\hspace*{-3mm} \mu h(\mathbf{X}+\mathbf{V}_{\scriptscriptstyle G}) - h(\mathbf{X}+\tilde{\mathbf{W}}_{\scriptscriptstyle G}) - h(\mathbf{W}_{\scriptscriptstyle G}) + h(\tilde{\mathbf{W}}_{\scriptscriptstyle G})\nonumber\\
\label{EI_eq51_2}
\hspace*{-3mm}& = &\hspace*{-3mm} \mu h(\mathbf{X}+\tilde{\mathbf{W}}_{\scriptscriptstyle G}+\hat{\mathbf{V}}_{\scriptscriptstyle G}) - h(\mathbf{X}+\tilde{\mathbf{W}}_{\scriptscriptstyle G}) - h(\tilde{\mathbf{W}}_{\scriptscriptstyle G}+\hat{\mathbf{W}}_{\scriptscriptstyle G})
+ h(\tilde{\mathbf{W}}_{\scriptscriptstyle G}) .
\end{eqnarray}

Since the expression will be minimized over $f_{\scriptscriptstyle X}(\mathbf{x})$, the last two terms in (\ref{EI_eq51_2}) are ignored,  and  by substituting $\mathbf{Y}$ and $\hat{\mathbf{X}}$ for $\mathbf{X}+\tilde{\mathbf{W}}_{\scriptscriptstyle G}+\hat{\mathbf{V}}_{\scriptscriptstyle G}$ and $\mathbf{X}+\tilde{\mathbf{W}}_{\scriptscriptstyle G}$, respectively, the inequality in (\ref{EI_eq50_1}) is equivalently expressed as the following variational problem:
\begin{eqnarray}
\label{EI_eq52_1}
\hspace*{-3mm}&&\hspace*{-3mm}\min_{f_{\scriptscriptstyle \hat{X}}, f_{\scriptscriptstyle Y}} \quad   \mu h(\mathbf{Y}) - h(\hat{\mathbf{X}}) -\mu\left(\mu-1\right)  h(\hat{\mathbf{V}}_{\scriptscriptstyle G})\nonumber\\
\hspace*{-3mm}&&\hspace*{-3mm}\text{s. t. }\quad \int \hspace*{-2mm} \int \hspace*{-1mm}f_{\scriptscriptstyle \hat{X}}(\mathbf{x}) f_{\scriptscriptstyle \hat{V}}(\mathbf{y}-\mathbf{x})d\mathbf{x}d\mathbf{y}-1 = 0, \nonumber\\
\hspace*{-3mm}&&\hspace{7mm}\int \hspace*{-2mm} \int \hspace*{-1mm}f_{\scriptscriptstyle \hat{X}}(\mathbf{x}) f_{\scriptscriptstyle \hat{V}}(\mathbf{y}-\mathbf{x}) \mathbf{x}\mathbf{x}^{\scriptscriptstyle T} d\mathbf{x}d\mathbf{y} -\boldsymbol{\Sigma}_{\scriptscriptstyle \hat{X}} \preceq \mathbf{0},\nonumber\\
\hspace*{-3mm}&&\hspace{7mm}\int \hspace*{-2mm} \int \hspace*{-1mm} f_{\scriptscriptstyle \hat{X}}(\mathbf{x}) f_{\scriptscriptstyle \hat{V}}(\mathbf{y}-\mathbf{x}) \mathbf{y}\mathbf{y}^{\scriptscriptstyle T} d\mathbf{x}d\mathbf{y} -{\boldsymbol{\Sigma}}_{\scriptscriptstyle Y^*}= \mathbf{0},\nonumber\\
\hspace*{-3mm} &&\hspace{7mm}\int \hspace*{-2mm} \int \hspace*{-1mm} f_{\scriptscriptstyle \hat{X}}(\mathbf{x}) f_{\scriptscriptstyle \hat{V}}(\mathbf{y}-\mathbf{x}) ( \mathbf{y}\mathbf{y}^{\scriptscriptstyle T} - \mathbf{x}\mathbf{x}^{\scriptscriptstyle T} - \left(\mathbf{y}-\mathbf{x}\right)\left(\mathbf{y}-\mathbf{x}\right)^{\scriptscriptstyle T} )d\mathbf{x}d\mathbf{y} = \mathbf{0},\nonumber\\
\hspace*{-3mm}&&\hspace{4mm}-\int \hspace*{-2mm} \int \hspace*{-1mm} f_{\scriptscriptstyle \hat{X}}(\mathbf{x}) f_{\scriptscriptstyle \hat{V}}(\mathbf{y}-\mathbf{x}) \log  f_{\scriptscriptstyle \hat{X}}(\mathbf{x})d\mathbf{x}d\mathbf{y}  \geq  p_{\scriptscriptstyle \hat{X}}\\
\hspace*{-3mm}&&\hspace{7mm}f_{\scriptscriptstyle Y}(\mathbf{y}) = \int f_{\scriptscriptstyle \hat{X}}(\mathbf{x})f_{\scriptscriptstyle \hat{V}}(\mathbf{y}-\mathbf{x}) d\mathbf{x},\nonumber
\end{eqnarray}
where $\hat{\mathbf{X}}=\mathbf{X}+\tilde{\mathbf{W}}_{\scriptscriptstyle G}$, $\mathbf{Y}=\hat{\mathbf{X}}+\hat{\mathbf{V}}_{\scriptscriptstyle G}$, $\mathbf{W}_{\scriptscriptstyle G} = \tilde{\mathbf{W}}_{\scriptscriptstyle G} + \hat{\mathbf{W}}_{\scriptscriptstyle G}$, $\mathbf{V}_{\scriptscriptstyle G} = \tilde{\mathbf{W}}_{\scriptscriptstyle G} + \hat{\mathbf{V}}_{\scriptscriptstyle G}$, $\boldsymbol{\Sigma}_{\scriptscriptstyle \hat{X}}=\boldsymbol{\Sigma} +\boldsymbol{\Sigma}_{\scriptscriptstyle \tilde{W}} $, $\boldsymbol{\Sigma}_{\scriptscriptstyle Y^*}= \boldsymbol{\Sigma}_{\scriptscriptstyle X^*} + \boldsymbol{\Sigma}_{\scriptscriptstyle V} $, and  $\boldsymbol{\Sigma}_{\scriptscriptstyle X^*}$ is the covariance matrix of the optimal solution $\mathbf{X}^*$.

The variational problem in (\ref{EI_eq52_1}) is exactly the same as the one in (\ref{EI_eq21_1}). Therefore, using the same method as in the proof of Theorem \ref{EI_thm2}, we obtain the following inequality (see the details  in the proof of Theorem \ref{EI_thm2}):
\begin{eqnarray}
\label{EI_eq53_1}
\hspace*{-4mm}&&\hspace*{-4mm}\mu h(\mathbf{X}+\tilde{\mathbf{W}}_{\scriptscriptstyle G}+\hat{\mathbf{V}}_{\scriptscriptstyle G})-h(\mathbf{X}+\tilde{\mathbf{W}}_{\scriptscriptstyle G}) - h(\tilde{\mathbf{W}}_{\scriptscriptstyle G}+\hat{\mathbf{W}}_{\scriptscriptstyle G}) + h(\tilde{\mathbf{W}}_{\scriptscriptstyle G})\nonumber\\
\hspace*{-4mm}&\geq&\hspace*{-3mm}\mu h(\mathbf{X}_{\scriptscriptstyle G}^*+\tilde{\mathbf{W}}_{\scriptscriptstyle G}+\hat{\mathbf{V}}_{\scriptscriptstyle G}) - h(\mathbf{X}_{\scriptscriptstyle G}^*+\tilde{\mathbf{W}}_{\scriptscriptstyle G}) - h(\tilde{\mathbf{W}}_{\scriptscriptstyle G}+\hat{\mathbf{W}}_{\scriptscriptstyle G})  +h(\tilde{\mathbf{W}}_{\scriptscriptstyle G}).
\end{eqnarray}

By appropriately choosing $\mathbf{X}_{\scriptscriptstyle G}^*$ and $\tilde{\mathbf{W}}_{\scriptscriptstyle G}$, the right-hand side of the equation in (\ref{EI_eq53_1}) is expressed as
\begin{eqnarray}
\label{EI_eq54_1}
\hspace*{-4mm}&& \hspace*{-4mm}\mu h(\mathbf{X}_{\scriptscriptstyle G}^*+\tilde{\mathbf{W}}_{\scriptscriptstyle G}+\hat{\mathbf{V}}_{\scriptscriptstyle G}) - h(\mathbf{X}_{\scriptscriptstyle G}^*+\tilde{\mathbf{W}}_{\scriptscriptstyle G}) - h(\tilde{\mathbf{W}}_{\scriptscriptstyle G}+\hat{\mathbf{W}}_{\scriptscriptstyle G}) + h(\tilde{\mathbf{W}}_{\scriptscriptstyle G})\nonumber\\
\hspace*{-4mm}& = &\hspace*{-4mm} \mu h(\mathbf{X}_{\scriptscriptstyle G}^*+\tilde{\mathbf{W}}_{\scriptscriptstyle G}+\hat{\mathbf{V}}_{\scriptscriptstyle G}) - h(\mathbf{X}_{\scriptscriptstyle G}^*+ \mathbf{W}_{\scriptscriptstyle G}) .
\end{eqnarray}
The equality in (\ref{EI_eq54_1}) is due to the equality condition of the data processing inequality in  \cite{Extrem:SPark}. For the completeness of the proof, we introduce a  technique, which is slightly different from the one in \cite{Extrem:SPark}.

To satisfy the equality in the equation (\ref{EI_eq54_1}), the equality condition in the following lemma must be satisfied.
\begin{lem}[Data Processing Inequality \cite{inf:cover}]\label{EI_lem1}
When three random vectors $\mathbf{Y}_1$, $\mathbf{Y}_2$, and $\mathbf{Y}_3$ represent a Markov chain  $\mathbf{Y}_1\rightarrow \mathbf{Y}_2 \rightarrow \mathbf{Y}_3$,
the following inequality is satisfied:
\begin{eqnarray}
\label{EI_eq55_1}
    I(\mathbf{Y}_1;\mathbf{Y}_3) \leq I(\mathbf{Y}_1;\mathbf{Y}_2).
\end{eqnarray}
The equality holds if and only if $I(\mathbf{Y}_1; \mathbf{Y}_2| \mathbf{Y}_3)=0$.
\end{lem}

In  Lemma \ref{EI_lem1}, $\mathbf{Y}_1$, $\mathbf{Y}_2$, and $\mathbf{Y}_3$ are defined as $\mathbf{X}_{\scriptscriptstyle G}^*$, $\mathbf{X}_{\scriptscriptstyle G}^*+\tilde{\mathbf{W}}_{\scriptscriptstyle G}$, and $\mathbf{X}_{\scriptscriptstyle G}^*+\tilde{\mathbf{W}}_{\scriptscriptstyle G}+\hat{\mathbf{W}}_{\scriptscriptstyle G}$, respectively. Therefore, the equality condition, $I(\mathbf{Y}_1; \mathbf{Y}_2| \mathbf{Y}_3)=0$ is expressed as
\begin{eqnarray}
\label{EI_eq56_1}
 \hspace{-2mm}  &   &  \hspace{-2mm}   I(\mathbf{Y}_1; \mathbf{Y}_2| \mathbf{Y}_3)    \nonumber \\
\hspace{-2mm}  & = &  \hspace{-2mm} h(\mathbf{Y}_1|\mathbf{Y}_3) - h(\mathbf{Y}_1|\mathbf{Y}_2, \mathbf{Y}_3)\nonumber\\
\hspace{-2mm} & = & \hspace{-2mm} \frac{1}{2}\log \left(2\pi e\right)^n \left|\boldsymbol{\Sigma}_{\scriptscriptstyle Y_1|Y_3}\right| - \frac{1}{2}\log \left(2\pi e\right)^n \left|\boldsymbol{\Sigma}_{\scriptscriptstyle Y_1|Y_2}\right|\nonumber\\
\hspace{-2mm} & = & \hspace{-2mm} \frac{1}{2}\log \left(2\pi e\right)^n \left|\boldsymbol{\Sigma}_{\scriptscriptstyle Y_1}- \boldsymbol{\Sigma}_{\scriptscriptstyle Y_1} \boldsymbol{\Sigma}_{\scriptscriptstyle Y_3}^{-1}\boldsymbol{\Sigma}_{\scriptscriptstyle Y_1}\right| - \frac{1}{2}\log \left(2\pi e\right)^n \left|\boldsymbol{\Sigma}_{\scriptscriptstyle Y_1}-\boldsymbol{\Sigma}_{\scriptscriptstyle Y_1} \boldsymbol{\Sigma}_{\scriptscriptstyle Y_2}^{-1} \boldsymbol{\Sigma}_{\scriptscriptstyle Y_1}\right|\nonumber\\
\hspace{-2mm} & = & \hspace{-2mm} \frac{1}{2}\log \left(2\pi e\right)^n \left|\boldsymbol{\Sigma}_{\scriptscriptstyle X^*}- \boldsymbol{\Sigma}_{\scriptscriptstyle X^*} \left(\boldsymbol{\Sigma}_{\scriptscriptstyle X^*}+\boldsymbol{\Sigma}_{\scriptscriptstyle \tilde{W}}+\boldsymbol{\Sigma}_{\scriptscriptstyle \hat{W}}\right)^{-1}\boldsymbol{\Sigma}_{\scriptscriptstyle X^*}\right|
 -  \frac{1}{2}\log \left(2\pi e\right)^n \left|\boldsymbol{\Sigma}_{\scriptscriptstyle X^*} -\boldsymbol{\Sigma}_{\scriptscriptstyle X^*} \left(\boldsymbol{\Sigma}_{\scriptscriptstyle X^*}+\boldsymbol{\Sigma}_{\scriptscriptstyle \tilde{W}}\right)^{-1} \boldsymbol{\Sigma}_{\scriptscriptstyle X^*}\right|\nonumber\\
\hspace{-2mm} & = & \hspace{-2mm} \frac{1}{2}\log \left(2\pi e\right)^n \left|\boldsymbol{\Sigma}_{\scriptscriptstyle X^*}\right|\left|I-  \left(\boldsymbol{\Sigma}_{\scriptscriptstyle X^*}+\boldsymbol{\Sigma}_{\scriptscriptstyle \tilde{W}}+\boldsymbol{\Sigma}_{\scriptscriptstyle \hat{W}}\right)^{-1}\boldsymbol{\Sigma}_{\scriptscriptstyle X^*}\right|
- \frac{1}{2}\log \left(2\pi e\right)^n \left|\boldsymbol{\Sigma}_{\scriptscriptstyle X^*}\right|\left|I -  \left(\boldsymbol{\Sigma}_{\scriptscriptstyle X^*}+\boldsymbol{\Sigma}_{\scriptscriptstyle \tilde{W}}\right)^{-1} \boldsymbol{\Sigma}_{\scriptscriptstyle X^*}\right|\nonumber\\
\hspace{-2mm} & = &\hspace{-2mm}  \frac{1}{2}\log \left(2\pi e\right)^n \left|I-  \left(\boldsymbol{\Sigma}_{\scriptscriptstyle X^*}+\boldsymbol{\Sigma}_{\scriptscriptstyle \tilde{W}}+\boldsymbol{\Sigma}_{\scriptscriptstyle \hat{W}}\right)^{-1}\boldsymbol{\Sigma}_{\scriptscriptstyle X^*}\right|
-  \frac{1}{2}\log \left(2\pi e\right)^n\left|I -  \left(\boldsymbol{\Sigma}_{\scriptscriptstyle X^*}+\boldsymbol{\Sigma}_{\scriptscriptstyle \tilde{W}}\right)^{-1} \boldsymbol{\Sigma}_{\scriptscriptstyle X^*}\right|\nonumber\\
\hspace{-2mm} & = & \hspace{-2mm} \frac{1}{2}\log \left(2\pi e\right)^n \left|I-  \left(\boldsymbol{\Sigma}_{\scriptscriptstyle X^*}+\boldsymbol{\Sigma}_{\scriptscriptstyle W}\right)^{-1}\boldsymbol{\Sigma}_{\scriptscriptstyle X^*}\right|
- \frac{1}{2}\log \left(2\pi e\right)^n\left|I -  \left(\boldsymbol{\Sigma}_{\scriptscriptstyle X^*}+\boldsymbol{\Sigma}_{\scriptscriptstyle \tilde{W}}\right)^{-1} \boldsymbol{\Sigma}_{\scriptscriptstyle X^*}\right|\nonumber\\
\label{EI_eq56_5}
\hspace{-2mm} & = & \hspace{-2mm} 0.
\end{eqnarray}
If $\left(\boldsymbol{\Sigma}_{\scriptscriptstyle X^*}+\boldsymbol{\Sigma}_{\scriptscriptstyle W}\right)^{-1}\boldsymbol{\Sigma}_{\scriptscriptstyle X^*}= \left(\boldsymbol{\Sigma}_{\scriptscriptstyle X^*}+\boldsymbol{\Sigma}_{\scriptscriptstyle \tilde{W}}\right)^{-1} \boldsymbol{\Sigma}_{\scriptscriptstyle X^*}$, the equality in (\ref{EI_eq56_5}) is satisfied, the equality condition in Lemma \ref{EI_lem1} holds, and therefore, the equality in (\ref{EI_eq54_1}) is proved. The validity of $\left(\boldsymbol{\Sigma}_{\scriptscriptstyle X^*}+\boldsymbol{\Sigma}_{\scriptscriptstyle W}\right)^{-1}\boldsymbol{\Sigma}_{\scriptscriptstyle X^*}= \left(\boldsymbol{\Sigma}_{\scriptscriptstyle X^*}+\boldsymbol{\Sigma}_{\scriptscriptstyle \tilde{W}}\right)^{-1} \boldsymbol{\Sigma}_{\scriptscriptstyle X^*}$ is proved by Lemma $8$ in \cite{Extrem:SPark}.

Therefore, $I(\mathbf{Y}_1; \mathbf{Y}_2| \mathbf{Y}_3)=0$, and from the equations in (\ref{EI_eq51_1}), (\ref{EI_eq53_1}), and (\ref{EI_eq54_1}), we obtain the following extremal entropy inequality:
\begin{eqnarray}
\label{EI_eq57_1}
\mu h(\mathbf{X}+\mathbf{V}_{\scriptscriptstyle G}) - h(\mathbf{X}+\mathbf{W}_{\scriptscriptstyle G})
\hspace*{-2mm}& \geq & \hspace*{-2mm}\mu h(\mathbf{X}+\mathbf{V}_{\scriptscriptstyle G}) - h(\mathbf{X}+\tilde{\mathbf{W}}_{\scriptscriptstyle G}) - h(\mathbf{W}_{\scriptscriptstyle G}) + h(\tilde{\mathbf{W}}_{\scriptscriptstyle G})\nonumber\\
\hspace*{-2mm}& = & \hspace*{-2mm}\mu h(\mathbf{X}+\tilde{\mathbf{W}}_{\scriptscriptstyle G}+\hat{\mathbf{V}}_{\scriptscriptstyle G}) - h(\mathbf{X}+\tilde{\mathbf{W}}_{\scriptscriptstyle G}) - h(\tilde{\mathbf{W}}_{\scriptscriptstyle G}+\hat{\mathbf{W}}_{\scriptscriptstyle G})
+ h(\tilde{\mathbf{W}}_{\scriptscriptstyle G})\nonumber\\
\hspace*{-2mm}&\geq&\hspace*{-2mm} \mu h(\mathbf{X}_{\scriptscriptstyle G}^*+\tilde{\mathbf{W}}_{\scriptscriptstyle G}+\hat{\mathbf{V}}_{\scriptscriptstyle G}) - h(\mathbf{X}_{\scriptscriptstyle G}^*+\tilde{\mathbf{W}}_{\scriptscriptstyle G}) - h(\tilde{\mathbf{W}}_{\scriptscriptstyle G}+\hat{\mathbf{W}}_{\scriptscriptstyle G})
 + h(\tilde{\mathbf{W}}_{\scriptscriptstyle G})\nonumber\\
\hspace*{-2mm}& = &\hspace*{-2mm} \mu h(\mathbf{X}_{\scriptscriptstyle G}^*+\tilde{\mathbf{W}}_{\scriptscriptstyle G}+\hat{\mathbf{V}}_{\scriptscriptstyle G}) - h(\mathbf{X}_{\scriptscriptstyle G}^*+\tilde{\mathbf{W}}_{\scriptscriptstyle G}) - h(\tilde{\mathbf{W}}_{\scriptscriptstyle G}+\hat{\mathbf{W}}_{\scriptscriptstyle G})
+ h(\tilde{\mathbf{W}}_{\scriptscriptstyle G})\nonumber\\
\hspace*{-2mm}& = &\hspace*{-2mm} \mu h(\mathbf{X}_{\scriptscriptstyle G}^*+\mathbf{V}_{\scriptscriptstyle G}) - h(\mathbf{X}_{\scriptscriptstyle G}^*+\mathbf{W}_{\scriptscriptstyle G}),\nonumber
\end{eqnarray}
and the proof is completed.

\end{proof}

%\newpage

\begin{IEEEbiographynophoto}{Sangwoo Park}
received the B.S. degree in electrical engineering from Chung-Ang University (CAU), Seoul, Korea, in 2004, and the M.S. and Ph.D. degrees in electrical engineering from Texas A\&M University, College Station, in
2008 and 2012, respectively. From 2004 to 2005, he  worked as a full-time assistant engineer for UMTS/WCDMA projects in Samsung Electronics. Currently, he is a research engineer at KT (Korea Telecom) in Korea. His research interests lie in wireless communications, information theory, and statistical signal processing.
\end{IEEEbiographynophoto}

\begin{IEEEbiographynophoto}{Erchin Serpedin}(F'13)
received the specialization degree in signal processing and transmission of information from Ecole Superieure \'{D}Electricite (SUPELEC), Paris, France, in 1992, the M.Sc. degree from the Georgia Institute of Technology, Atlanta, in 1992, and the Ph.D. degree in electrical engineering from the University of Virginia,
Charlottesville, in January 1999. He is currently a professor in the Department of Electrical and Computer
Engineering at Texas A\&M University, College Station. He is the author of two research monographs, one edited textbook, 100 journal papers and 150 conference papers, and has served as associate editor for about 10 journals  such as IEEE Transactions on Information Theory, IEEE Transactions on Communications, Signal Processing (Elsevier), IEEE Transactions on Signal Processing, IEEE Transactions on Wireless Communications, IEEE Communications Letters, IEEE Signal Processing Letters, Phycom,  EURASIP Journal on Advances in Signal Processing, and EURASIP Journal on Bioinformatics and Systems Biology. His research interests include  signal processing, wireless communications, computational statistics, bioinformatics and systems biology.
\end{IEEEbiographynophoto}

\begin{IEEEbiographynophoto}{Khalid Qaraqe}(M'97-S'00 ) received with honors the B.S. degree in EE from the University of Technology, Baghdad, Irak, in 1986. He received the M.S. degree in EE from the University of Jordan, Jordan, in 1989, and he earned his Ph.D. degree in EE from Texas A\&M University, College Station, TX, in 1997. From 1989 to 2004, Dr. Qaraqe  held a variety of positions in many companies. He has over 15 years of experience in the telecommunications industry.  Dr. Qaraqe has worked for Qualcomm, Enad Design Systems, Cadence Design Systems/Tality Corporation, STC, SBC and Ericsson.   He has worked on numerous GSM, CDMA, WCDMA projects and   has experience in product development, design, deployment, testing and integration.  Dr. Qaraqe joined Texas A\&M University at Qatar, in July 2004, where he is now a professor.
Dr. Qaraqe research interests include communication theory and its application to design and performance analysis of cellular systems and indoor communication systems. Particular interests are in the development of 3G UMTS, cognitive radio systems,  broadband  wireless communications and diversity techniques.
\end{IEEEbiographynophoto}


\begin{thebibliography}{1}

\bibitem{inf:cover}
T.~M. Cover and J.~A. Thomas, \textit{Elements of Information Theory (2nd edition)}, New York: Wiley, 2006.


\bibitem{Extremal:Liu}
T.~Liu and P. Viswanath, ``An Extremal Inequality Motivated by Multiterminal Information-Theoretic Problems," \emph{IEEE Trans. Inform. Theory}, vol. 53, no. 5, pp. 1839 - 1851, May 2007.

%\bibitem{CalVariations:Gelfand}
%I. M. Gelfand and S. V. Fomin, \textit{Calculus of Variations}, New York: Dover, 1991.

\bibitem{image}
G. Aubert  and P.  Kornprobst,  \textit{Mathematical Problems in Image Processing: Partial Differential Equations and the Calculus of Variations}. Applied Mathematical Sciences vol. 147. Springer Verlag. New York, 2006.

\bibitem{palomar0}
G. Scutari, D. Palomar, F. Facchinei, and J.-S. Pang, ``Convex Optimization, Game Theory, and Variational Inequality Theory,''  \textit{IEEE Signal Processing Magazine}, vol. 27, no. 3, pp. 35-49, May 2010.

%\bibitem{palomar1}
%G. Scutari, D. Palomar, J.-S. Pang, and F. Facchinei, ``Flexible Design of Cognitive Radio Wireless Systems,'' \textit{IEEE Signal Processing Magazine}, %vol. 26, no. 5, pp. 107-123, Sep. 2009.

%\bibitem{palomar2}
%J.-S. Pang, G. Scutari, D. P. Palomar, and F. Facchinei, ``Design of Cognitive Radio Systems Under Temperature-Interference Constraints: A Variational %Inequality Approach,'' \textit{IEEE Trans. on Signal Processing,} vol. 58, no. 6, pp. 3251-3271, Jun. 2010.

\bibitem{debbah1}
M. Debbah and R. Muller, ``MIMO Channel Modelling and the Principle of Maximum Entropy,'' \textit{IEEE Trans.  Inform. Theory}, vol. 51, no. 5, pp. 1667-1690, May 2005.

%\bibitem{debbah2}
%M. Guillaud, M. Debbah and A. Moustakas, ``A Characterization of Maximum Entropy Spatially Correlated Wireless Channel Models,'' \textit{Information Theory %Workshop}, May 2008, Porto, Portugal, pp. 56-60.

%\bibitem{debbah3}
%M. Guillaud, M. Debbah and A. Moustakas, ``Modelling the Multiple-Antenna Wireless Channel Using Maximum Entropy Methods,'' \textit{27th International %Workshop on Bayesian Inference and Maximum Entropy Methods in Science and Engineering 2007}, Jul. 2007, New York, USA, pp. 435-442.

\bibitem{radar1}
D. F. Delong, Jr., and E. M. Hofstetter,  ``On the Design of Optimum Radar Waveforms for Clutter Rejection,'' \textit{IEEE Trans. Inform. Theory,} vol. 13, no. 3, pp. 454-463, Jul. 1967.

\bibitem{radar2} L. J. Spafford, ``Optimum Radar Signal Processing in Clutter'', \textit{IEEE Trans. Inform. Theory,} vol. 14, no. 5, pp. 734-743, Sep. 1968.

%\bibitem{kay}
%S. Kay and J. H. Thanos, ``Optimal transmit signal design for active sonar/radar,'' \textit{2002 IEEE International Conference on Acoustics, Speech, and %Signal Processing (ICASSP),}  Orlando, Florida, 2002, vol. II, pp. 1513-1516.

\bibitem{jaynes}
E. T. Jaynes, ``On the Rationale of Maximum Entropy Methods," \textit{Proc. of the IEEE}, vol. 70, no. 9, pp. 939-952, Sep. 1982.

\bibitem{WANoise:Cover}
S.~N.~Diggavi and T.~M.~Cover, ``The worst additive noise under a covariance constraint,'' \textit{IEEE Trans. Inform. Theory}, vol. 47, no. 7, pp. 3072 - 3081, Nov. 2001.

\bibitem{EPI:Rioul}
O.~Rioul, ``Information Theoretic Proofs of Entropy Power Inequalities,'' \textit{IEEE Trans. Inform. Theory}, vol. 57, no. 1, pp. 33 - 55, Jan. 2011.

\bibitem{Extrem:SPark}
S. Park, E. Serpedin, and K. Qaraqe ``An Alternative Proof of an Extremal Entropy Inequality,'' arXiv:1201.6681.

\bibitem{CalVariations:Gelfand}
I. M. Gelfand and S. V. Fomin, \textit{Calculus of Variations}, New York: Dover, 1991.

\bibitem{ConsOpt:Gregory}
J. Gregory, \textit{Constrained Optimization in the Calculus of Variations and Optimal Control Theory}, New York: Van Nostrand Reinhold, 1992.

\bibitem{Cal:Hans}
H. Sagan, \textit{Introduction to the Calculus of Variations}, New York: Dover, 1992.

\bibitem{MinFisher:Bercher}
J. Bercher and C. Vignat, ``On minimum Fisher information distributions with restricted support and fixed variance,'' \textit{Inform. Sci.,} vol. 179, no. 22, pp. 3832-3842, Nov. 2009

\bibitem{CapBroad:Shamai}
H.~Weingarten, Y.~Steinberg, and S.~Shamai, ``The Capacity Region of the Gaussian Mutiple-Input Multiple-Output Broadcast Channel,''
\textit{IEEE Trans. Inform. Theory}, vol. 52, no. 9, pp. 3936 - 3964, Sep. 2006.

\bibitem{EPI:Dembo}
A.~Dembo, T.~M.~Cover, and J.~A.~Thomas, ``Information theoretic inequalities,'' \textit{IEEE Trans. Inform. Theory}, vol. 37, no. 6, pp. 1501 - 1518, Nov. 1991.

\bibitem{WAN:Ihara}
S. Ihara, ``On the capacity of channels with additive non-Gaussian noise,'' \textit{Inform. Contr.,} vol. 37, no. 1, pp. 34-39, Apr. 1978.

\bibitem{CapBroad:Bergmans}
P.~P.~Bergmans, ``A Simple Converse for Broadcast Channels with Additive White Gaussian Noise,'' \textit{IEEE Trans. Inform. Theory}, vol. 20, no. 2, pp. 279 - 280, Mar. 1974.

\bibitem{SecrecyCap:Liu}
T.~Liu and S.~Shamai (Shitz), ``A Note on the Secrecy Capacity of the Multiple-Antenna Wiretap Channel,'' \textit{IEEE Trans. Inform. Theory}, vol. 55, no. 6, pp. 2547 - 2553, Jun. 2009.

\bibitem{tieliu1}
R. Liu, T. Liu, H. Poor, and S. Shamai, ``A Vector Generalization of Costa's Entropy-Power Inequality with Applications,'' \textit{IEEE Trans. on Inform. Theory,} vol. 56, no. 4, pp. 1865-1879, Apr. 2010.

\bibitem{Infor:Shannon}
C.~E.~Shannon, ``A Mathematical Theory of Communication,'' \textit{Bell System Tech. J.}, vol. 27, pp. 623-656, Oct. 1948.

\bibitem{guo}
S. Verdu and D. Guo, ``A simple proof of the entropy power inequality,'' \textit{IEEE Trans. Inform. Theory,} vol. 52, no. 5, pp. 2165-2166, May 2006.

\bibitem{Rate:Oohama}
Y. Oohama, ``The rate-distortion function for the quadratic Gaussian CEO problem,'' \textit{IEEE Trans. Inform. Theory}, vol. 44, no. 3, pp. 1057 - 1070, May 1998.

\bibitem{LargeCRB:Stoica}
P. Stoica and P. Babu, ``The Gaussian Data Assumption Leads to the Largest Cram\'{e}r-Rao Bound,'' \textit{IEEE Signal Process. Mag.}, vol. 28, no. 3, pp. 132-133, May 2011.

\bibitem{TrainingSeq:Stoica}
P. Stoica and O. Besson, ``Training Sequence Design for Frequency Offset and Frequency-Selective Channel Estimation,'' \textit{IEEE Trans. Commun.}, vol. 51, no. 11, pp. 1910-1917, Nov. 2003.

\bibitem{park:ieeespm}
S. Park, E. Serpedin, and K. Qaraqe, ``Gaussian Assumption: The Least Favorable but the Most Useful," \textit{IEEE Signal Process. Mag.}, vol. 30, no. 3, pp. 183-186, May 2013.

\bibitem{esbook}
E. Serpedin and Q. Chaudhari,  \textit{Synchronization of Wireless Sensor Networks:
 Parameter Estimation, Performance Benchmarks and Protocols,} Cambridge University Press, August 2009.

\bibitem{BroadCast:Korner}
K. Marton, ``A Coding Theorem for the Discrete Memoryless Broadcast Channel,'' \textit{IEEE Trans. Inform. Theory}, vol. 25, no. 3, pp. 306 ?311, May 1979.

\bibitem{costa}
M. H. M. Costa, ``A new entropy power inequality,'' \textit{IEEE Trans. Inform. Theory,} vol. 31, no. 6, pp. 751-760, Nov. 1985.

\bibitem{palomar3}
M. Payaro, M. Gregori, and D. Palomar, ``Yet Another Power Entropy Inequality with an Application," \textit{2011 International Conference on Wireless Communications and Signal Processing (WCSP)}, Nanjing, China, Nov. 2011, pp. 1-5.

\bibitem{palomar4}
M. Payaro and D. Palomar, ``A Multivariate Generalization of Costa's Entropy Power Inequality," \textit{IEEE International Symposium in Information Theory 2008 (ISIT 2008)}, Toronto, Canada, Jul. 2008, pp. 1088 - 1092.

\bibitem{Cheong}
S. K. Leung-Yan-Cheong and M. E. Hellman, ``The Gaussian wire-tap channel,'' \textit{IEEE Trans. Inform. Theory}, vol. 24, no. 4, pp. 451 - 456, Jul. 1978.
\end{thebibliography}
\end{document}